\documentclass[11pt,a4paper]{article}

\usepackage[T1]{fontenc}
\usepackage[utf8]{inputenc}
\usepackage{lmodern}
\usepackage[margin=1in]{geometry}
\AtBeginDocument{%
  \DeclareFontShape{T1}{lmr}{m}{scit}{<->ssub*lmr/m/scsl}{}%
  \DeclareFontShape{T1}{lmr}{bx}{sc}{<->ssub*lmr/m/sc}{}%
}
\usepackage{amsmath}
\usepackage{mathtools}
\usepackage{amsfonts}
\usepackage{amssymb}
\usepackage{amsthm}
\usepackage{graphicx}
\usepackage{caption}
\usepackage[ruled,vlined]{algorithm2e}
\usepackage{tikz}
\usetikzlibrary{arrows.meta,calc,positioning}
\usepackage{xcolor}
\usepackage{paralist}
\usepackage{authblk}
\usepackage[numbers,sort&compress]{natbib}
\usepackage{xurl}
\usepackage[hidelinks]{hyperref}
\usepackage{cleveref}

\theoremstyle{definition}
\newtheorem{definition}{Definition}[section]
\definecolor{revieworange}{RGB}{190,85,0}
\definecolor{topofull}{RGB}{171,201,234}
\definecolor{topomixed}{RGB}{227,238,251}
\definecolor{toporegular}{RGB}{177,211,145}
\definecolor{topoambiguous}{RGB}{193,108,42}
\definecolor{topomerge}{RGB}{178,55,48}
\definecolor{topointerface}{RGB}{45,45,45}

\tikzset{
  topo grid/.style={draw=black!75, line width=0.45pt},
  topo center/.style={draw=black, line width=0.95pt},
  topo iface/.style={draw=topointerface, line width=1.1pt, line cap=round},
  topo orient/.style={topo iface, -{Latex[length=2.5mm,width=1.6mm]}},
  topo merge/.style={draw=topomerge, line width=1.15pt},
  topo candidate/.style={draw=black!45, dashed, line width=0.55pt},
  topo resolved/.style={draw=black, line width=0.9pt, -{Latex[length=2.3mm,width=1.5mm]}},
  topo graph node/.style={
    draw=black!80,
    rounded corners=1.3pt,
    minimum width=5.4mm,
    minimum height=5.4mm,
    inner sep=0pt,
    font=\scriptsize
  },
  topo title/.style={font=\scriptsize\bfseries, align=center},
  topo tiny/.style={font=\scriptsize},
}

\definecolor{algpolyfill}{RGB}{231,238,246}
\definecolor{algpolyedge}{RGB}{95,103,115}
\definecolor{algclipfill}{RGB}{241,96,89}
\definecolor{algaccent}{RGB}{49,82,136}
\definecolor{algdark}{RGB}{35,35,35}
\definecolor{algsoft}{RGB}{178,186,196}

\tikzset{
  alg poly/.style={draw=algpolyedge, fill=algpolyfill, line join=round, line width=0.8pt},
  alg clip/.style={draw=none, fill=algclipfill, fill opacity=0.92},
  alg core/.style={draw=none, fill=algaccent, fill opacity=0.58},
  alg cap/.style={draw=none, fill=cyan!40!white, fill opacity=0.9},
  alg wedge/.style={draw=none, fill=algpolyedge!42, fill opacity=0.55},
  alg iface/.style={draw=algdark, line width=1.15pt, line cap=round},
  alg iface dashed/.style={draw=algdark, line width=1.0pt, line cap=round, dashed},
  alg circle/.style={draw=algsoft, line width=0.85pt},
  alg guide/.style={draw=algsoft, line width=0.7pt, dashed},
  alg point/.style={circle, fill=algdark, inner sep=1.2pt},
  alg vertex/.style={circle, fill=white, draw=algdark, line width=0.5pt, inner sep=0.9pt},
  alg cusp/.style={circle, fill=algclipfill, draw=white, line width=0.4pt, inner sep=1.7pt},
  alg center/.style={circle, fill=algaccent, draw=white, line width=0.4pt, inner sep=1.4pt},
  alg label/.style={font=\scriptsize},
  alg tiny/.style={font=\scriptsize},
}

\hypersetup{
  pdftitle={Feature-preserving interface reconstruction using piecewise circles and cusps},
  pdfauthor={Jerry Liu, Kenneth Weiss, Jin Yao},
  pdfsubject={Preprint},
  pdfkeywords={Volume-of-Fluid, interface reconstruction, circular facets, topological consistency}
}

\title{Feature-preserving interface reconstruction\\using piecewise circles and cusps}
\author[1]{Jerry Liu\thanks{Corresponding author: \href{mailto:jerrywliu@stanford.edu}{\texttt{jerrywliu@stanford.edu}}.\protect\newline
Code released publicly at \url{https://github.com/jerrywliu/interface-reconstruction}.}}
\author[2]{Kenneth Weiss}
\author[2]{Jin Yao}
\affil[1]{Institute for Computational and Mathematical Engineering, Stanford University}
\affil[2]{Lawrence Livermore National Laboratory}

\date{\small\textit{Preprint}\\[3pt]October 28, 2026}

\begin{document}
\maketitle

\begin{abstract}
We present a geometric interface reconstruction method for Volume-of-Fluid (VOF) simulations that captures both curved interfaces and sharp features using only volume fraction data.
We represent interfaces with linear facets, circular arcs, and corner facets formed by two lines or arcs joined within a single cell.
We first propagate local orientation constraints through a graph and merge cells to resolve ambiguities, treating topological consistency as a first-class constraint.
An optional conservative refinement improves $C^0$ continuity while preserving local volume constraints.
Users can independently enable circular facets, corner facets, cell merging, and continuity refinement to balance accuracy and computational cost.
In its simplest configuration, our method provides a fast local line-fitting scheme competitive with state-of-the-art PLIC methods.
On standard static benchmarks, the full configuration reconstructs identifiable lines, circles, and sufficiently resolved corners to numerical precision on coarse, perturbed Cartesian meshes.
On varying-curvature interfaces, it achieves third-order convergence in facet-gap error and approximately first-order convergence in curvature error.
These results demonstrate global volume conservation and accurate reconstruction of curved and corner features.

\end{abstract}


\section{Introduction}

Accurately capturing material-interface geometry is fundamental to simulations of multiphase flow. Representative applications include free-surface and interfacial flows~\citep{Hirt1981,Scardovelli1999,Tryggvason2001} and multimaterial hydrodynamics~\citep{Benson1992}. Interfaces in these settings can develop sharp geometric features, whose resolution is important for preserving volume and interface topology as the geometry evolves~\citep{Rider1998,Garimella2005}.

The Volume-of-Fluid (VOF) method remains a widely adopted interface-capturing approach due to its strict mass conservation and ability to naturally handle topology changes~\citep{Hirt1981,Rider1998,Scardovelli1999}. In VOF, the interface is represented implicitly through a field of volume fractions over a computational mesh. However, reconstructing a geometric interface from this field is an underdetermined problem: many distinct interface geometries can yield the same volume fractions~\citep{Huettenberger2013Topological}. As a result, \emph{interface reconstruction}, the task of estimating the interface within each mixed cell given volume fraction data, requires additional assumptions about the interface's local shape and orientation. Ideally, a robust reconstruction scheme must be accurate, topologically consistent, and adaptable to complex geometries.

Piecewise linear interface construction (PLIC) methods include Youngs' method~\citep{Youngs1987,ParkerYoungs1992}, the least-squares LVIRA formulation~\citep{Puckett1991,Puckett1997}, and its efficient Cartesian specialization ELVIRA~\citep{Pilliod1992,Pilliod2004}. These methods represent the interface by a straight segment, or \emph{linear facet}, in each mixed cell, with its orientation inferred from a local volume-fraction stencil. The strongest classical variants attain second-order accuracy for smooth interfaces on Cartesian meshes~\citep{Pilliod2004}. However, a single linear facet cannot represent curvature or a corner within a cell, and independently reconstructed facets need not meet continuously or assign shared mesh vertices to the same phase~\citep{Garimella2005}.

Higher-order VOF reconstructions have introduced parabolic, circular, and split facets to improve smooth-interface geometry and continuity~\citep{Renardy2002,Evrard2017,Remmerswaal2021ParabolicIR,Maity2020,Maity2024PiecewiseCI,VANDEREIJK2024112656}. These methods differ in the information and intermediate geometry they require: some refine PLIC-derived geometry, while others use precomputed normals, height functions, or coupled level-set data~\citep{Evrard2017,Maity2020,Remmerswaal2021ParabolicIR,Maity2024PiecewiseCI,Chen2024ACI}. To our knowledge, no prior framework directly reconstructs line, circular, and corner primitives from local zeroth moments while treating topology identification as a first-order constraint.

To address these challenges, we introduce a new geometric interface reconstruction method that represents the interface using piecewise circular arcs and corner facets. Unlike prior work that augments linear reconstructions with higher-order fitting, our method is designed from the ground up to operate with high-order primitives. We formulate interface reconstruction as a local geometric optimization problem, using only volume fractions to match interface geometry across neighboring cells (\Cref{sec:method_overview}).

A central design principle of our approach is to identify a consistent local interface walk before fitting geometry.
We identify a local interface orientation from the volume fraction pattern in a local stencil (\Cref{fig:regular_cases}), and use a cell merging procedure to resolve additional ambiguities and enforce topological consistency of the reconstructed material body.
The topology identification procedure and experiments in this paper focus on perturbed Cartesian 2D meshes, though in principle the ideas can be extended to general polygonal meshes.
Once we have determined an interface orientation, we use three ordered adjacent polygonal support cells to fit each interface facet.
Finally, we apply an optional conservative post-processing correction to improve $C^0$ continuity by adjusting shared edge crossings and neighboring facet curvatures while retaining the local volume constraints.

Our \emph{modular} framework consists of four components, each of which can be enabled or disabled independently, allowing users to trade off computational cost and complexity of the representation for accuracy and robustness:
\begin{itemize}
    \item \textbf{Merging.} We merge neighboring cells to resolve ambiguous interface orientations in under-resolved regions (\Cref{sec:topology-identification}).
    \item \textbf{Circular facets.} We reconstruct interfaces using circular arcs to accurately capture local curvature (\Cref{sec:fit-circular}). We note that our method recovers linear representations as a special case via arcs with zero curvature.
    \item \textbf{Corner facets.} We detect and represent sharp cusps by fitting two (linear or circular) facets, joined at a shared point, within a single cell (\Cref{sec:fit-linear-corner,sec:fit-curved-corner}).
    \item \textbf{$C^0$ enforcement.} We iteratively refine shared edge crossings from adjacent facets along with the neighboring facet curvatures under local volume constraints, enforcing a conservative $C^0$-continuous interface when admissible (\Cref{sec:c0}).
\end{itemize}

In the simplest configuration, our algorithm uses local linear facets and omits circles, corners, and $C^0$ enforcement, matching the performance of prior PLIC methods. At the other end of the spectrum, our algorithm combines merging, line and arc fitting, and explicit corners to resolve regions with high interfacial curvature or sharp features. All configurations are volume conservative and use only cell-wise volume fractions and local mesh geometry.

In summary, our work makes four main contributions:
\begin{itemize}
    \item We formulate a geometric reconstruction framework that directly fits line, circular, and corner primitives using only zeroth-moment VOF data within a local neighborhood. Our approach has \emph{perfect reconstruction} for locally linear and circular interfaces with cusps.
    \item We separate topology identification from facet fitting, allowing \emph{topological consistency} to be enforced as a first-order constraint.
    \item We implement our method as a modular family of reconstructions, where merging, circle-fitting, corner-fitting, and $C^0$ enforcement can each be enabled or disabled independently. This allows our framework to transition between simpler PLIC-like operations and higher-fidelity reconstructions.
    \item We systematically ablate the effects of each module on static benchmarks, demonstrating third-order convergence in facet-gap error and first-order convergence in curvature error on varying-curvature interfaces.
\end{itemize}

The remainder of the paper is organized as follows.
	\Cref{sec:background} reviews prior work and defines our two motivating desiderata for interface reconstruction: \emph{perfect reconstruction} (\Cref{def:perfect_reconstruction}) and \emph{topological consistency} (\Cref{def:topological_consistency}).
	\Cref{sec:method_overview} provides an overview of our reconstruction algorithm, which directly addresses both desiderata through topology identification (\Cref{sec:topology-identification}) and facet fitting (\Cref{sec:facet-fitting}).
	\Cref{sec:results} evaluates the method on static tests and we conclude with a discussion of limitations and directions for future work.

\section{Background}
\label{sec:background}

\subsection{Problem Setup}
\label{subsec:problem_setup}

Given only volume fraction data on a polygonal mesh, an interface reconstruction approach attempts to recover a curve that represents the interfaces between immiscible fluid phases. We focus on the two-dimensional, two-phase setting.
Let $D \subset \mathbb{R}^2$ denote the spatial domain, partitioned into a collection of convex polygonal cells $\{P_i\}_{i=1}^n$.
At any given time, the domain is divided into two disjoint fluid regions: $\Omega_A \subset D$ occupied by fluid A, and $\Omega_B = D \setminus \Omega_A$ occupied by fluid B.
We focus on reconstructing the interface $\partial \Omega_A$, and assume without loss of generality that the volume fractions $v_i \in [0,1]$ represent the fraction of cell $P_i$ occupied by fluid A:
\[
v_i = \frac{|\Omega_A \cap P_i|}{|P_i|}.
\]

Our goal is to reconstruct an approximation of the interface $\partial \Omega_A$ using only the volume fraction data and the geometry of the mesh.
For simplicity, we drop the phase subscript and refer to cells with $v_i=1$ as ``full'' and cells with $v_i=0$ as ``empty''.
In each \emph{mixed} cell (i.e., a cell with $0 < v_i < 1$), we aim to construct a geometric \emph{facet} $f_i \subset P_i$ (i.e., a portion of our interface representation within a mesh cell) that partitions the cell in a way that is consistent with the prescribed volume fraction. Let $V(f_i)$ denote the volume fraction induced by the facet $f_i$, and let $V(f) = \{V(f_i)\}$ denote the collection of volume fractions across the mesh. An \emph{interface reconstruction method} then defines a mapping
\[
\mathrm{IR} : (\{v_i\}, \{P_i\}) \mapsto \{f_i\},
\]
where the output $\{f_i\}$ should collectively approximate the true interface.

Since many different configurations of $\{f_i\}$ may induce the same volume fractions $\{v_i\}$, particularly on coarse meshes, the reconstruction problem is underdetermined. A robust reconstruction method must therefore introduce additional structure, such as geometric priors or topological constraints, to resolve this ambiguity in a principled way.

\subsection{Desiderata}
\label{subsec:desiderata}

Our interface reconstruction method is motivated by two important properties of interest to visualization and analysis: \emph{perfect reconstruction} and \emph{topological consistency}.

\paragraph{Perfect reconstruction}

Intuitively, an interface reconstruction method has the perfect reconstruction property when it can \emph{exactly} recover any interface defined by its represented primitives. We formalize this as follows:

Let $\mathcal{F}$ be a class of admissible interface primitives (e.g., line segments, circular arcs), and let $\{P_i\}_{i=1}^n$ be a collection of convex polygonal cells covering the domain $D \subset \mathbb{R}^2$. Suppose $\Omega \subset D$ is a fluid region whose boundary satisfies
\[
\partial \Omega = \bigcup_{i=1}^n f_i, \quad \text{where } f_i := \partial \Omega \cap P_i \in \mathcal{F},
\]
and assume the volume fractions $v_i \in [0,1]$ are given by
\[
v_i = \frac{|\Omega \cap P_i|}{|P_i|}.
\]

\begin{definition}[Perfect reconstruction]
\label{def:perfect_reconstruction}
Let $\mathcal{F}$ be a class of interface facet primitives. An interface reconstruction method
\[
\mathrm{IR}_{\mathcal{F}} : [0,1]^n \times \{P_i\} \to \mathcal{F}^n
\]
is said to satisfy the \emph{perfect reconstruction} property on an admissible instance if the volume fractions, mesh geometry, and assumed interface topology uniquely identify facets $f_i := \partial \Omega \cap P_i \in \mathcal{F}$, and the method recovers that identified boundary:
\[
\mathrm{IR}_{\mathcal{F}}(\{v_i\}, \{P_i\}) = \{f_i\}.
\]
\end{definition}

\begin{figure}[t]
  \centering
  \includegraphics[width=0.98\textwidth]{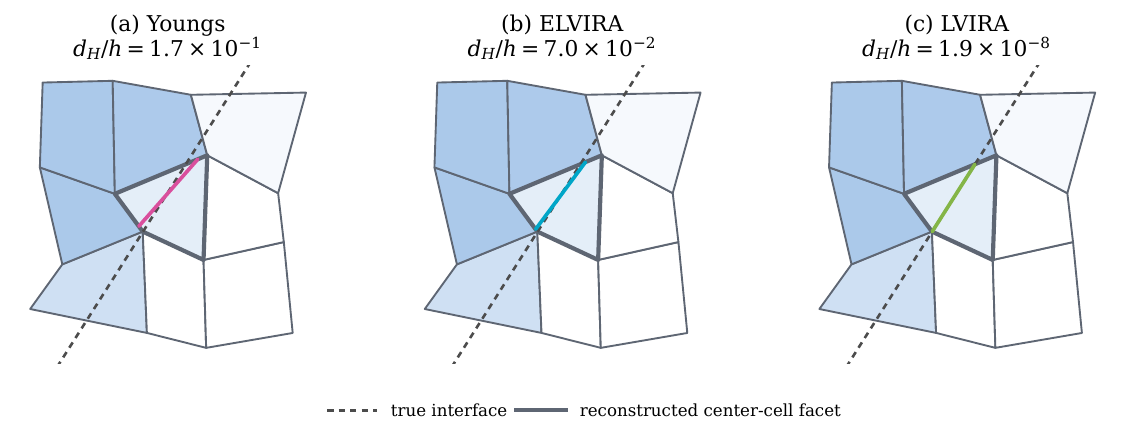}
  \caption{\textbf{Perfect reconstruction of a linear interface from zeroth moments on a perturbed Cartesian grid.} Cell shading encodes the prescribed volume fraction, the dashed black line is the generating interface, and the colored solid segment is the reconstruction in the outlined center cell. All three methods use information from the displayed $3\times3$ stencil. Youngs' method (a) and ELVIRA (b) conserve the center-cell volume but select incorrect normals on perturbed grids. LVIRA's least-squares stencil optimization (c) recovers the generating line to numerical tolerance and thus achieves \emph{perfect reconstruction} of linear interfaces. Here $d_H$ is the center cell's Hausdorff error and $h$ is the center cell's width.}
  \label{fig:perfect_reconstruction_plic}
\end{figure}

For classical PLIC methods, the admissible facet family is the set of piecewise linear interfaces. \Cref{fig:perfect_reconstruction_plic} isolates this requirement on a perturbed Cartesian stencil: while Youngs' method and ELVIRA both reproduce the prescribed center-cell volume, only LVIRA recovers the generating line. Our circle and corner fitting stages (\Cref{sec:fit-circular,sec:fit-linear-corner,sec:fit-curved-corner}) are designed around the same \emph{perfect reconstruction} objective by solving directly for the primitives that match the available volumes, while enlarging the admissible facet family beyond straight segments.\footnote{Perfect reconstruction is relative to the supplied information. Zeroth-moment VOF can recover only interfaces identifiable from cell volumes. Moment-of-Fluid methods add higher-moment information and can therefore distinguish richer subcell geometries~\citep{Dyadechko2008Reconstruction,Shashkov2023Moments}.
This work focuses on the zeroth-moment setting.}

\paragraph{Topological consistency}

We next describe the notion of \emph{topological consistency}. Intuitively, this condition guarantees local phase-label agreement by making sure that every mesh vertex is assigned to one phase consistently across all incident cells. The goal is to ensure that the reconstructed interface produces globally coherent fluid regions, without introducing spurious disconnections or overlaps. Following the vertex-based consistency criterion of Garimella et al.~\citep{Garimella2005}, we first define a discrete labeling of mesh vertices induced by the fluid region:

\begin{definition}[Vertex coloring]
\label{def:vertex_coloring}
Let $\Omega \subset D$ be a fluid region whose boundary $\partial \Omega$ is a finite union of pairwise disjoint, closed simple curves, and assume that $\partial \Omega$ does not pass exactly through any mesh vertex.\footnote{This assumption precludes the case where pure cells occupied by opposite materials are edge-adjacent.
This case implies an interface along their shared mesh edge and can be represented explicitly as a special case in our algorithm by adding a facet along that edge.} For a given cell $P_i$ and corresponding interface facet $f_i = \partial \Omega \cap P_i$, oriented so that the interior of $\Omega$ lies to the left of $f_i$, we define a \emph{vertex coloring} function
\[
g_i : \text{vertices}(P_i) \to \{0, 1\}
\]
by setting $g_i(p) = 1$ if $p$ lies on the full side of $f_i$, and $g_i(p) = 0$ otherwise.
\end{definition}

Note that this labeling is well-defined as long as the interface does not intersect any vertex exactly.

\begin{definition}[Topological consistency]
\label{def:topological_consistency}
An interface reconstruction method $\mathrm{IR}_{\mathcal{F}}$ is said to satisfy the \emph{topological consistency} property if, for every vertex $p$ shared by multiple cells $P_{i_1}, \dots, P_{i_k}$, the corresponding vertex colorings all agree:
\[
g_{i_1}(p) = \dots = g_{i_k}(p).
\]
\end{definition}

Classical PLIC methods typically struggle to enforce this property because they assume a linear segment in each cell and therefore often cannot simultaneously satisfy both volume conservation and global topological constraints, particularly in regions of high curvature. Garimella et al.~\cite{Garimella2005} explicitly add a post-processing step to resolve topological inconsistencies. However, their reconstruction still retains one linear segment per mixed cell and therefore cannot represent an intra-cell curved segment or cusp. The same issue can persist in higher-order reconstructions, especially in methods that refine PLIC-derived linear segments and so can inherit their inconsistent topologies.

\begin{table}
\begin{center}
\small
\caption{Comparison of representative VOF interface reconstruction methods. Checkmarks denote capabilities explicitly guaranteed by the cited formulation, and blanks denote the absence of such a guarantee. Perfect reconstruction denotes exact recovery, within the method's stated mesh class, of interfaces that are uniquely identifiable and expressible using its geometric primitive family. The final column indicates whether the reconstruction formulation applies directly to the perturbed Cartesian grids considered here, rather than being restricted to standard Cartesian grids. }
\begin{tabular}{l|c|c|c|c|c}
  \textbf{Methods} & \begin{tabular}{@{}c@{}}\textbf{Geometric} \\ \textbf{primitives}\end{tabular} & \begin{tabular}{@{}c@{}}\textbf{Perfect} \\ \textbf{reconstruction}\end{tabular} & \begin{tabular}{@{}c@{}}\textbf{Topological} \\ \textbf{consistency}\end{tabular} &\textbf{$C^0$} & \begin{tabular}{@{}c@{}}\textbf{Perturbed} \\ \textbf{Cartesian grid}\end{tabular} \\
  \hline Youngs~\cite{Youngs1987,ParkerYoungs1992} & Lines &            & & & \\
  \hline ELVIRA~\cite{Pilliod1992,Pilliod2004} & Lines & \checkmark & & & \\
  \hline LVIRA~\cite{Puckett1991,Puckett1997,Pilliod2004} & Lines & \checkmark & & & \checkmark \\
  \hline BLIC~\cite{VANDEREIJK2024112656} & Up to two lines & & & & \\
  \hline Garimella et al.~\cite{Garimella2005} & Lines & \checkmark & \checkmark & & \checkmark \\
  \hline Evrard et al.~\cite{Evrard2017} & Parabolas & & & & \checkmark \\
  \hline QUASI~\cite{Diwakar2009AQS} & Quadratic splines & & & \checkmark & \\
  \hline PLVIRA~\cite{Remmerswaal2021ParabolicIR} & Parabolas & & & & \\
  \hline PCIC~\cite{Maity2020} & Circular arcs & & & \checkmark & \\
  \hline H-PCIC~\cite{Maity2024PiecewiseCI} & Circular arcs & & & \checkmark & \\
  \hline Ours           & Lines, circles, cusps & \checkmark & \checkmark & \checkmark & \checkmark \\
\end{tabular}
\label{tab:methods_compare}
\end{center}
\end{table}

\subsection{Related Work}
\label{subsec:related_work}

Interface reconstruction methods span a broad spectrum of geometric complexity and information requirements. In the classical VOF setting~\cite{Hirt1981}, the earliest geometric reconstructions were SLIC-type methods~\cite{Noh1976}, which restrict the interface to axis-aligned segments. PLIC methods generalize these via a single line segment per mixed cell~\cite{Rider1998, Scardovelli1999}.

The main distinction across PLIC methods is how each linear facet's orientation is chosen. Youngs' method~\cite{Youngs1987,ParkerYoungs1992} estimates the interface normal from finite differences of the volume fraction field. LVIRA~\cite{Puckett1991,Puckett1997} instead minimizes a local least-squares residual on a $3\times 3$ stencil defined by the cell and its neighbors.
ELVIRA~\cite{Pilliod1992,Pilliod2004} further gives an efficient specialization to Cartesian meshes by reducing that search to a finite candidate set in closed form.
Garimella et al.~\cite{Garimella2005} extend linear reconstruction to general polygonal meshes and explicitly repair inconsistent vertex classifications, while Scheufler and Roenby~\cite{Scheufler2019} iteratively reconstruct a local distance function to improve interface normals and positions on general meshes.

Another line of work seeks to enrich the reconstructed interfaces with higher-order nonlinear primitives (such as parabolic, circular, and quadratic spline representations) in order to improve curvature representation and continuity. However, these higher-order methods do not simultaneously provide all of the desiderata in \Cref{subsec:desiderata} from zeroth moments alone. Several initialize from or enrich a preliminary PLIC reconstruction rather than solving directly for a nonlinear primitive from local volume fractions. QUASI progressively converts a Youngs' PLIC interface into a quadratic spline~\cite{Diwakar2009AQS}, PCIC fits piecewise circular arcs to points derived from preliminary PLIC segments~\cite{Maity2020}, and the BLIC method of {van der Eijk} and Wellens builds upon an initial PLIC segment~\cite{VANDEREIJK2024112656} by creating up to two connected linear segments within a cell. Other methods supply geometric information separately from the nonlinear fit: Evrard et al.~\cite{Evrard2017} solve for parabolas in a frame chosen from a precomputed interface normal, PLVIRA and H-PCIC use height-function information~\cite{Remmerswaal2021ParabolicIR,Maity2024PiecewiseCI}, and Chen et al.~\cite{Chen2024ACI} use coupled level-set data. These dependencies can improve accuracy on varying-curvature interfaces, but they do not directly target exact recovery of an identifiable nonlinear primitive from zeroth moments. Furthermore, without a separate topology-identification stage, they do not guarantee consistent phase labels at shared vertices. To our knowledge, our method is the first zeroth-moment VOF reconstruction framework that directly targets exact recovery of linear and circular primitives, with cusps where appropriate, while enforcing topological consistency as a first-order constraint.

Moment-of-Fluid (MoF) methods are a separate line of work that seeks to produce higher-fidelity interfaces by augmenting volume fractions with additional data, such as centroids or other higher-order moments. These methods can reconstruct substantially more complex subcell geometry than classical VOF alone~\cite{Dyadechko2008Reconstruction,Shashkov2023Moments,Spainhour2025MultiplaneMOF}, but require computing and storing higher-order moments.

Our method remains in the minimal VOF setting and uses higher-order linear and circular facets with cusps to expand the admissible local geometry beyond linear representations. However, unlike methods that first estimate a line and then refine it, ours \emph{directly} solves for nonlinear primitives from local volume constraints. We use circles because in two dimensions they are coordinate-free three-parameter primitives with constant curvature, whereas a general parabola requires an additional axis or frame choice. By combining lines and circular arcs with explicit corners, our method also targets sharp features that smooth high-order reconstructions struggle to represent.
Table~\ref{tab:methods_compare} compares representative VOF approaches against our perfect reconstruction and topological consistency desiderata.

\section{Interface Reconstruction Framework}
\label{sec:method_overview}

In this section, we describe the high-level structure of our 2D interface reconstruction method. The method uses only volume-fraction data and combines geometric fitting over neighboring cells with graph-based propagation of local orientation constraints to target \emph{perfect reconstruction} and \emph{topological consistency} (as defined in~\Cref{subsec:desiderata}). Its geometric stages are modular, allowing line, circle, corner, merging, and continuity components to be selected according to the desired accuracy and cost.

We achieve our desiderata by decomposing the interface reconstruction into two tractable subproblems:

\begin{itemize}
    \item \textbf{Identify interface topology.} First, for every mixed cell in the mesh, we determine the topology of the nearby interface elements and establish the local orientation of the interface (\Cref{sec:topology-identification}).
    We note that for topologically Cartesian meshes with sufficient resolution relative to the interface's curvature, the local interface topologies reduce to three distinct cases that our algorithm detects.
    In rare cases without sufficient mesh resolution, ambiguities may remain. In these cases, we propose an optional \emph{cell-merging} phase, which trades off local interface resolution for resolving topological constraints.

    \item \textbf{Fit interface facets.} Next, for each mixed cell, we fit a geometric primitive to match its local volume fractions and interface topology (\Cref{sec:facet-fitting}).
    We focus on three geometric primitives: \emph{line segments}, \emph{circular arcs}, and \emph{corners}, which consist of two lines or circular arcs within a single cell that intersect at a cusp.
    We augment PLIC schemes with circles and corners to be able to represent interfaces with curvature and sharp discontinuities, respectively.
    For cells with identified interface topology, we use an oriented triplet of adjacent mixed cells to fit a facet.
    In rare cases when a cell remains unfit, our method uses a volume-conserving linear fallback (\Cref{sec:failsafe}).
    After fitting, an optional conservative post-processing correction iteratively refines adjacent facet endpoints and curvatures under local volume constraints, improving $C^0$ continuity when an admissible refinement is found (\Cref{sec:c0}).
\end{itemize}

This structure generalizes prior PLIC schemes and allows a smooth progression between simpler, low-order methods and high-fidelity, third-order reconstructions.
In our experimental results (\Cref{sec:results}), we show that each component of our reconstruction algorithm contributes to the overall performance. 
Merging resolves orientation ambiguities when the local VOF data are too under-resolved to support a unique topology,
circular facets add the extra degree of freedom needed to capture high curvature regions,
corner facets resolve discontinuities that no single smooth primitive can capture, 
and the optional $C^0$ correction improves continuity while preserving local volume constraints.
Our linear fallbacks provide a volume-conserving reconstruction in rare cases where the topology or geometry cannot be fit.
Algorithm~\ref{alg:full} summarizes our reconstruction framework and its optional stages.

\begin{algorithm}[tbp]
\DontPrintSemicolon
\caption{Feature-preserving interface reconstruction.}
\label{alg:full}
\KwIn{Mesh cells $\{P_i\}$, volume fractions $\{v_i\}$, topology variant: $\{$\textsc{PerCell} or \textsc{GraphCoordinated}$\}$, \\
  module toggles: $\{$\textsc{Corners}, \textsc{Circles}, \textsc{$C^0$}$\}$ }
\KwOut{Facets $\{f_i\}$ forming the reconstructed interface}

\BlankLine
\textbf{1. Topology stage (Section~\ref{sec:topology-identification})}\;
$\mathcal{M} \leftarrow \{i : 0 < v_i < 1\}$\;
Construct edge-adjacency among $\mathcal{M}$\;
\eIf{the topology variant is \textsc{PerCell}}{
    Orient each regular mixed cell independently using its vertex coloring\;
    Mark non-regular mixed cells for PLIC fallback\;
}{
    Selectively merge compatible single-neighbor mixed-cell pairs\;
    Identify regular components and orient each $(P_L,P_R)$ by vertex coloring\;
    Build the neighbor-candidate graph and initialize a queue of ambiguous components\;
    \While{the queue makes progress}{
        Propagate reciprocal left/right assignments and prune incompatible candidates\;
        Use the local full/empty pattern whenever it determines a regular orientation\;
        Selectively merge compatible adjacent components when propagation leaves ambiguity\;
        Re-run local propagation after an orientation or merge event\;
    }
    Split any still-unresolved component into base cells, attach centered $3\times3$ stencils, and mark them for PLIC fallback\;
}

\BlankLine
\textbf{2. Facet-fitting stage (Section~\ref{sec:facet-fitting})}\;
For each oriented component, accept a \textbf{line} when its two-neighbor support matches the target-cell volume\;
\If{\textsc{Corners} enabled}{
    Test unfit cells for \textbf{line--line corners} between accepted supports\;
}
\If{\textsc{Circles} enabled}{
    Fit \textbf{circular arcs} in the remaining oriented cells\;
}
\If{\textsc{Corners} and \textsc{Circles} enabled}{
    Test remaining/transition cells for \textbf{line--arc or arc--arc corners}\;
}
\ForEach{cell still lacking a facet}{
    \eIf{the cell has an oriented pair $(P_L,P_R)$}{
        Fit a \emph{linear segment}\;
    }{
        Apply the LVIRA fallback on the centered $3\times3$ stencil\;
    }
}

\BlankLine
\textbf{3. Conservative $C^0$ post-processing correction (Section~\ref{sec:c0})}\;
\If{\textsc{$C^0$} enabled}{
    Iteratively refine neighboring facet endpoints and curvatures under local volume constraints\;
    Retain admissible $C^0$-continuous refinements; otherwise preserve the original facets\;
}

\BlankLine
\Return reconstructed facets $\{f_i\}$\;

\end{algorithm}

\section{Identifying Interface Topology}
\label{sec:topology-identification}

We now describe the topology identification stage of the reconstruction pipeline. For clarity, we specialize our discussion to two-material flows over perturbed Cartesian meshes in 2D, though we note that the methods can be naturally extended to polygonal meshes. 

The objective is to determine, for each mixed cell $P$, a consistent orientation and an ordered pair of neighboring mixed cells $(P_L,P_R)$ so that traversing ``to the right'' along facets induces a full walk around the boundary of the fluid's connected component. Our key insight is that establishing this orientation converts otherwise independent volume constraints during the facet fitting phase into a series of topologically consistent optimization problems. As a result, our topology identification stage coordinates reconstructions across adjacent cells; provides enough degrees of freedom to select among linear, circular arc, and corner primitives; and aligns neighboring facets for the optional $C^0$ correction. Assuming sufficient resolution, the resulting interface orientation also enables a uniquely solvable geometric fit in~\Cref{sec:facet-fitting}.

In this section, we propose two alternative methods for identifying interface topology. Both are assembled from local cell--neighbor operations. The \emph{per-cell} variant classifies each target cell independently, without cross-cell propagation or merging. The \emph{graph-coordinated} variant iteratively propagates orientation constraints between adjacent cells and selectively merges compatible adjacent cells to resolve topological ambiguities. Our per-cell variant admits a fully cell-parallel algorithm, but does not fully leverage all the local information provided by the volume fractions, while our graph-coordinated variant can resolve more under-resolved topologies but may require multiple passes.

\subsection{Assumptions}
We assume the true interface is a finite union of pairwise disjoint, closed simple curves. Furthermore, we assume that the mesh is sufficiently resolved so that, in typical mixed cells, the interface intersects the cell boundary in exactly two edges. Under these conditions, most mixed cells are \emph{regular}: they have exactly two edge-adjacent mixed neighbors, and the remaining edge-adjacent neighbors satisfy one of three consistent Cartesian full/empty patterns (see~\Cref{fig:regular_cases}).\footnote{Even with sufficiently fine meshing, there are cases where mixed cells have fewer or more than two edge-adjacent mixed neighbors. Our merging procedure in~\Cref{sec:neighbor-graph-heuristic} and PLIC fallback are used to handle these non-regular cases.}
For such cells, the local orientation $(P_L,P_R)$ is determined unambiguously by the induced vertex coloring (\Cref{def:vertex_coloring}).

\subsection{Identifying Interface Topology for Regular Cells}

\begin{figure}[t]
  \centering
  \resizebox{0.78\textwidth}{!}{\includegraphics{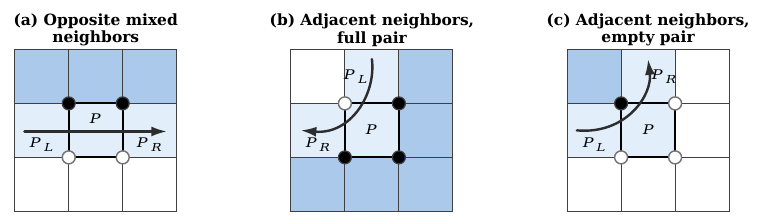}}
  \caption{\textbf{The three regular mixed-cell topologies on a Cartesian grid, up to rotation and reflection.} In each case, the central cell $P$ has exactly two mixed neighbors $(P_L,P_R)$. Dark blue cells are full, light blue cells are mixed, and white cells are empty. Black and white corner markers indicate the induced vertex coloring, and the black interface sketch shows the interface's local topology that is compatible with the vertex coloring.
	 Diagram code prepared with assistance from OpenAI Codex.}
  \label{fig:regular_cases}

\end{figure}

We start by labeling each mesh cell $P$ as empty, mixed, or full according to whether $v_P=0$, $0<v_P<1$, or $v_P=1$, respectively.
Our goal is to establish a vertex coloring, in the sense of \Cref{def:vertex_coloring}, that is mutually consistent with the full/mixed/empty cell labels. On a Cartesian grid, these constraints reduce to one of three colorings shown in \Cref{fig:regular_cases}, up to rotation and reflection. The two mesh edges shared with the resulting neighbors $P_L$ and $P_R$ are then the candidate interface-crossing edges.

\begin{samepage}
Concretely, our procedure for identifying regular cells involves three checks:
\begin{itemize}
  \item \textbf{Two mixed neighbors:} the target cell has exactly two edge-adjacent mixed neighbors.
  \item \textbf{Compatible vertex coloring:} the local full/mixed/empty pattern admits one of the three consistent colorings in \Cref{fig:regular_cases}.
  \item \textbf{Consistent orientation:} we order the two mixed neighbors as $(P_L,P_R)$ so that full cells lie to the left of the directed walk $P_L \to P \to P_R$, and empty cells lie to the right.
\end{itemize}
\end{samepage}

\subsection{Per-Cell Orientation Without Merging}
The per-cell variant identifies local interface topology for all regular mixed cells in parallel via the rule above:
\begin{enumerate}
  \item Detect and orient every regular mixed cell and record its $(P_L,P_R)$.
  \item For any mixed cell that is not regular, defer to a robust PLIC fallback for that cell during facet fitting.
\end{enumerate}
Each orientation decision depends only on the target cell and its immediate neighbors. This makes the variant parallel and inexpensive, but it may underutilize available information when a few cells break regularity because of local under-resolution. We find that under-resolved cells occur rarely, and primarily only near regions with high curvature or sharp cusps.

\subsection{Graph-Coordinated Orientation with Cell Merging}

The graph-coordinated variant uses the same regular-cell orientation rule, but further identifies topologies in under-resolved regions by propagating reciprocal orientation constraints between adjacent cells and selectively merging cells.

\begin{figure}[t]
  \centering
  \resizebox{0.72\textwidth}{!}{\includegraphics{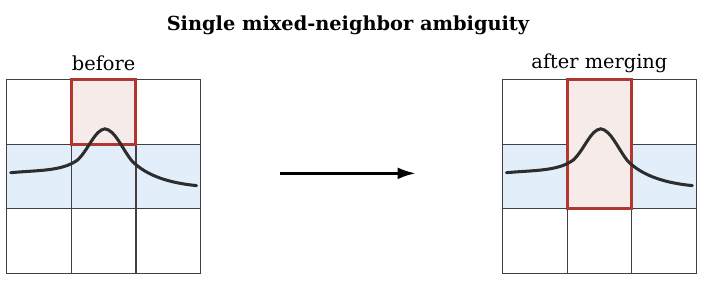}}
  \caption{\textbf{An example of cell merging.} Before merging, the highlighted mixed cell has only one mixed neighbor. Their union forms a single merged cell with two exterior mixed neighbors and a regular local interface walk. The black curve represents the generating interface. Diagram code prepared with assistance from OpenAI Codex.}
  \label{fig:merging_ambiguous_cases}
\end{figure}

Merging coalesces the volumes and adjacencies of the participating cells, trading a small amount of local geometric expressivity for a simpler interface walk while preserving the total phase volume. The simplest motivating case is a cell with a single mixed neighbor (see~\Cref{fig:merging_ambiguous_cases}). The interface may enter and exit through the same side of this cell, as near an under-resolved cusp. Merging it with its sole mixed neighbor forms a larger regular cell with two mixed neighbors.

Other ambiguities may occur in mixed cells with more than two candidate neighbors. In these cases, we use a neighbor-candidate graph to guide our merging strategy, as described next. An example of such an ambiguity resolved by merging is shown and discussed in \Cref{fig:appendix_actual_multineighbor_merge}.

\subsubsection{Neighbor-candidate graph and greedy merging}
\label{sec:neighbor-graph-heuristic}
There are multiple possible ways to decide which cells to merge when ambiguities arise. We describe a simple, modular heuristic based on a \emph{neighbor-candidate graph}.

\paragraph{Neighbor-candidate graph}
Let $G=(\mathcal{V},\mathcal{E})$ denote the neighbor-candidate graph. $G$ is a dual graph of our computational mesh, where a node $u\in\mathcal{V}$ represents a reconstruction cell (either a mixed mesh cell or a merged union of mixed cells) and an edge $(u,w)\in\mathcal{E}$ connects adjacent reconstruction cells whose shared mesh edge may be crossed by the reconstructed interface. We call the set of graph edges incident to $u$ its \emph{candidate edge set}.

\begin{figure}[t]
  \centering
  \resizebox{0.94\textwidth}{!}{\includegraphics{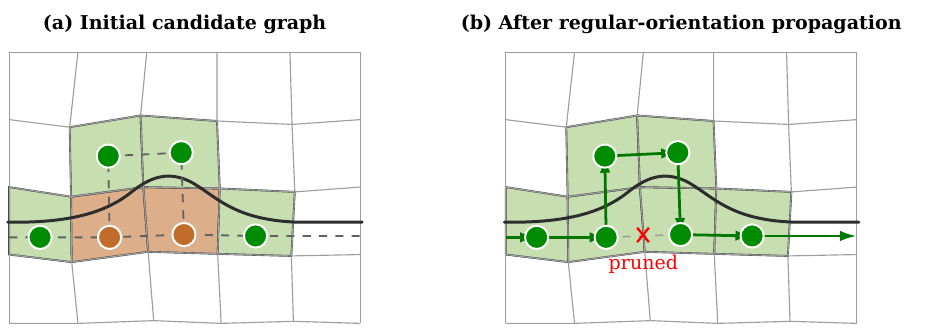}}
  \caption{\textbf{Neighbor-candidate graph overlaid on a perturbed Cartesian mesh.} Thin gray lines are mesh edges, green and orange cells are regular and initially ambiguous cells, respectively, and the black curve is the generating interface. Circular graph nodes and dashed edges denote mixed cells and their candidate interface continuations within the neighbor-candidate graph.
  (a) The two orange cells admit both a direct continuation and a path through the two upper regular cells.
  (b) Green arrows show the interface walk selected by regular-orientation propagation. The direct orange--orange edge is pruned because the upper path already gives each orange node its two reciprocal interface neighbors. Diagram code prepared with assistance from OpenAI Codex.}
  \label{fig:orientation_dependencies}
\end{figure}

Figure~\ref{fig:orientation_dependencies} depicts this graph over a perturbed Cartesian mesh. Note that regular reconstruction cells correspond to nodes of degree two. Any other degree indicates a locally ambiguous interface continuation.
In this example, propagation of constraints from the regular cells in green fixes two reciprocal neighbor connections per orange cell.
Since this already fixes the degree-two interface walk, it allows us to prune the direct edge between the orange cells.

Our algorithm repeatedly propagates local constraints as follows:
\begin{itemize}
  \item \emph{Pruning:} once a cell is oriented with respect to the interface, its reciprocal left/right assignments remove incompatible candidate edges from adjacent cells. The oriented cell remains part of the interface walk and the reduced candidate edge sets may be used to determine nearby orientations.
  \item \emph{Merging:} when a local ambiguity cannot be resolved by propagation, a pair of compatible adjacent cells may be replaced by a single merged cell.
  This operation contracts the pair of nodes in the neighbor-candidate graph and takes the union of their edges.
  \item \emph{Termination and fallback:} propagation and local merging continue while they produce new information. If a complete pass makes no further progress, resolution of the remaining ambiguous cells is deferred.
\end{itemize}

\subsubsection{Fallback and limitations}
Our focus on a single connected facet per mixed cell means that cases where $\partial\Omega \cap P$ is disconnected or when the interface is under-resolved relative to the mesh resolution may not be perfectly recovered.
In these cases, topological ambiguities may remain after the orientation algorithm.
We thus implement a locally conservative PLIC fallback (\Cref{sec:failsafe}): each unresolved merged component is split back into its original cells and reconstructed independently using LVIRA.

Additionally, since merging reduces local degrees of freedom, in extreme cases a poor merge can theoretically over-constrain the reconstructed geometry.
In practice, we find that our neighbor graph algorithm in \Cref{sec:neighbor-graph-heuristic} performs well on the studied problems (\Cref{sec:results}).
However, for sufficiently low mesh resolution, local greedy merging may be ineffective. We relegate more advanced backtracking or lookahead-based merging strategies to future work.

\section{Fitting Interface Facets}
\label{sec:facet-fitting}

After topology identification, each successfully oriented mixed cell $P_M$ has an ordered neighbor pair $(P_L,P_R)$ and corresponding volume fractions $(v_L,v_M,v_R)$. The ordering is chosen so that fluid interiors are oriented consistently: starting from any mixed cell and repeatedly advancing to its right neighbor traces a counterclockwise walk along the boundary of the target interface.\footnote{In rare cases, a cell whose orientation remains unresolved instead is routed to an LVIRA fallback described in \Cref{sec:failsafe}. Across our static benchmark suite, these cases together account for less than $0.1\%$ of mixed cells. \Cref{tab:fallback_incidence} gives the detailed breakdown.}

This section describes the second phase of our algorithm, \emph{facet fitting}. For each oriented triplet $(P_L, P_M, P_R)$, we solve a local geometric reconstruction problem and assign the resulting primitive as the reconstructed interface facet in $P_M$. The admissible primitives are one or two connected line segments or circular arcs per cell, with corners introduced only when needed to represent sharp features.

Our facet fitting algorithms retain the least-squares objectives as in LVIRA and PLVIRA~\citep{Puckett1991,Puckett1997,Pilliod2004,Remmerswaal2021ParabolicIR}, where the goal of each geometric reconstruction problem is to minimize the volume-fraction mismatch over a set of support cells.
The key difference is that our topology identification stage reduces the geometric fitting support from a centered $3\times3$ stencil as in (P)LVIRA to the three ordered mixed cells $(P_L, P_M, P_R)$.
Intuitively, this is because the full and empty neighbors of regular cells have already been used as a hard constraint to identify the interface orientation.

\begin{figure}[t]
  \centering
  \resizebox{0.98\textwidth}{!}{\includegraphics{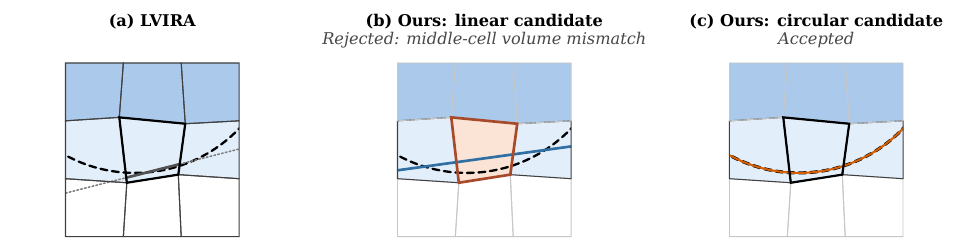}}
  \caption{\textbf{Perturbed Cartesian and oriented interface supports.}
  All three panels use volume fractions sampled from the same generating circular interface (dashed black curve). Dark blue, light blue, and white denote full, mixed, and empty cells, respectively.
  (a) \textit{LVIRA.} The gray line conserves the center cell's volume while minimizing volume-fraction mismatch over the whole $3\times3$ stencil.
  (b) \textit{Ours (linear).} Our topology identification stage first uses the full and empty cells to determine the interface orientation.
  With this orientation, the blue line is the unique candidate that conserves volume in the two outer mixed support cells, in light blue.
  However, its induced volume fraction in the orange middle cell differs from the prescribed value.
  This suggests the candidate linear facet has too few degrees of freedom, so the candidate is rejected.
  (c) \textit{Ours (circular).} The additional curvature degree of freedom allows the circular facet candidate, in orange, to satisfy all three mixed cell volume constraints. The facet coincides with the generating arc to numerical tolerance. Diagram code prepared with assistance from OpenAI Codex.}
  \label{fig:oriented-stencil-comparison}
\end{figure}

Reducing the support cells reduces the number of constraints per geometric optimization problem.
Beyond simplifying each problem, the reduced constraints also allow us to determine whether a particular region of the interface warrants a higher-order circular arc or corner, or when a simple linear facet suffices.
This adaptivity represents the main advantage provided by our topology identification algorithm.
We describe the progressive facet fitting strategy in \Cref{sec:progressive-fitting} and provide a visual example in \Cref{fig:oriented-stencil-comparison}.

\subsection{Progressive fitting strategy}
\label{sec:progressive-fitting}

Our algorithm progressively fits increasingly complex facets to match the true local order of the interface:
\begin{enumerate}
  \item linear facets,
  \item corners with linear edges,
  \item circular facets, and
  \item corners with curved (circular) edges.
\end{enumerate}

This progression is motivated by three observations:
\begin{itemize}
  \item \textbf{Linear regions.} A line has two degrees of freedom. Given three mixed support cells, the two outer volumes can be used to determine the candidate line, while the middle cell volume can be used to test whether that line remains consistent across the full oriented triplet (\Cref{fig:oriented-stencil-comparison}).

  \item \textbf{Smooth curved regions.} For a locally smooth, well-resolved interface, the three volume fractions and the phase orientation uniquely determine a signed circle with center $(x,y)$ and signed radius $r$. Note that a line is recovered in the zero-curvature limit.

  \item \textbf{Sharp regions.} Near a cusp, a single smooth primitive struggles to preserve all volume fractions. Changes in the normals and curvatures of neighboring facets can be used to identify where a corner facet is appropriate.
\end{itemize}

The cumulative stages are illustrated in panels (c)--(f) of \Cref{fig:staged_reconstruction}.
\Cref{sec:fit-linear} through \Cref{sec:fit-curved-corner} describe the problem formulation, area evaluation, fitting algorithm, and acceptance criterion for each primitive family.
In extreme cases, if none of these primitive stages succeeds, we use the appropriate linear facet fallback described in \Cref{sec:failsafe}.

\begin{figure}[!t]
  \centering
  \resizebox{0.44\textwidth}{!}{\includegraphics{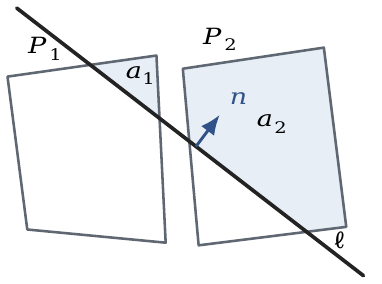}}
  \caption{\textbf{Linear facet fitting.}
  For strictly separated convex support cells with $0<v_1,v_2<1$, the prescribed volume fractions determine a unique line oriented from $P_1$ to $P_2$, with fluid on its left (\Cref{sec:fit-linear}).
  The blue regions are the fluid portions of polygons $P_1$ and $P_2$, measured on the left side of the oriented line $\ell$ with normal $n$. Diagram code prepared with assistance from OpenAI Codex.}
  \label{fig:linear_facet_fitting}
\end{figure}

\begin{figure}[p]
  \centering
  \includegraphics[width=0.80\textwidth]{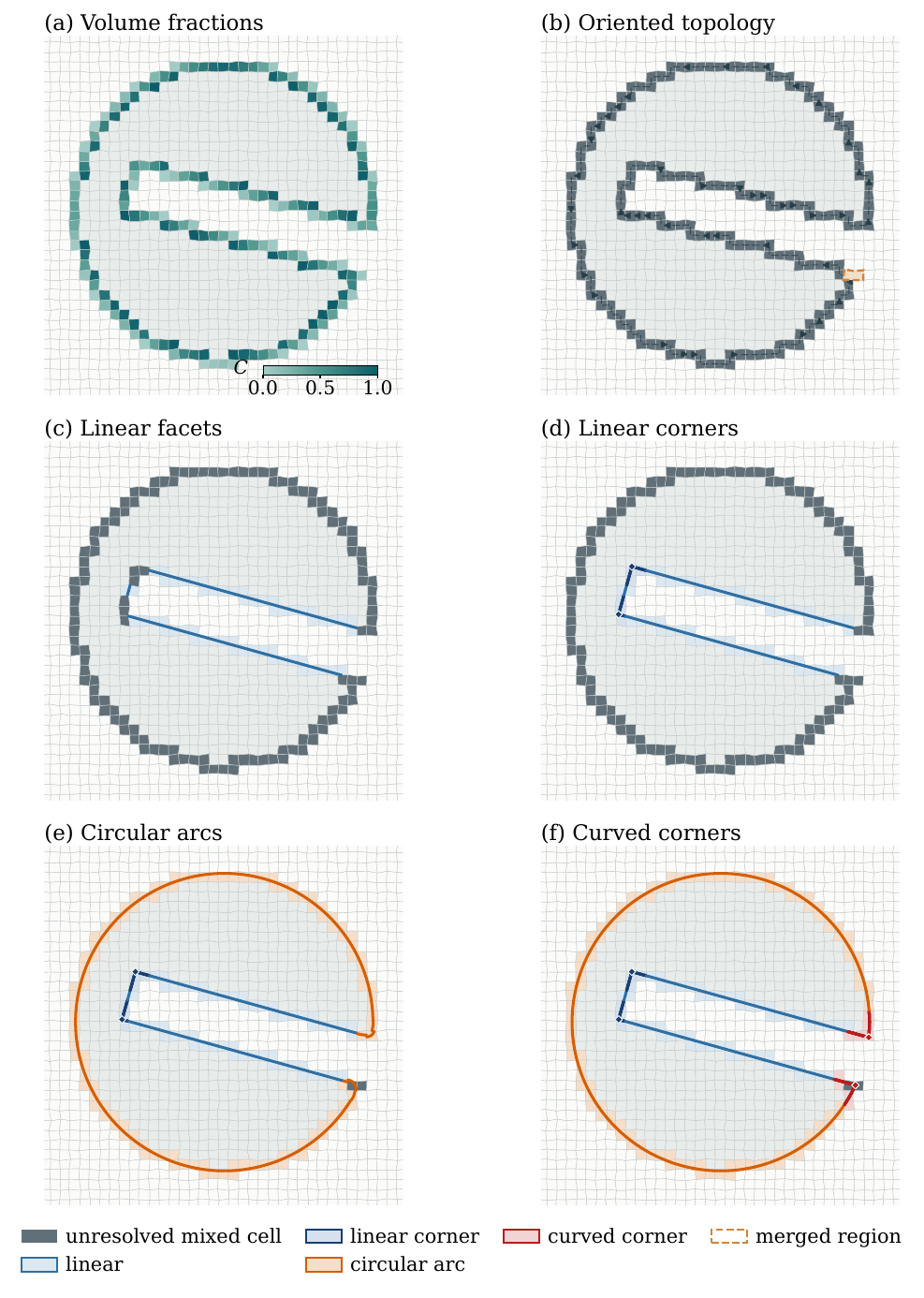}
  \caption{\textbf{Illustration of our reconstruction algorithm on the Zalesak disk over a perturbed Cartesian grid.} At each stage, colors identify the accepted primitive type, while dark mixed cells represent unresolved mixed cells. (a) The input consists of cell volume fractions. Full and empty cells are shown lightly, while mixed cells are shaded according to their volume fractions. (b) The topology identification stage (\Cref{sec:topology-identification}) organizes the mixed cells into an oriented cycle, indicated by arrows. The dashed orange boundary marks a merged region. (c) Cells satisfying the linear acceptance criterion (\Cref{sec:fit-linear}) receive linear facets. (d) Line--line corner facets resolve sharp features between accepted linear facets (\Cref{sec:fit-linear-corner}). (e) Circular arcs are then fitted on the remaining smooth portions of the interface (\Cref{sec:fit-circular}). (f) Curved corner facets resolve sharp line--arc and arc--arc transitions (\Cref{sec:fit-curved-corner}).
  An optional conservative $C^0$ correction can subsequently improve continuity between adjacent facets (\Cref{sec:c0}).}
  \label{fig:staged_reconstruction}
\end{figure}

\subsection{Linear facets}
\label{sec:fit-linear}
\label{appendix_linear_facets}

\paragraph{Problem formulation}
We denote the linear facet fitting problem by \textsc{FitLinearFacet}. Its inputs are two pairs $(P_1,v_1)$ and $(P_2,v_2)$, comprising support cells and their corresponding volume fractions, and its output is an oriented support line $\ell=(p_\ell,n_\ell)$, represented by a point $p_\ell$ and unit normal $n_\ell$. The operation seeks the line satisfying
\[
\frac{|P_1\cap H_\ell|}{|P_1|}=v_1,
\qquad
\frac{|P_2\cap H_\ell|}{|P_2|}=v_2.
\]
Here, $H_\ell$ denotes the half-plane to the left of $\ell$.
For two strictly separated convex support cells with $0<v_1,v_2<1$, the prescribed volume fractions uniquely determine an oriented support line when traversal from $P_1$ to $P_2$ is required and fluid lies on its left~\citep[Theorem~2]{Barany2008Slicing}.
The orientation is selected by the ordered neighbors produced during topology identification.
\Cref{fig:linear_facet_fitting} illustrates an example of the linear facet fitting problem.

\paragraph{Area evaluation}
The enclosed area is evaluated by clipping each polygon against $H_\ell$ and applying Gauss' area formula to the resulting polygon. The intersections of $\ell$ with the cell boundaries provide the endpoints of the linear facet candidate.

\paragraph{Fitting algorithm}
\textsc{FitLinearFacet} parameterizes the line by one support point associated with each cell and updates both points along the current normal. A centered finite-difference approximation of the $2\times2$ area Jacobian defines a Newton step for these two scalar displacements. If the Jacobian is singular, we instead hold the second point fixed while displacing the first point along the normal to match $v_1$, then hold the updated first point fixed while similarly displacing the second point to match $v_2$. The solve terminates when both volume-fraction residuals are below $\varepsilon_{\mathrm{supp}}$.

\paragraph{Acceptance criterion}
For an oriented triplet $(P_L,P_M,P_R)$, we apply \textsc{FitLinearFacet} to $(P_L,v_L)$ and $(P_R,v_R)$ to construct a linear facet candidate for $P_M$. The candidate is accepted when its support line intersects $P_M$ and reproduces $v_M$ within $\varepsilon_{\mathrm{lin}}$. If rejected, the candidate's orientation and support geometry initialize the subsequent circular and corner fitting stages. The numerical values of both tolerances used in our experiments are listed in \Cref{tab:reconstruction_parameters}.

Algorithm~\ref{alg:fit-linear} in Appendix~\ref{appendix_geometric_fitting_details} summarizes the linear facet fitting algorithm.

\begin{figure}[!t]
  \centering
  \begin{minipage}[c]{0.58\textwidth}
    \centering
    \resizebox{0.8\linewidth}{!}{\includegraphics{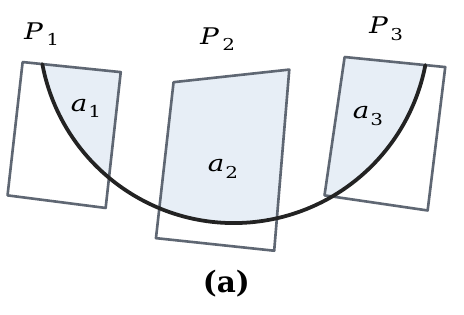}}
  \end{minipage}\hspace{0.02\textwidth}
  \begin{minipage}[c]{0.36\textwidth}
    \centering
    \resizebox{0.8\linewidth}{!}{\includegraphics{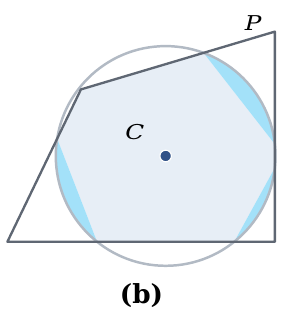}}
  \end{minipage}
  \caption{\textbf{Circular facet fitting and area evaluation.} (a) Three neighboring volume fractions determine a locally circular interface. The blue regions are the fluid portions of polygons $P_1$, $P_2$, and $P_3$ selected by the signed circle. (b) The disk $C$ is clipped by polygon $P$. The light-blue core is assembled from polygon vertices and circle--edge intersection points, while the teal pieces contain circular arcs that complete $P \cap C$. Diagram code prepared with assistance from OpenAI Codex.}
  \label{fig:circular_facet_fitting}
\end{figure}

\subsection{Circular facets}
\label{sec:fit-circular}
\label{appendix_circular_facets}

\paragraph{Problem formulation}
We denote the circular facet fitting problem by \textsc{FitCircularFacet}.
Its inputs are three pairs $(P_1,v_1)$, $(P_2,v_2)$, and $(P_3,v_3)$, comprising ordered support cells and their corresponding volume fractions.
Its output is a signed circle, represented by its center $c=(x,y)$ and signed radius $r$.
The induced volume fraction in support cell $P_i$ is
\[
\widehat v_i(x,y,r)=\frac{|P_i\cap C(x,y,r)|}{|P_i|}.
\]
The problem seeks parameters that minimize
\begin{equation}
L(x,y,r)=\frac12\sum_{i=1}^{3}\left(\widehat v_i(x,y,r)-v_i\right)^2,
\label{eqn:circular_facet_objective}
\end{equation}
where $C(x,y,r)$ denotes the phase region selected by the signed circle.
A signed circle has three degrees of freedom, and from our experiments we find that the three volume-fraction constraints from a regular cell typically identify a unique solution.
The fitting problem is illustrated in \Cref{fig:circular_facet_fitting}(a).

\paragraph{Area evaluation}
The area cut from a polygon by a circle is decomposed into a polygonal core and circular-segment corrections.
The core is formed from polygon vertices and circle--edge intersections and is evaluated with Gauss' formula.
Circular-segment areas between consecutive intersection points are computed by subtracting the area of the triangle formed by the two intersection points and the circle center from the corresponding circular-sector area.
The polygon--circle area evaluation is illustrated in \Cref{fig:circular_facet_fitting}(b).

\paragraph{Fitting algorithm}
\textsc{FitCircularFacet} initializes the solve from the oriented linear facet candidate fitted to $P_1$ and $P_3$ and considers both direct and complemented phase orientations.
The geometry is rotated so that this support line is horizontal.
In this frame, the circle is parameterized by its two intercepts $q_1$ and $q_3$ on the outer supports and by a positive radius magnitude $R$, with center
\[
\left(\frac{q_1+q_3}{2},
\sqrt{R^2-\frac{(q_3-q_1)^2}{4}}\right).
\]
Appendix~\ref{appendix_geometric_fitting_details} derives this expression from the chord midpoint and the Pythagorean theorem.

The three volume-fraction constraints are enforced by alternating scalar solves: the radius is adjusted to match the middle cell, the first intercept is updated to match the first support, the radius is readjusted, and the second intercept is updated to match the third support.
Radius updates use a binary expanding and halving search, while intercept updates use finite-difference Newton steps constrained by $|q_3-q_1|\leq2R$.
The solve terminates when all three residuals are below $\varepsilon_{\mathrm{arc}}$.
Algorithm~\ref{alg:fit-circular} summarizes the procedure.

\paragraph{Acceptance criterion}
For an oriented triplet $(P_L,P_M,P_R)$, we apply \textsc{FitCircularFacet} to $(P_L,v_L)$, $(P_M,v_M)$, and $(P_R,v_R)$ to construct a circular facet candidate for $P_M$. The candidate is accepted when all three volume-fraction residuals are below $\varepsilon_{\mathrm{arc}}$.
The numerical tolerance used in our experiments is listed in \Cref{tab:reconstruction_parameters}.

\subsection{Corner facets}
\label{sec:fit-linear-corner}
\label{sec:fit-curved-corner}
\label{appendix_corner_facet}

\paragraph{Problem formulation}
We denote the corner facet fitting problem by \textsc{FitCornerFacet}.
Its inputs are pairs $\{(P_i,v_i)\}_{i=1}^m$ for a consecutive sequence of unresolved cells and the nearest accepted linear or circular facet on each side.
Its output is a cusp $p_c$ and two branches, where a branch is the portion of a line or circular arc on one side of the cusp.
A straight corner has line--line branches, while a curved corner has line--arc or arc--arc branches.
Clipping the branches to the target cells gives the corner facet candidates per mixed cell.
The method starts by extending the lines or circles containing the neighboring accepted facets and uses their intersections as cusp proposals.
It tests each proposal against the relevant target volume fractions and may refine the selected cusp to reduce their collective residual.

\textit{Straight corners.}
For a line--line corner, the cusp proposal is the intersection of the lines containing the nearest accepted linear facets on the two sides of the unresolved cells.

\textit{Curved corners.}
For a line--arc or arc--arc corner, the cusp proposals are the corresponding line--circle or circle--circle intersections.
When two intersections exist, a local proximity criterion selects the appropriate one.

\begin{figure}[t]
  \centering
  \resizebox{0.4\textwidth}{!}{\includegraphics{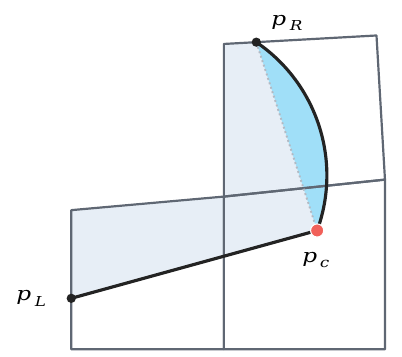}}
  \caption{\textbf{Corner facet fitting and area evaluation.} A corner facet joins two branches at a cusp point $p_c$. Here, a line--arc corner joins the line segment from $p_L$ to $p_c$ and the circular arc from $p_c$ to $p_R$. The pale blue region marks the fluid side of the oriented interface.
The darker blue circular segment between the arc and the gray dotted chord $p_c p_R$ isolates the arc's area contribution. Diagram code prepared with assistance from OpenAI Codex.}
  \label{fig:corner_facet_fitting}
\end{figure}

\Cref{fig:corner_facet_fitting} illustrates an example of the corner facet fitting problem for a line--arc junction.

\paragraph{Area evaluation}
For each target cell, the enclosed fluid area combines the contributions from the two corner branches.

\textit{Straight corners.}
For a line--line corner, the fluid area combines a half-plane clipping with the wedge between the two rays extending from the cusp.
The corner orientation determines whether the wedge is added or removed.

\textit{Curved corners.}
If either branch is circular, its circular-segment contribution is added to the straight-corner area calculation.
The circular branch in \Cref{fig:corner_facet_fitting} illustrates the sector-minus-triangle area contribution.

\paragraph{Fitting algorithm}
Corner fitting attempts to fit a proposed cusp to an entire connected block of unresolved mixed cells between the nearest accepted facets.
For either corner type, an optional refinement adjusts the cusp to improve local volume conservation while keeping the neighboring facet endpoints, branch types, and curvatures fixed.

\paragraph{Acceptance criterion}
A corner candidate must form a valid two-branch geometry and reproduce the volume fractions of the cells it spans sufficiently closely.

\textit{Straight corners.}
A line--line corner facet candidate is rejected if the two lines do not intersect, $|\sin\theta|\leq\tau_{\parallel}$ for the angle $\theta$ between their tangents, or neither branch intersects the target cell.
Otherwise, it is accepted when its volume-fraction residual is below $\varepsilon_{\mathrm{SC}}$.

\textit{Curved corners.}
A line--arc or arc--arc corner facet candidate is rejected if $|\sin\theta|\leq\tau_{\parallel}$.
Otherwise, the shared corner is accepted for the block when its induced area is well defined in every intervening cell, the cusp lies in their union, and the geometric mean of the volume-fraction residuals is below $\varepsilon_{\mathrm{CC}}$.

The numerical tolerances used in our experiments are listed in \Cref{tab:reconstruction_parameters}.

\subsection{Linear fallback}
\label{sec:failsafe}

If none of the preceding primitive stages produces an accepted facet, the mixed cell receives a volume-conserving linear facet.

There are two fallback cases. If the topology identification stage (\Cref{sec:topology-identification}) provides an oriented support pair, we use \textsc{FitLinearFacet} from \Cref{sec:fit-linear} to determine the line orientation and then position the line to reproduce the target volume fraction. For a dead-end cell with only one identified adjacent mixed cell, the target and adjacent cells form this support pair instead.

If the orientation remains unresolved after topology identification, we instead use LVIRA. Any merged cell is first split into its constituent mixed cells. LVIRA then estimates the linear facet's normal independently for each original cell from its centered $3\times3$ stencil, as in \Cref{fig:perfect_reconstruction_plic}(c), and positions the line to reproduce that cell's volume fraction.

These fallbacks are rare in our experiments: the \textsc{FitLinearFacet} fallback is used in $0.0025\%$ of mixed cells and the LVIRA fallback is used in $0.0809\%$ of mixed cells.
A detailed breakdown across our static benchmarks is provided in \Cref{tab:fallback_incidence}.

\subsection{\texorpdfstring{$C^0$}{C0} continuity enforcement}
\label{sec:c0}

Because adjacent facets are fitted under separate cell volume constraints, their endpoints may not coincide exactly.
We therefore provide an optional conservative $C^0$ post-processing correction for eligible line and arc facets.
Explicit corner facets are left unchanged. The correction has two steps.

\begin{enumerate}
  \item \textbf{Parallel local pass.}
  On a smooth, well-resolved interface, adjacent facets already have small endpoint gaps that decrease at approximately third order (\Cref{sec:results-ellipses}).
  For each eligible join, we thus propose the midpoint of the two reconstructed endpoints as their shared crossing.
  Each participating facet's curvature is then refit through its proposed endpoints while preserving its cell's volume fraction.
  We accept the proposal only if the adjusted facet remains within its reconstruction cell and its normalized volume-fraction residual is at most $\varepsilon_{C^0}$.

  \item \textbf{Small-component joint refinement.} The joins whose endpoint gap remains above $\varepsilon_{\mathrm{gap}}$ form connected components (\Cref{tab:reconstruction_parameters}).
  For each component, we next perform a constrained nonlinear solve that jointly adjusts the shared crossings and facet curvatures, while fixing the exterior endpoints, to preserve each participating volume fraction.
\end{enumerate}

We note two limitations with our $C^0$ post-processing correction.
First, if no admissible joint solution is found, the conservative input facets are retained.
Second, the $C^0$ post-processing step does not impose a global curvature-sign condition and may introduce local concavity on an otherwise convex interface.
In our experiments, such occurrences are rare and typically weak.

Our main benchmark results report reconstructions before this optional correction unless explicitly stated.
We report additional results in \Cref{sec:results-component-effects} and Appendix~\ref{appendix_c0_study}.

\section{Experimental Results}
\label{sec:results}

We evaluate our method on five static benchmark families, isolating the effect of each of our modular contributions.
The line and circle benchmarks assess perfect reconstruction with linear and circular facets,
the square benchmark isolates linear corners,
the ellipse benchmark measures convergence as curvature varies,
and the Zalesak benchmark combines straight, curved, and mixed line--arc corners.
On selected benchmarks, we also compare our two topology identification methods (per-cell and graph-coordinated) and evaluate the effectiveness of our $C^0$ refinement step.

Throughout this experimental section, references to our linear, circular, and corner-aware reconstructions use the graph-coordinated topology identification method and omit the optional $C^0$ refinement step unless explicitly specified.
We report additional experimental results in Appendix~\ref{appendix_static_full_panels}.

\subsection{Shared Experimental Details}
\label{sec:shared-experimental-details}

All benchmark geometries are defined on the fixed domain $[0,100]^2$, and $N$ denotes the number of cells per side. Table~\ref{tab:static_benchmark_definitions} summarizes the benchmark geometries and sampling.

\begin{table}
\begin{center}
  \centering
  \small
  \captionof{table}{Geometry, sampling, and number of distinct instances used for the static benchmarks.}
  \label{tab:static_benchmark_definitions}
  \begin{tabular}[t]{p{0.13\textwidth}|p{0.22\textwidth} p{0.44\textwidth} r}
    \hline
    \textbf{Benchmark} & \textbf{Geometry} & \textbf{Sampling and reference construction} & \shortstack{\textbf{Distinct}\\\textbf{instances}} \\
    \hline
    \textbf{Line} & Straight interface & $25$ equally spaced orientations $\theta\in[0,2\pi)$, reference point sampled uniformly in $[50,51]^2$ & $750$ \\
    \textbf{Square} & Side length $s\in[10,30]$ & $25$ equally spaced side lengths, $\theta\sim U[0,\pi/2)$ and center sampled uniformly in $[50,51]^2$ & $625$ \\
    \textbf{Circle} & $R=10$, $\kappa=0.1$ & Center sampled uniformly in $[50,51]^2$ & $750$ \\
    \textbf{Ellipse} & $a=30$, $b\in[10,20]$ & $25$ equally spaced aspect ratios $a/b\in[1.5,3]$; $\theta\sim U[0,\pi/2)$ and center sampled uniformly in $[50,51]^2$ & $750$ \\
    \textbf{Zalesak} & $R=15$, slot width $W=5$, slot height $y_{\mathrm{top}}=10$ & $\theta\sim U[0,\pi/2)$ and center sampled uniformly in $[50,51]^2$ & $625$ \\
    \hline
  \end{tabular}
\end{center}
\end{table}

\paragraph{Sweeps}
All quantitative sweeps vary both the number of cells per side of the mesh, $N$ (equivalently $h^{-1}$, where $h=100/N$ is the nominal Cartesian spacing), and the mesh perturbation magnitude.
Lines, circles, and ellipses use $N\in\{32,50,64,100,128,150\}$, while squares and Zalesak disks use $N\in\{50,64,100,128,150\}$.
We parameterize the mesh perturbation magnitude by $w$, where $w=0$ denotes an unperturbed Cartesian mesh.
For each $w$, every interior mesh vertex receives independent coordinate displacements $\delta x,\delta y\sim U[-wh,wh]$ (boundary vertices remain fixed).
All methods use the same seeded mesh for a given $(N,w)$.
For each benchmark and each resolution, we sweep $w\in\{0,0.05,0.1,0.2,0.3\}$.
For each geometry, method, and sweep setting, we report the median and interquartile range over $25$ benchmark configurations, which vary interface placement and, where applicable, interface orientation or shape parameter.
We introduce broken y-axes in our plots to separate numerical floor errors, observed when a method achieves perfect reconstruction, from larger errors.

Each method is thus evaluated on $6\times5\times25=750$ reconstructions for lines, circles, and ellipses and on $5\times5\times25=625$ reconstructions for squares and Zalesak.
Across all method--benchmark combinations, the primary sweep comprises $970$ run settings and $24{,}250$ case reconstructions.

All computations use double precision. Volume fractions within $\varepsilon_{\mathrm{mix}}=10^{-10}$ of zero or one are classified as empty or full, respectively. This classification tolerance sets the numerical floor for the results in this section. The numerical parameters are summarized in \Cref{tab:reconstruction_parameters}.

\paragraph{Metrics}
Throughout our experiments, we report three complementary metrics of reconstruction fidelity and continuity: Hausdorff distance, curvature error, and facet gap.
We define these quantities formally below.

Let $\Gamma$ denote the reference interface and let $\Gamma_h=\bigcup_{F\in\mathcal{F}_h}F$ denote the union of the reconstructed facets.

\textit{Hausdorff distance.}
We measure interface-position error using the symmetric Hausdorff distance
\[
d_H(\Gamma,\Gamma_h)
=\max\!\left\{
\sup_{\mathbf{x}\in\Gamma}\operatorname{dist}(\mathbf{x},\Gamma_h),
\sup_{\mathbf{y}\in\Gamma_h}\operatorname{dist}(\mathbf{y},\Gamma)
\right\}.
\]
In our experiments, each directed supremum is estimated adaptively from $100$ initial samples per source facet.
Distances from a point to a target line, circular arc, or parabolic/quadratic representation (from our higher-order baselines) are evaluated analytically.
Intuitively, the Hausdorff distance measures the largest discrepancy between the reconstructed and generating interfaces.

\textit{Curvature error.}
We measure curvature error using the arc-length-weighted mean absolute difference between reconstructed and reference curvatures:
\[
e_{\kappa}
=\frac{1}{L_h}\sum_{F\in\mathcal{F}_h}
\int_F
\left|\kappa_F(\mathbf{x})
-\kappa\!\left(\pi_\Gamma(\mathbf{x})\right)\right|\,\mathrm{d}s,
\qquad
L_h=\sum_{F\in\mathcal{F}_h}\operatorname{length}(F),
\]
where $\pi_\Gamma(\mathbf{x})$ is a nearest point on the analytic reference interface and $\kappa_F$ is the unsigned curvature of the reconstructed facet.
Thus, lines contribute zero curvature, circles contribute the reciprocal of their radius, and parabolic or quadratic facets contribute their pointwise curvature.
Intuitively, this metric measures the local mismatch between reconstructed and generating interface curvatures, averaged over the length of the reconstructed interface.

\textit{Facet gap.}
For the ordered set $\mathcal{J}$ of adjacent reconstructed facet pairs, we measure interface discontinuity using the mean facet gap
\[
g_h=\frac{1}{|\mathcal{J}|}\sum_{(i,j)\in\mathcal{J}}
\min_{\mathbf{p}\in\partial F_i,\,\mathbf{q}\in\partial F_j}
\|\mathbf{p}-\mathbf{q}\|_2.
\]
Here, $\partial F_i$ denotes the endpoints of facet $F_i$.
Intuitively, the facet gap measures how closely neighboring facets meet, and is zero when the reconstructed interface is $C^0$.

All metric plots are rendered programmatically from saved case-level records using reproducible Python plotting code prepared with assistance from OpenAI Codex.

Whenever we report an observed convergence order for one of these metrics, we perform a least-squares fit to log error vs. log $N$ over a resolution sweep, and we report the negative slope.

\paragraph{Baselines}
For our linear-only method (turning off circular and corner facets), we include comparisons with Youngs' method~\citep{Youngs1987,ParkerYoungs1992}, ELVIRA~\citep{Pilliod1992,Pilliod2004}, and LVIRA~\citep{Puckett1991,Puckett1997}.
Beyond the line benchmark, we omit Youngs because we find that ELVIRA and LVIRA generally yield lower median facet-gap errors across the displayed resolutions. Implementation details for these PLIC
baselines appear in Appendix~\ref{appendix_plic_baselines}.

For our method with circular facets, we additionally reproduce and compare with PLVIRA~\citep{Remmerswaal2021ParabolicIR}, PCIC~\citep{Maity2020}, and QUASI~\citep{Diwakar2009AQS}.
These three methods are formulated only for structured Cartesian meshes.
As such, we restrict this higher-order comparison to the unperturbed Cartesian meshes ($w=0$) and to the circle, ellipse, and Zalesak benchmarks.
Details about our reproductions are included in Appendix~\ref{appendix_external_baseline_scope}.

\subsection{Lines Benchmark}
\label{sec:results-lines}

\begin{figure}[p]
  \centering
  \includegraphics[width=0.92\textwidth]{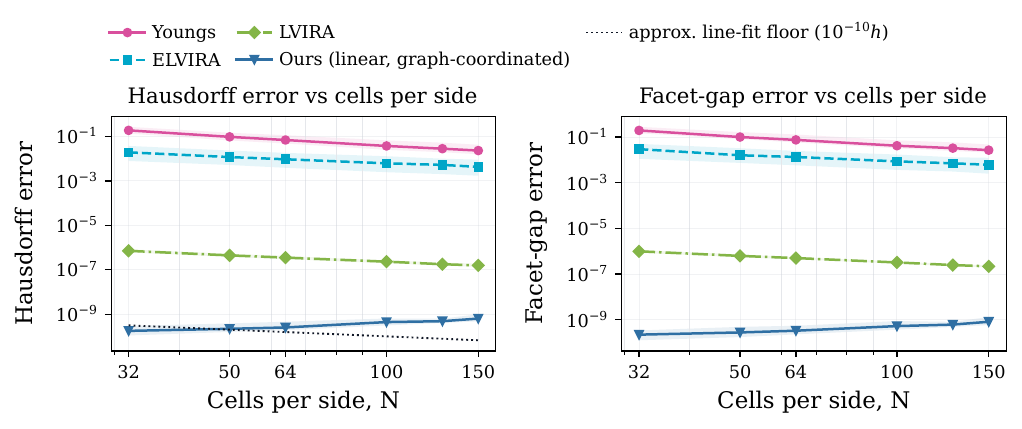}

  \includegraphics[width=0.86\textwidth]{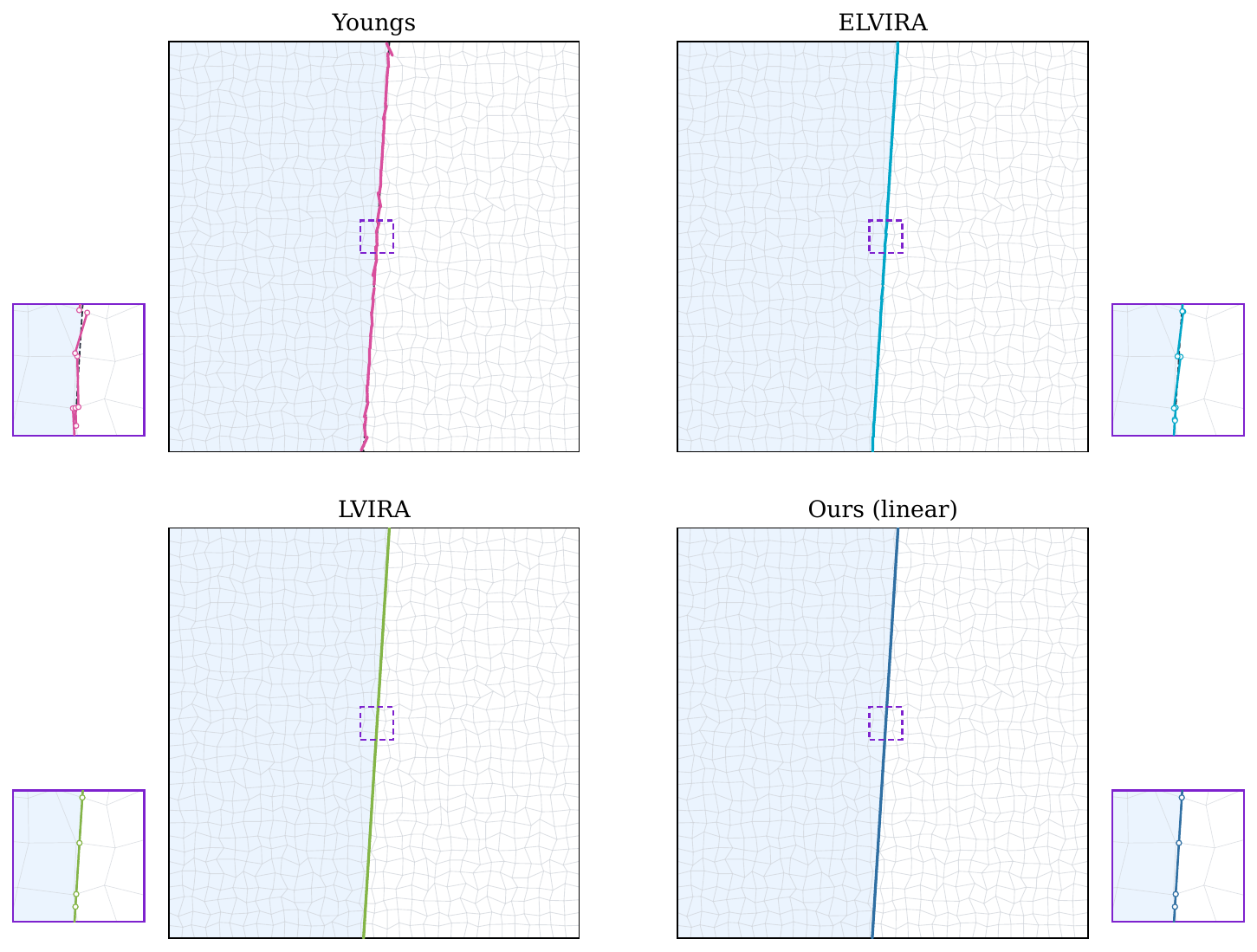}
  \caption{\textbf{Line benchmark.} Upper panels: Hausdorff distance and facet gap versus cells per side, $N$. At each $N$, curves show the median and shaded bands show the interquartile range over all $125$ combinations of perturbation magnitude and benchmark configuration. Our graph-coordinated linear method recovers the generating line to the numerical floor. LVIRA achieves perfect reconstruction on interior linear interfaces, but its $3\times3$ stencil is truncated at domain boundaries, which introduces the residual errors seen in the global metrics. Youngs and ELVIRA do not exactly recover linear interfaces once the Cartesian mesh is perturbed. Lower panels: representative reconstructions at $N=32$ and $w=0.3$. The dashed black line is the generating interface, and the spyglasses magnify the local reconstruction.}
  \label{fig:results_lines}
  \label{fig:results_lines_representative}
\end{figure}

\paragraph{Setup}
The line benchmark (\Cref{fig:results_lines}) assesses whether a reconstruction method recovers a linear interface from its volume fractions on perturbed Cartesian meshes. We use $25$ equally spaced orientations over $[0,2\pi)$. For each orientation, the generating line passes through a point sampled uniformly from $[50,51]^2$.

\paragraph{Results}
Our linear method solves directly for the line shared by neighboring mixed cells and remains at the numerical floor in both Hausdorff distance and facet gap (at least $240\times$ and $260\times$ lower than LVIRA, respectively). LVIRA is the strongest PLIC baseline: it recovers interior linear interfaces, but its error concentrates at mesh boundaries where its $3\times3$ stencil is truncated, as illustrated by the boundary spyglass in \Cref{fig:results_lines}. On perturbed meshes, ELVIRA and Youngs exhibit errors in the predicted interface normal throughout the domain. Across the plotted resolutions, their median Hausdorff errors are $27{,}000$--$30{,}000\times$ and $150{,}000$--$270{,}000\times$ those of LVIRA, respectively. We report further results across mesh resolutions and perturbation magnitudes in \Cref{fig:appendix_lines_all_methods}.

\subsection{Squares Benchmark}
\label{sec:results-squares}

\begin{figure}[p]
  \centering
  \includegraphics[width=0.92\textwidth]{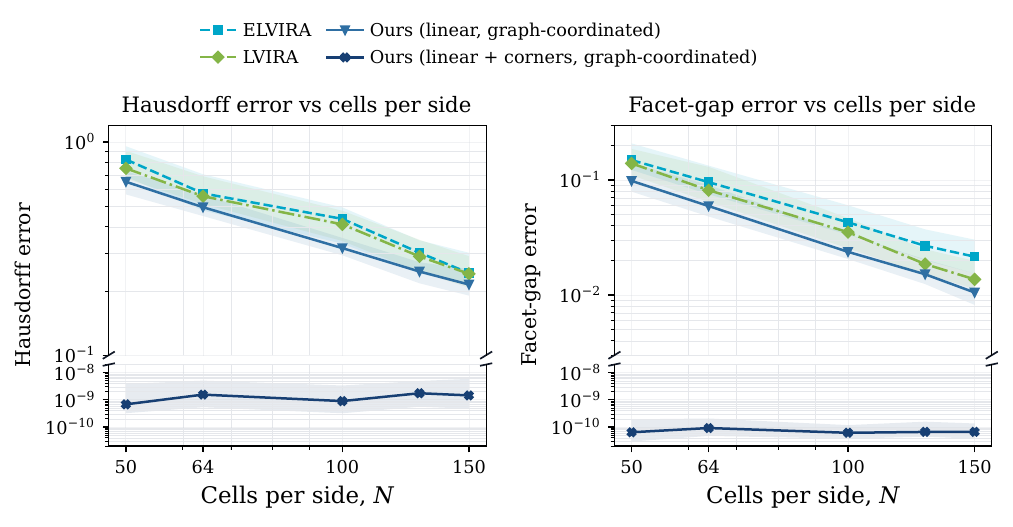}

  \vspace{5mm}
  \includegraphics[width=0.81\textwidth]{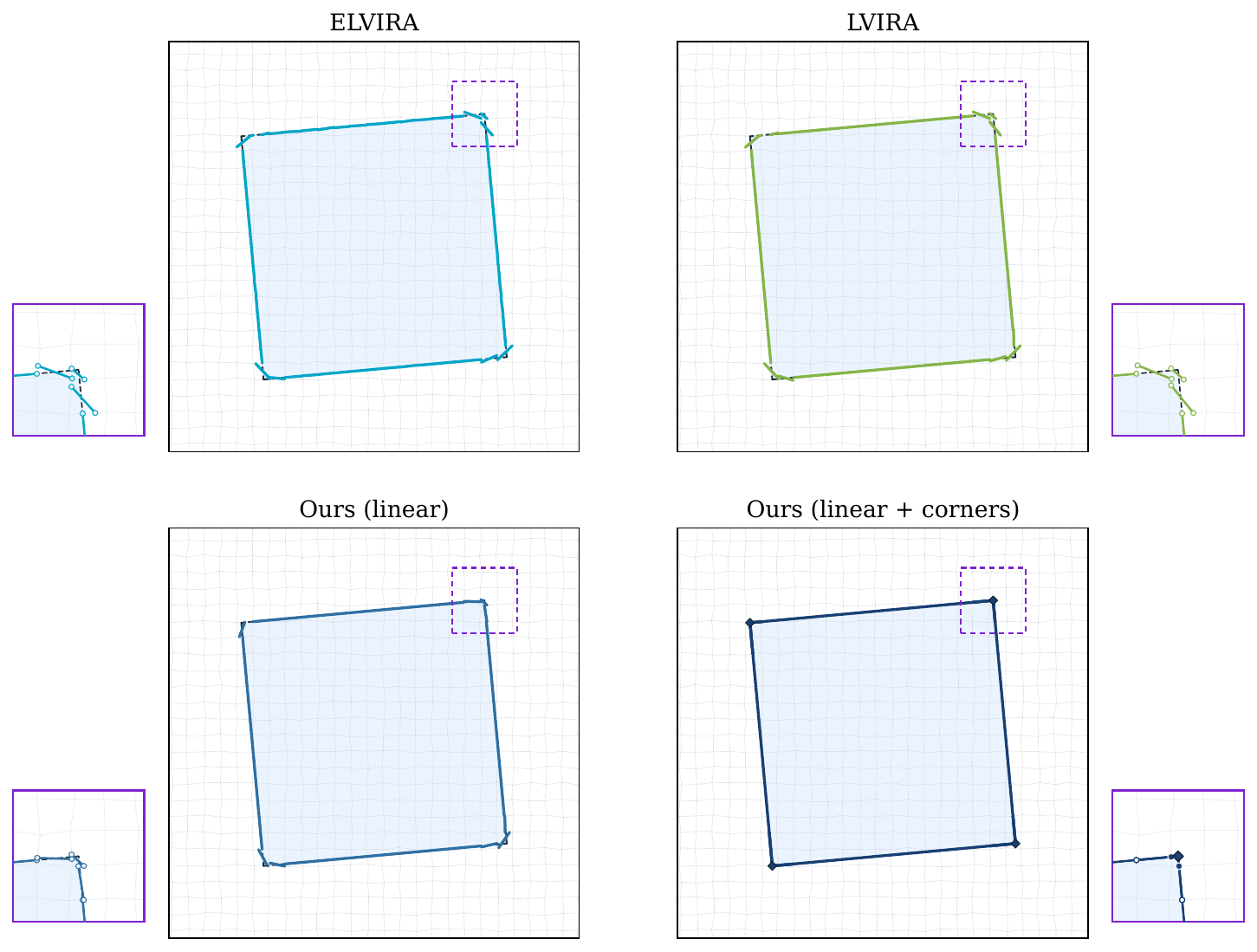}
  \caption{\textbf{Square benchmark.} Upper panels: Hausdorff distance and facet gap versus cells per side, $N$. At each $N$, curves show the median and shaded bands show the interquartile range over all $125$ combinations of perturbation magnitude and benchmark configuration. ELVIRA and LVIRA recover the straight sides well, but a single linear facet cannot represent both incident edges within a corner cell. Our graph-coordinated linear method's reconstruction has more localized corner error, consistent with fitting the smaller support of three oriented mixed cells rather than a full $3\times3$ stencil, but has the same fundamental limitation as (E)LVIRA due to its linear facets. Adding line--line corner primitives recovers the identifiable square geometry to the numerical floor. Lower panels: a representative reconstruction at $N=50$ and $w=0.1$. The dashed black curve is the generating square. The spyglasses show how each linear method rounds the vertices and how the corner primitive restores them. Diamonds mark corner tips, and open circles mark intersections of corner branches with cell boundaries.
  Broken y-axes separate the numerical floor from larger error ranges, and curves are not connected across omitted intervals.
  }
  \label{fig:results_squares}
  \label{fig:results_squares_representative}
\end{figure}

\paragraph{Setup}
The square benchmark (\Cref{fig:results_squares}) assesses the reconstruction of sharp linear corners, which a single PLIC facet cannot represent within a corner cell. We use $25$ equally spaced side lengths from $10$ to $30$. For each side length, the square center is sampled uniformly from $[50,51]^2$ and its rotation is sampled uniformly from $[0,\pi/2)$.

\paragraph{Results}
Across the plotted resolutions, our linear-only method achieves $11$--$23\%$ lower median Hausdorff distance and $18$--$33\%$ lower median facet gap than LVIRA. For both methods, the remaining error concentrates near the square vertices, where a single linear facet cannot represent both incident edges within one cell. Adding explicit corner primitives, our method recovers the square to the numerical floor in both metrics. Results across mesh resolutions and perturbation magnitudes are reported in \Cref{fig:appendix_squares_all_methods}.

\subsection{Circles Benchmark}
\label{sec:results-circles}

\begin{figure}[p]
  \centering
  \includegraphics[width=0.88\textwidth]{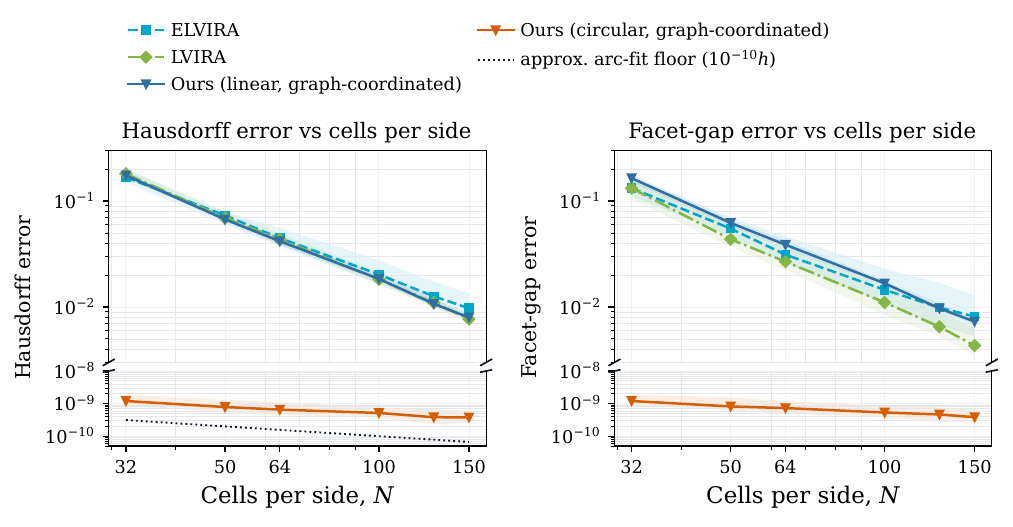}

  \includegraphics[width=0.69\textwidth]{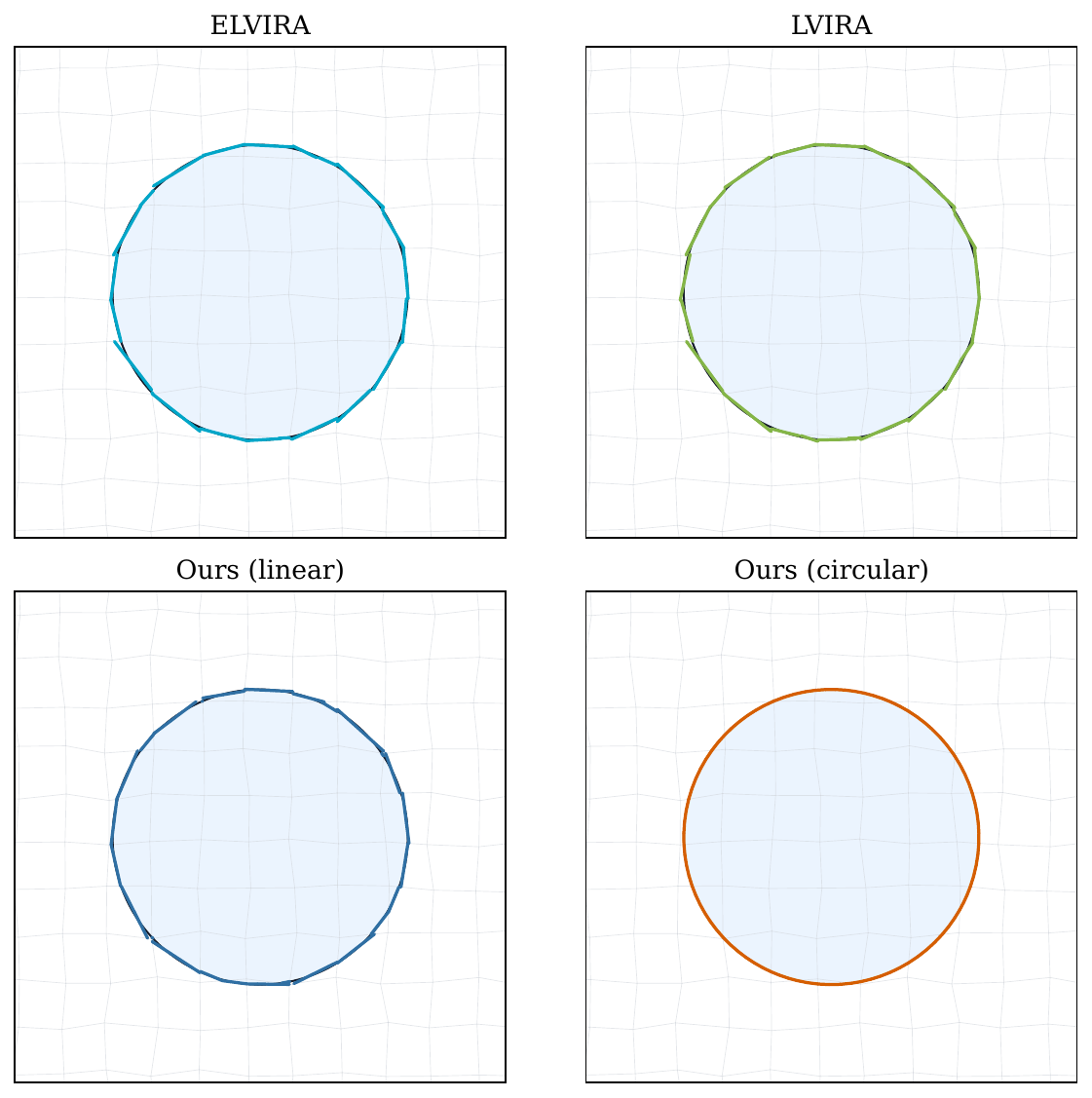}
  \caption{\textbf{Circle benchmark.} Upper panels: Hausdorff distance and facet gap versus cells per side, $N$. At each $N$, curves show the median and shaded bands show the interquartile range over all $125$ combinations of perturbation magnitude and benchmark configuration. ELVIRA, LVIRA, and our graph-coordinated linear method all approximate the circle well with linear facets. Our linear method and LVIRA converge at approximately second order in both metrics, as shown in the resolution study in \Cref{fig:appendix_circle_linear_extended}. Adding circular facets recovers the constant-curvature interface to the numerical floor. Lower panels: a representative reconstruction at $N=32$ and $w=0.1$. The dashed black curve is the generating circle. The spyglasses show the faceting retained by the linear methods and the fitted circular arcs.
  Broken y-axes separate the numerical floor from larger error ranges, and curves are not connected across omitted intervals.
  }
  \label{fig:results_circles}
  \label{fig:results_circles_representative}
\end{figure}

\paragraph{Setup}
The circle benchmark (\Cref{fig:results_circles}) is the constant-curvature analogue of the line benchmark. We reconstruct $25$ circles of fixed radius $R=10$, with centers sampled uniformly from $[50,51]^2$ to vary their placement relative to the mesh.

\paragraph{Results}
When restricted to linear facets, our graph-coordinated method has lower median Hausdorff error than ELVIRA at five of the six resolutions we study, and remains within $4\%$ of LVIRA at every resolution. The resolution study in \Cref{fig:appendix_circle_linear_extended} shows that both ours and LVIRA have approximately second-order convergence in Hausdorff error and facet gap. LVIRA has a slightly smaller facet gap, consistent with its broader $3\times3$ stencil smoothing neighboring facet orientations and thus improving inter-facet continuity within the reconstructed interface. However, our graph-coordinated linear facet method has a slightly better Hausdorff error for perturbed meshes. This indicates that LVIRA's better continuity does not translate directly to better reconstruction fidelity overall. We present additional results on perturbed meshes in~\Cref{fig:appendix_circles_all_methods}.

\begin{figure}[p]
  \centering
  \includegraphics[width=0.96\textwidth]{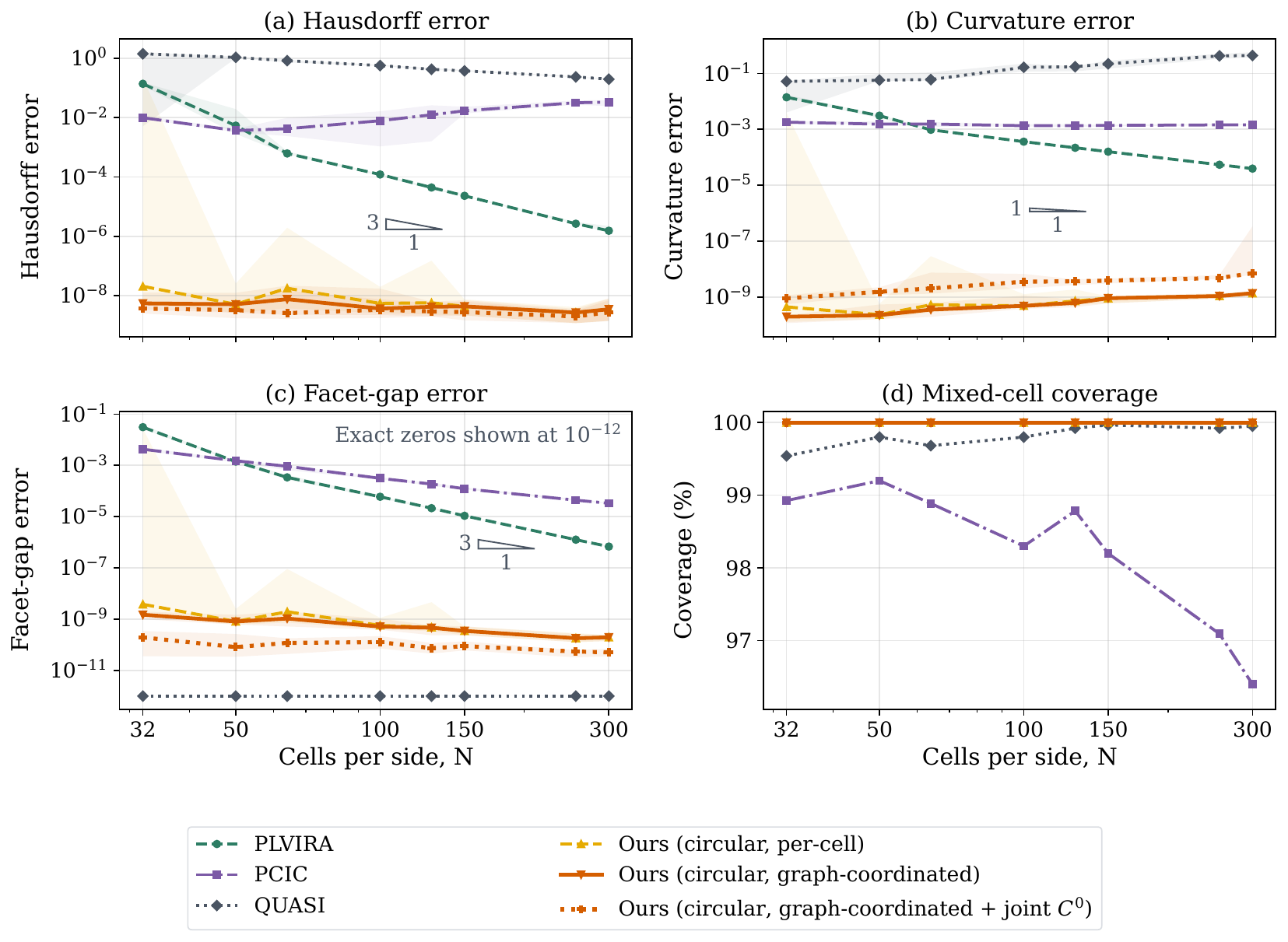}
  \caption{\textbf{Higher-order circle comparison on Cartesian meshes.}
  Panels show Hausdorff error, curvature error, mean facet gap, and reconstructed mixed-cell coverage versus cells per side, $N$.
  At each $N$, curves show the median and shaded bands show the interquartile range over $25$ benchmark configurations.
  Our three circular variants remain at the numerical floor in all three error metrics.
  At $N=300$, the per-cell and graph-coordinated circular variants have median Hausdorff and curvature errors $445\times$ and $28{,}000\times$ lower than those of PLVIRA, the strongest reproduced baseline at this resolution.
  Coverage below $100\%$ denotes unresolved mixed cells, and exact zero gaps are displayed at the plotting floor.
  Baseline implementation details are given in Appendix~\ref{appendix_external_baseline_scope}.}
  \label{fig:results_circles_higher_order}
  \label{fig:appendix_circles_higher_order}
\end{figure}

With circular facets, our method recovers the generating circle to the numerical floor in both Hausdorff distance and facet gap on the perturbed meshes. Evaluating on Cartesian meshes to compare against higher-order baselines (\Cref{fig:results_circles_higher_order}), we find that our method also has lower median Hausdorff and curvature errors than PLVIRA, PCIC, and QUASI at every resolution in the Cartesian comparison through $N=300$. At $N=300$, its median Hausdorff and curvature errors are $3.46\times10^{-9}$ and $1.39\times10^{-9}$, compared to $1.54\times10^{-6}$ and $3.96\times10^{-5}$ for PLVIRA, the strongest external baseline at that resolution. These errors are lower by $445\times$ and $28{,}000\times$, respectively.

\subsection{Ellipses Benchmark}
\label{sec:results-ellipses}

\begin{figure}[p]
  \centering
  \includegraphics[width=0.88\textwidth]{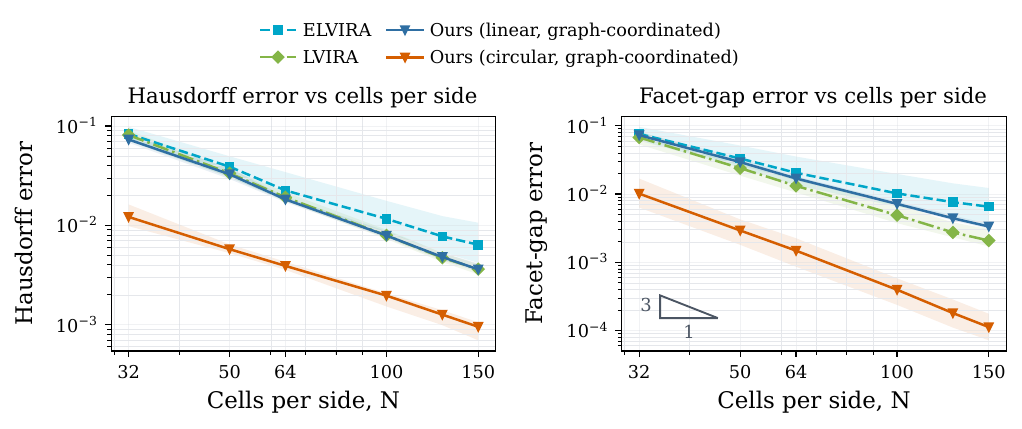}

  \includegraphics[width=0.80\textwidth]{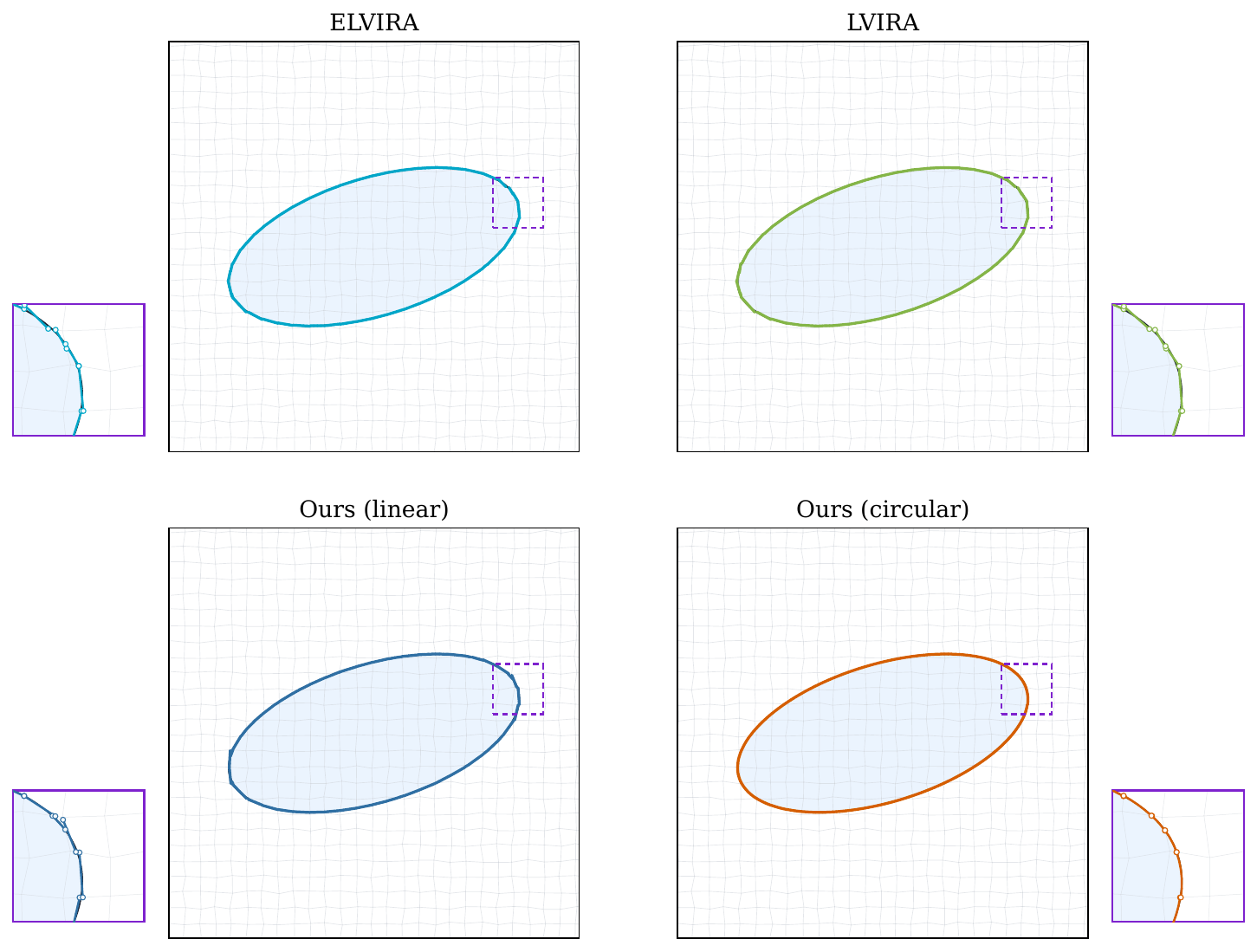}
  \caption{\textbf{Ellipse benchmark.} Upper panels: Hausdorff distance and facet gap versus cells per side, $N$. At each $N$, curves show the median and shaded bands show the interquartile range over all $125$ combinations of perturbation magnitude and benchmark configuration. LVIRA and our graph-coordinated linear method have nearly identical Hausdorff errors, with ours lower at five of the six resolutions. LVIRA has a smaller facet gap, consistent with its broader stencil producing smoother neighboring facet orientations, but our linear method's Hausdorff error matches LVIRA's, indicating that the two methods have similar overall reconstruction fidelity. Adding circular facets reduces both errors and gives an observed facet-gap convergence order of approximately $2.92$. Lower panels: a representative reconstruction at $N=32$ and $w=0.1$. The dashed black curve is the generating ellipse. The spyglasses magnify the maximum-curvature region where the linear methods retain visible faceting.}
  \label{fig:results_ellipses}
  \label{fig:results_ellipses_representative}
\end{figure}

\paragraph{Setup}
The ellipse benchmark (\Cref{fig:results_ellipses}) assesses circular facets when curvature varies and exact recovery is no longer possible. We fix the major semiaxis at $a=30$ and use $25$ equally spaced aspect ratios $a/b\in[1.5,3]$. For each aspect ratio, the center is sampled uniformly from $[50,51]^2$ and the rotation $\theta$ is sampled uniformly from $[0,\pi/2)$. The reference curvature at parameter $\phi$ is
\[
\kappa(\phi)=\frac{ab}{\big(a^2\sin^2(\phi-\theta)+b^2\cos^2(\phi-\theta)\big)^{3/2}}.
\]

\paragraph{Results}
With linear facets, our graph-coordinated method has lower median Hausdorff error than LVIRA at five of the six plotted resolutions and remains within $2\%$ at every resolution. As in the circle benchmark, this difference is consistent with LVIRA's broader $3\times3$ stencil producing smoother interfaces  between neighboring facets rather than a closer fit to the generating interface. Adding circular facets reduces median Hausdorff distance by $3.7$--$6.7\times$ and median facet gap by $6.7$--$19\times$ relative to LVIRA across the plotted resolutions (\Cref{fig:results_ellipses}). For our circular facet method, facet gap converges at approximately third order, while curvature error converges at approximately first order. We present additional results on perturbed meshes in~\Cref{fig:appendix_ellipses_all_methods}.

\begin{figure}[p]
  \centering
  \includegraphics[width=0.96\textwidth]{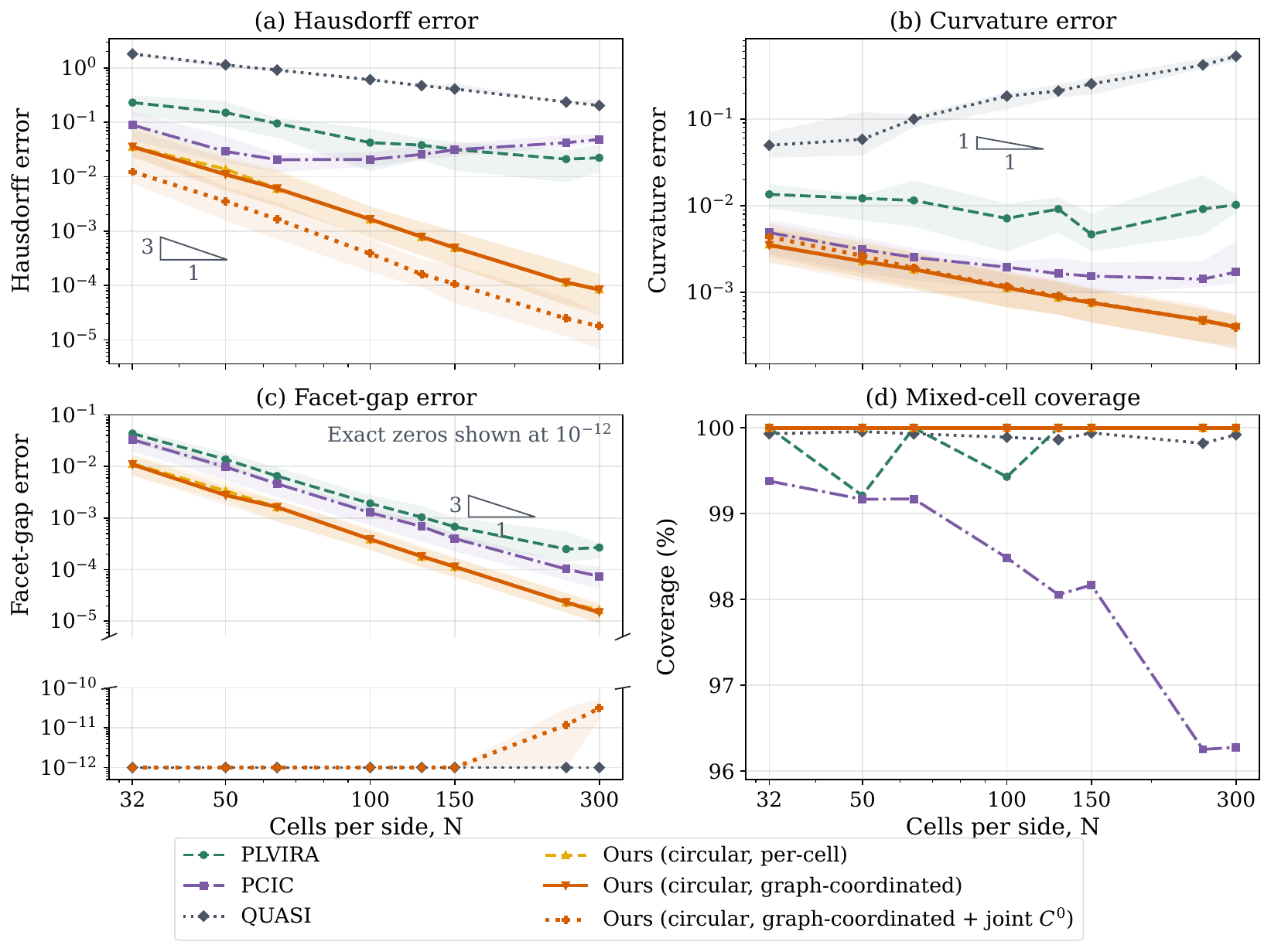}
  \caption{\textbf{Higher-order ellipse comparison on Cartesian meshes.} At each $N$, we report the median and interquartile range over all $25$ benchmark configurations. Panels show Hausdorff distance, curvature error, facet gap, and reconstructed mixed-cell coverage. The per-cell and graph-coordinated circular variants both give approximately third-order facet-gap convergence and first-order convergence in curvature error. Joint $C^0$ refinement gives approximately third-order Hausdorff convergence and drives the facet gap to the numerical floor without changing the convergence order of curvature error. At $N=300$, the full method has $1{,}240\times$ lower Hausdorff error than PLVIRA and $4.3\times$ lower curvature error than PCIC. The slight increase in facet gap at the two finest resolutions remains at the numerical floor and reflects accumulated floating-point error. QUASI has zero median facet gap, but its curvature error increases over the plotted resolutions. The coverage panel reports the fraction of mixed cells for which each method returns a facet. All variants of our method achieve $100\%$ coverage. Reproduction details appear in Appendix~\ref{appendix_external_baseline_scope}.
  Broken y-axes separate the numerical floor from larger error ranges, and curves are not connected across omitted intervals.
  }
  \label{fig:results_ellipses_higher_order}
\end{figure}

To compare with higher-order baselines, we next restrict the ellipse benchmark to Cartesian meshes and extend the resolution study through $N=300$.
As shown in \Cref{fig:results_ellipses_higher_order}, our per-cell and graph-coordinated circular variants have nearly identical convergence rates on Cartesian ellipses: roughly third order in Hausdorff error ($2.79$ and $2.76$, respectively) and first order in curvature error ($0.97$ for both variants). Joint $C^0$ refinement further raises the Hausdorff order to $2.99$ and drives the facet gap error to the numerical floor, while the curvature error converges at approximately first order.

\Cref{fig:results_ellipses_higher_order} also compares our approach with PLVIRA, PCIC, and QUASI.
All three of our circular variants have lower median Hausdorff and curvature errors at every resolution through $N=300$.
At the highest resolution, our best-performing method, the full graph-coordinated method with joint $C^0$ refinement, yields median Hausdorff and curvature errors of $1.80\times10^{-5}$ and $3.98\times10^{-4}$.
These are more than $1{,}200\times$ lower than PLVIRA and $4.3\times$ lower than PCIC, respectively, the strongest baselines for these metrics.
We show representative low-resolution ellipse reconstructions for PLVIRA, PCIC and QUASI in~\Cref{fig:appendix_maity_ellipse}.

\subsection{Zalesak Disk Benchmark}
\label{sec:results-zalesak}

\begin{figure}[p]
  \centering
  \includegraphics[width=0.92\textwidth]{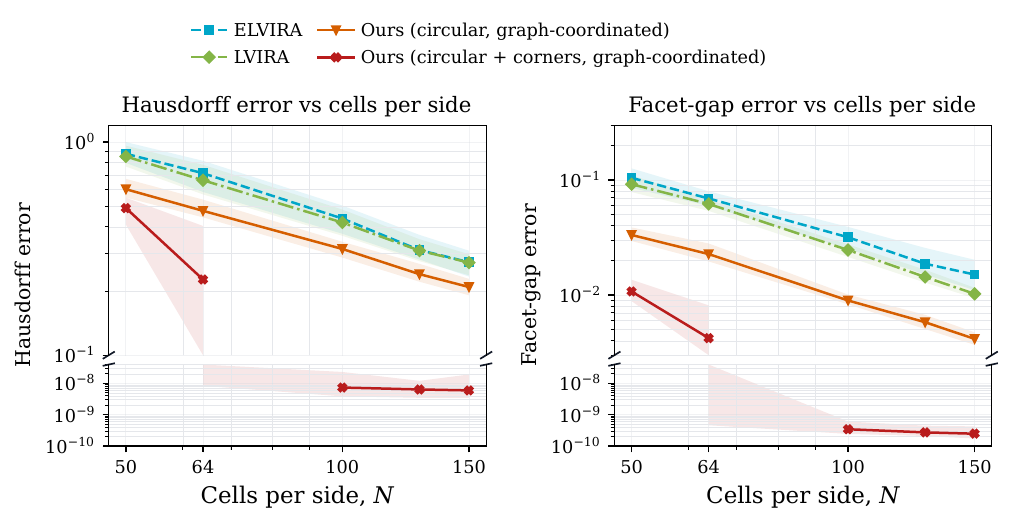}

  \includegraphics[width=0.84\textwidth]{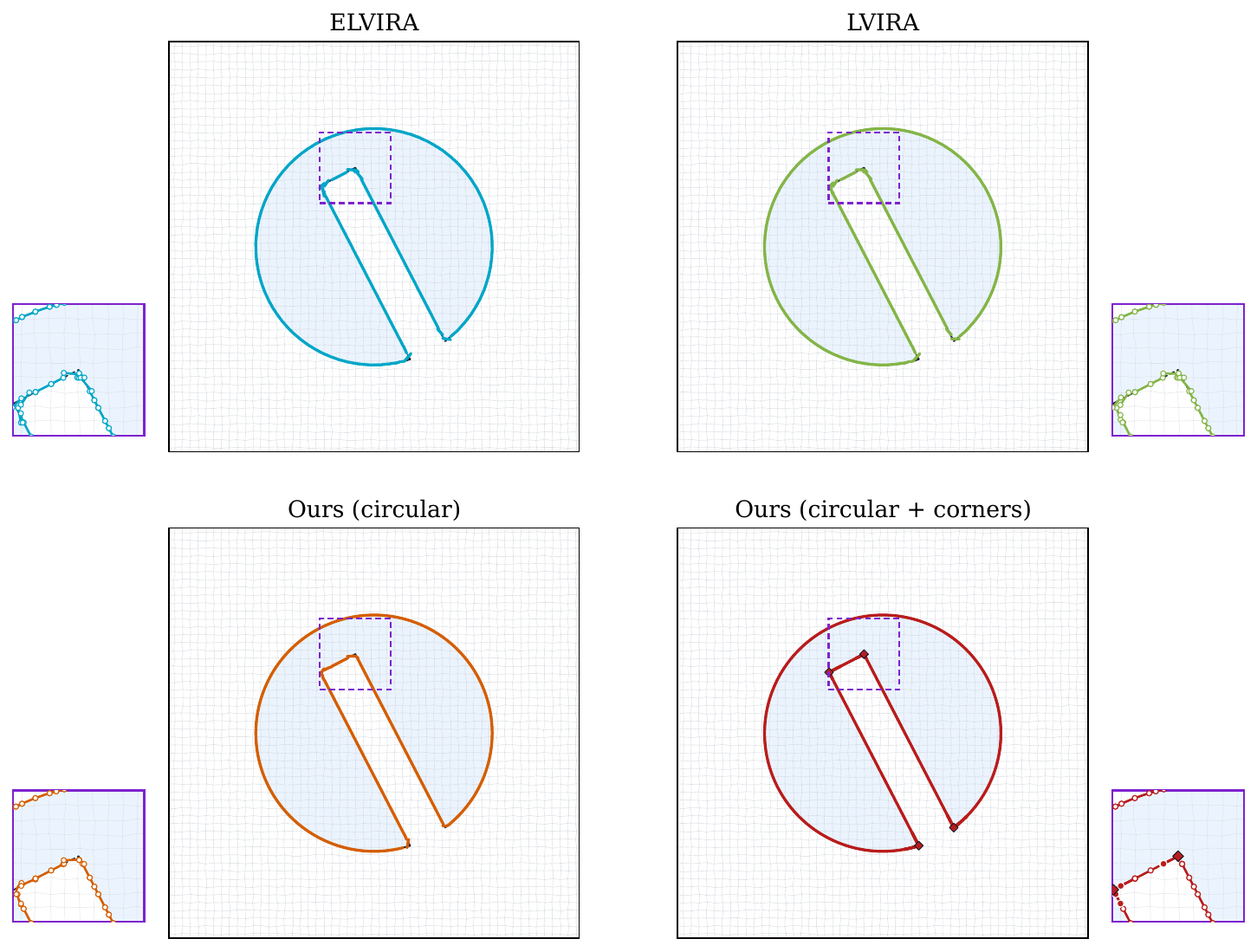}
  \caption[Zalesak benchmark]{\textbf{Zalesak benchmark.} Upper panels: Hausdorff distance and facet gap versus cells per side, $N$. At each $N$, curves show the median and shaded bands show the interquartile range over all $125$ combinations of perturbation magnitude and benchmark configuration. The linear methods approximate the straight slot walls and circular boundary, but their largest errors remain near the sharp line--arc transitions. Circular facets improve the smooth parts of the interface and smoothen the representations near the corners but still cannot capture corners accurately. Adding explicit line--arc corner primitives brings both the median and the full interquartile range to the numerical floor from $N=100$ onward. At $N=50$ and $64$, some corner configurations remain topologically ambiguous, which triggers linear fallbacks and raises the error. Lower panels: a representative reconstruction at $N=100$ and $w=0.1$. The dashed black curve is the generating interface. The spyglasses magnify a slot corner. Diamonds mark corner tips, and open circles mark intersections of corner branches with cell boundaries.
  Broken y-axes separate the numerical floor from larger error ranges, and curves are not connected across omitted intervals.
  }
  \label{fig:results_zalesak}
  \label{fig:results_zalesak_representative}
\end{figure}

\paragraph{Setup}
The Zalesak benchmark (\Cref{fig:results_zalesak}) combines constant-curvature arcs, straight slot walls, and sharp line--arc transitions, and therefore tests the full sequence of geometric primitives in our method. Following the slotted-disk benchmark of Zalesak~\citep{Zalesak1979}, we use disk radius $R=15$, slot width $W=5$, and slot-top height $y_{\mathrm{top}}=10$ relative to the disk center. Across the $25$ configurations, the center is sampled uniformly from $[50,51]^2$ and the geometry is rotated by an angle sampled uniformly from $[0,\pi/2)$. Its exact phase area is
\[
A=\pi R^2-\left(Wy_{\mathrm{top}}+q\sqrt{R^2-q^2}+R^2\arcsin(q/R)\right),
\qquad q=\min(W/2,R).
\]

\paragraph{Results}
With circular facets but no explicit corners, our graph-coordinated method reduces median Hausdorff distance by $22$--$30\%$ and median facet gap by $2.5$--$2.8\times$ relative to LVIRA across the plotted resolutions (\Cref{fig:results_zalesak}). The remaining error is concentrated near the slot corners, where one circular arc cannot represent a sharp line--arc transition.

Adding line--arc corner primitives, our method reaches the numerical floor in median Hausdorff distance and facet gap from resolution $N=100$ onward. At $N=50$ and $64$, regions near the slot corners remain under-resolved by the graph-coordinated topology identification algorithm, so linear facet fallbacks are introduced that lower the overall reconstruction fidelity. Complete results across mesh resolutions and perturbation magnitudes appear in \Cref{fig:appendix_zalesak_all_methods} and additional reconstructions across resolutions and mesh families appear in \Cref{fig:appendix_zalesak_resolution}.
On Cartesian meshes at $N=300$, our graph-coordinated corner-aware method has median Hausdorff error $1.34\times10^{-8}$, compared with $0.125$ for PLVIRA, the strongest higher-order baseline at this resolution, a reduction by approximately $9.4\times10^6$.
Joint $C^0$ refinement further lowers the median Hausdorff error to $8.56\times10^{-9}$.
A representative Cartesian comparison with PLVIRA, PCIC, and QUASI is shown in \Cref{fig:appendix_zalesak_higher_order_gallery}.

\subsection{Effects of Modular Components}
\label{sec:results-component-effects}

\paragraph{Graph coordination and cell merging}

Graph coordination improves reconstruction on perturbed meshes.
To assess topology identification on the exactly representable line benchmark, we examine two complementary diagnostics: shared-vertex conflicts, where facets incident to the same mesh vertex assign it different material labels, and the high-error fraction, defined as the fraction of cases with mean facet gap significantly above the numerical floor (specifically, $g_h>10^{-7}$).

On the exactly representable line benchmark, increasing the perturbation from $w=0$ to $w=0.3$ raises the fraction of per-cell cases with shared-vertex conflicts from $0\%$ to $29.3\%$ and the high-error fraction from $0\%$ to $45.3\%$.
With graph coordination, these fractions are only $2.7\%$ and $3.3\%$ at $w=0.3$, and the median errors remain at the numerical floor.
Graph coordination also reduces the facet gap for circular facets on ellipses.
At $w=0.3$, graph coordination lowers the median facet gap from $1.68\times10^{-3}$ to $6.40\times10^{-4}$ (\Cref{fig:appendix_ellipses_all_methods}).

Selective cell merging is crucial for perfect interface recovery near sharp features, but is rarely needed.
Across the complete benchmark suite, only $0.419\%$ of mixed cells participate in a merge (\Cref{tab:merging_incidence}).
Merging occurs in $48.0\%$ of square instances and $57.4\%$ of Zalesak instances, but involves only $1.26\%$ and $0.87\%$ of their mixed cells, respectively.
Most merges are also small.
The median affected instance contains a single two-cell merge, and ambiguities involving three or more candidate neighbors affect only $0.0248\%$ of mixed cells.
A representative multi-neighbor case appears in \Cref{fig:appendix_actual_multineighbor_merge}.

\begin{table}[t]
\centering
\begin{minipage}{0.86\textwidth}
  \centering
  \small
  \caption{Incidence of cell merging in the static benchmark suite.}
  \label{tab:merging_incidence}
  {\renewcommand{\arraystretch}{1.08}
  \setlength{\tabcolsep}{8pt}
  \begin{tabular}{l|rr}
    \hline
    \raisebox{0.7ex}{\textbf{Benchmark}} & \shortstack{\textbf{Instances with}\\\textbf{a merge}} & \textbf{\boldmath$\displaystyle\frac{\text{\# merged mixed cells}}{\text{\# all mixed cells}}$} \\[1.0ex]
    \hline
    \textbf{Lines}    & $1.47\%$ & $0.023\%$ \\
    \textbf{Circles}  & $2.53\%$ & $0.072\%$ \\
    \textbf{Ellipses} & $1.87\%$ & $0.023\%$ \\
    \textbf{Squares}  & $48.0\%$ & $1.26\%$ \\
    \textbf{Zalesak}  & $57.4\%$ & $0.87\%$ \\
    \hline
    \textbf{All}      & $20.1\%$ & $0.419\%$ \\
    \hline
  \end{tabular}}
\end{minipage}
\end{table}

\paragraph{\texorpdfstring{$C^0$ refinement}{C0 refinement}}
We next consider the benefits of our $C^0$ refinement strategy and compare the two correction steps discussed in~\Cref{sec:c0}.
The parallel local pass closes $89.89\%$ of eligible joins in the complete ellipse and Zalesak study.
The remaining joins form small connected components with a median of $3$ facets, and $90\%$ contain at most $5$ facets.
Joint refinement raises the observed $C^0$ rate to $99.81\%$ under the $\varepsilon_{\mathrm{gap}}=10^{-8}$ endpoint-gap tolerance.
Stagewise rates and component sizes appear in \Cref{tab:appendix_c0_stage_incidence} and Figure~\ref{fig:appendix_c0_component_sizes}.

Joint refinement lowers the median Hausdorff error on the ellipse benchmark in all $30$ perturbation--resolution settings and does not worsen the median Hausdorff error on the Zalesak benchmark in any of the $25$ settings, with $24$ improvements and one tie.
Across these settings, the median facet gap falls from $2.59\times10^{-3}$ to zero for ellipses and from $3.19\times10^{-3}$ to $8.79\times10^{-11}$ for Zalesak, which is at the numerical floor.
In the perturbed-mesh ellipse study, we apply joint refinement to our graph-coordinated linear reconstruction.
The refinement introduces curvature while preserving each participating cell's volume fraction.
Across the six resolutions in \Cref{fig:appendix_ellipses_c0_metrics}, it lowers the median Hausdorff error by $1.4\times$--$2.0\times$ and drives the median facet gap from $3.36\times10^{-3}$--$7.32\times10^{-2}$ to zero.
In the Cartesian higher-order ellipse study, we apply joint refinement to our graph-coordinated circular reconstruction.
At $N=300$, joint refinement lowers the median Hausdorff error from $8.40\times10^{-5}$ to $1.80\times10^{-5}$.
Across the full Cartesian resolution sweep, it increases the observed Hausdorff convergence order from $2.76$ to $2.99$, while closing nearly every eligible facet gap, and curvature error remains approximately first order.
Local concavity remains rare, affecting $0.047\%$ of reconstructed arcs in the Cartesian ellipse study.
We present additional aggregate results and representative reconstructions in Appendix~\ref{appendix_c0_study}.

\paragraph{Observed primitive use}
Table~\ref{tab:primitive_incidence} summarizes the distributions of final primitives across our experiments.
Circles and ellipses use arcs in more than $99.95\%$ of mixed cells, squares use line--line corners in $12.57\%$, and only the Zalesak benchmark involves every primitive type.

\begin{table}[t]
\centering
\begin{minipage}{\textwidth}
  \centering
  \footnotesize
  \caption{Fraction of original mixed cells represented by each final geometric primitive.}
  \label{tab:primitive_incidence}
  {\setlength{\tabcolsep}{3.8pt}
  \begin{tabular}{l|rrrrrr}
    \hline
    \textbf{Benchmark} & \textbf{Mixed cells} & \textbf{Line} & \textbf{Arc} & \textbf{Line--line} & \textbf{Line--arc} & \textbf{Arc--arc} \\
    \hline
    \textbf{Lines}    & $95{,}040$  & $100.000\%$ & $0$ & $0$ & $0$ & $0$ \\
    \textbf{Circles}  & $52{,}855$  & $0.021\%$ & $99.979\%$ & $0$ & $0$ & $0$ \\
    \textbf{Ellipses} & $120{,}180$ & $0.048\%$ & $99.952\%$ & $0$ & $0$ & $0$ \\
    \textbf{Squares}  & $62{,}581$  & $87.426\%$ & $0$ & $12.574\%$ & $0$ & $0$ \\
    \textbf{Zalesak}  & $113{,}042$ & $31.805\%$ & $60.119\%$ & $2.539\%$ & $5.529\%$ & $0.008\%$ \\
    \hline
    \textbf{All} & $443{,}698$ & $41.869\%$ & $54.300\%$ & $2.420\%$ & $1.409\%$ & $0.002\%$ \\
    \hline
  \end{tabular}}
\end{minipage}
\end{table}

\paragraph{Fallback use}
\textsc{FitLinearFacet} is used when an oriented higher-order fit fails, whereas LVIRA is used when orientation remains unresolved.
As shown in \Cref{tab:fallback_incidence}, these fallbacks account for only $0.0834\%$ of mixed cells across the complete suite.

\begin{table}[t]
\centering
\begin{minipage}{0.94\textwidth}
  \centering
  \small
  \captionof{table}{Incidence of \textsc{FitLinearFacet} and LVIRA fallbacks. Percentages in the cell columns are relative to all mixed cells for that benchmark, and percentages in the instance columns are relative to all benchmark instances.}
  \label{tab:fallback_incidence}
  {\setlength{\tabcolsep}{5pt}
  \begin{tabular}{l|rrrr}
    \hline
    \textbf{Benchmark} & \textbf{\textsc{FitLinearFacet} cells} & \textbf{Instances} & \textbf{LVIRA cells} & \textbf{Instances} \\
    \hline
    \textbf{Lines}    & $0$ & $0$ & $293\;(0.3083\%)$ & $2\;(0.267\%)$ \\
    \textbf{Circles}  & $1\;(0.0019\%)$ & $1\;(0.133\%)$ & $4\;(0.0076\%)$ & $2\;(0.267\%)$ \\
    \textbf{Ellipses} & $10\;(0.0083\%)$ & $8\;(1.07\%)$ & $12\;(0.0100\%)$ & $6\;(0.800\%)$ \\
    \textbf{Squares}  & $0$ & $0$ & $22\;(0.0352\%)$ & $13\;(2.08\%)$ \\
    \textbf{Zalesak}  & $0$ & $0$ & $28\;(0.0248\%)$ & $16\;(2.56\%)$ \\
    \hline
    \textbf{All}      & $11\;(0.0025\%)$ & $9\;(0.257\%)$ & $359\;(0.0809\%)$ & $39\;(1.11\%)$ \\
    \hline
  \end{tabular}}
\end{minipage}
\end{table}


\section{Conclusion}
\label{sec:conclusion}

We presented a geometric interface reconstruction method that combines graph-based propagation of local orientation constraints with progressive fitting of line, circular arc, and corner primitives, achieving our dual goals of \emph{topological consistency} and \emph{perfect reconstruction}.
Across static reconstruction benchmarks, the method recovers identifiable lines, circles, and sufficiently resolved corners to numerical precision.
Circular facets improve reconstruction of smooth interfaces relative to piecewise-linear reconstruction, achieving third-order convergence of facet gaps and approximately first-order convergence of curvature error on varying-curvature interfaces.
An optional conservative $C^0$ post-processing correction further closes endpoint gaps by iteratively refining shared edge crossings and neighboring facet curvatures, while preserving local and global volume conservation.

Several directions remain.
The topology and orientation procedure we propose warrants further analysis, including formal guarantees and more global search strategies such as backtracking when local propagation remains ambiguous.
We are also eager to explore extensions of our method to multi-material interfaces, general polygonal meshes (including non-convex cells), and three dimensions.
Incorporating Moment-of-Fluid data could further expand the class of identifiable subcell geometries and further resolve ambiguous topologies in low-resolution regimes.
Finally, coupling the method to VOF advection would determine how these static reconstruction gains translate to transport accuracy.

\clearpage

\section*{CRediT authorship contribution statement}

\textbf{Jerry Liu:} Conceptualization, Methodology, Software, Validation, Investigation, Visualization, Writing -- original draft, Writing -- review \& editing.
\textbf{Kenneth Weiss:} Conceptualization, Supervision, Writing -- review \& editing.
\textbf{Jin Yao:} Conceptualization, Methodology, Supervision, Writing -- review \& editing.

\section*{Declaration of competing interest}

The authors declare that they have no known competing financial interests or personal relationships that could have appeared to influence the work reported in this paper.

\section*{Data and code availability}

The source code, experiment scripts, and benchmark data are available in the versioned software and data release~\citep{Liu2026InterfaceCodeData}.
The source code is distributed under the Apache License 2.0, and the research data are distributed under the Creative Commons Attribution 4.0 International License.

\section*{Acknowledgments}

This work was performed under the auspices of the U.S. Department of Energy by Lawrence Livermore National Laboratory under Contract DE-AC52-07NA27344. Jerry Liu was supported by the Department of Energy Computational Science Graduate Fellowship under Award DE-SC0023112.

\clearpage

\appendix
\makeatletter
\@addtoreset{equation}{section}
\@addtoreset{figure}{section}
\@addtoreset{table}{section}
\makeatother
\renewcommand{\theequation}{\thesection.\arabic{equation}}
\renewcommand{\thefigure}{\thesection.\arabic{figure}}
\renewcommand{\thetable}{\thesection.\arabic{table}}
\renewcommand{\theHequation}{appendix.\thesection.\arabic{equation}}
\renewcommand{\theHfigure}{appendix.\thesection.\arabic{figure}}
\renewcommand{\theHtable}{appendix.\thesection.\arabic{table}}

\section{Implementation details and numerical parameters}
\subsection{PLIC Baselines}
\label{appendix_plic_baselines}

The line-only baselines reconstruct the center cell of a Cartesian $3\times3$ stencil and shift every candidate line to reproduce that cell's prescribed volume. They differ only in how they choose the interface normal.

\subsubsection{Youngs}
Let $a_{i,j}=|P_{i,j}|v_{i,j}$ denote the phase area in cell $P_{i,j}$.
Our Youngs normal estimate~\citep{Youngs1987,ParkerYoungs1992} is
\[
\begin{aligned}
 g_x &= \sum_{m=-1}^{1} w_m\bigl(a_{i+1,j+m}-a_{i-1,j+m}\bigr),
 &g_y &= \sum_{m=-1}^{1} w_m\bigl(a_{i+m,j+1}-a_{i+m,j-1}\bigr),\\
 n_Y &= \frac{(g_x,g_y)}{\sqrt{g_x^2+g_y^2}},
 &(w_{-1},w_0,w_1)&=(1,2,1).
\end{aligned}
\]
On a uniform Cartesian mesh, the same normal is obtained by replacing $a_{i,j}$ with $v_{i,j}$.
The line is then translated along $n_Y$ to reproduce the center-cell volume fraction.

\subsubsection{LVIRA}
LVIRA chooses the normal angle by minimizing the volume-fraction discrepancy over the stencil~\cite{Puckett1991,Puckett1997,Pilliod2004}. With $n(\theta)=(\cos\theta,\sin\theta)$ and the line offset fixed by the center-cell constraint, its objective is
\[
E_2(\theta)^2 = \sum_{k,l=-1}^{1}\left(\widetilde v_{i+k,j+l}(\theta)-v_{i+k,j+l}\right)^2,
\]
where $\widetilde v$ is induced by the candidate line. We minimize this one-dimensional objective with a derivative-free local search initialized from the Youngs and ELVIRA orientations. Both half-plane orientations are tested, and the search stops when its angular step is below $\varepsilon_{\theta}$. The tolerance is listed in \Cref{tab:reconstruction_parameters}.

\subsubsection{ELVIRA}
ELVIRA replaces the continuous search with Cartesian candidates obtained from backward, centered, and forward differences of the stencil's row and column sums~\cite{Pilliod1992,Pilliod2004}. Each of the six slope families is shifted to match the center cell and evaluated with the same least-squares residual, and the best candidate is retained. A generating linear interface on a Cartesian mesh (without boundary effects) produces a zero-residual candidate~\cite{Pilliod2004}. Thus, ELVIRA achieves the perfect reconstruction property of linear interfaces on non-perturbed Cartesian meshes. As illustrated in \Cref{fig:perfect_reconstruction_plic}, this property does not generally extend to perturbed Cartesian meshes.

\subsection{Higher-Order Reconstruction Baselines}
\label{appendix_external_baseline_scope}

The PLVIRA, PCIC, and QUASI comparisons use static reconstructions on uniform, axis-aligned Cartesian meshes ($w=0$).
We implement the interface reconstruction stages described in the cited papers.
We describe our choices for implementation details that the source papers leave ambiguous:

\subsubsection{PLVIRA}
\paragraph{Method}
PLVIRA reconstructs one parabola in each mixed cell~\citep{Remmerswaal2021ParabolicIR}.
It first estimates the local curvature from neighboring Cartesian volume fractions using generalized height functions.
The estimator first calculates curvature from three adjacent interface heights.
When complete height columns are unavailable, it fits a quadratic to other consistent height samples, including samples obtained from preliminary PLIC segments.
If the final curvature fit is degenerate, the estimator returns the zero-curvature limit.
Holding this curvature fixed, it varies the parabola orientation and translates the interface to preserve the target-cell volume fraction.
The selected orientation minimizes the squared volume-fraction error over the complete $3\times3$ stencil.

\paragraph{Reproduction}
We implement this curvature estimator and restricted parabolic fit on uniform Cartesian meshes.
The curvature calculation may inspect cells beyond the $3\times3$ fitting stencil.
If the required neighboring data are unavailable, we report the cell as unsupported.

\subsubsection{PCIC}
\paragraph{Method}
PCIC reconstructs one circular arc in each mixed cell~\citep{Maity2020}.
It begins with PLIC segments in a Cartesian neighborhood and samples points along those segments.
A weighted circle fit through these points determines the candidate center and radius.
A final conservative correction recovers the target-cell volume fraction either by translating the center at fixed radius or by adjusting the radius at fixed center.

\paragraph{Reproduction}
Since the paper does not identify which conservative correction produced its reported bare-PCIC results, we implemented and evaluated both variants.
In a preliminary matched study over five benchmarks, we found that center translation left almost $5\times$ fewer mixed cells unresolved compared to radius adjustment.
We therefore use center translation in the reported comparisons.
We omit the optional smoothing and $C^0$ passes because their geometric details and update order are not reported sufficiently to reproduce them precisely.
Cases with no admissible crossing or conservative correction are counted as unsupported.

\subsubsection{QUASI}
\paragraph{Method}
QUASI begins with a Youngs PLIC reconstruction in each mixed cell and replaces each line with an area-preserving quadratic segment~\citep{Diwakar2009AQS}.
It first moves neighboring endpoints to establish a shared crossing where possible.
It then sweeps over adjacent cells and adjusts their shared endpoints while refitting both quadratic segments to preserve their local volume fractions.
The method aims to establish continuous endpoints and matching tangent directions between neighboring segments.

\paragraph{Reproduction}
The paper does not discuss how to select among multiple admissible roots, handle a missing root, order the updates, or stop the continuity sweeps.
We select the admissible root on the shared edge that is nearest the current endpoint.
If no such root exists, we retain the original area-preserving quadratic.
We apply deterministic in-place updates and perform curvature correction before the continuity sweeps.
The sweeps stop after ten passes or when the maximum endpoint displacement falls below $10^{-11}$.

\subsection{Geometric facet fitting details}
\label{appendix_geometric_fitting_details}

This subsection gives more geometric details about the \textsc{FitLinearFacet} and \textsc{FitCircularFacet} algorithms described in \Cref{sec:fit-linear,sec:fit-circular}.

\subsubsection{\textsc{FitLinearFacet}}
\label{appendix_fit_linear_details}
For the two support cells, we define the dimensionless volume-fraction residuals $e_i=\widehat v_i-v_i$, $i=1,2$, where $\widehat v_i$ is the volume fraction induced by the candidate line.
The line solver uses coupled Newton steps for normal displacements of two support points, with alternating scalar searches when the Jacobian is singular.
The $2\times2$ Jacobian is estimated by centered finite differences with step $\Delta_{\mathrm{FD}}=10^{-8}$ in coordinate units, perturbing one support point at a time along the current normal while holding the other point fixed.

\begin{algorithm}[H]
\small
\DontPrintSemicolon
\caption{\textsc{FitLinearFacet}: two-cell half-plane fitting}
\label{alg:fit-linear}
\KwIn{Ordered supports $(P_1,v_1),(P_2,v_2)$; tolerance $\varepsilon_{\mathrm{supp}}$; maximum iterations $K_{\max}$}
\KwOut{Oriented support line $\ell$, or failure}
Initialize $p_1,p_2$ at the ordered cell centroids\;
\For{$k=1,\ldots,K_{\max}$}{
  Form the oriented line $\ell$ through $p_1,p_2$ and its unit normal $n$\;
  Evaluate $e_i=|P_i\cap H_\ell|/|P_i|-v_i$, $i=1,2$\;
  \If{$\max_i|e_i|<\varepsilon_{\mathrm{supp}}$}{\Return $\ell$\;}
  For $j=1,2$, evaluate residual vectors $e^{j,\pm}$ after moving only $p_j$ by $\pm\frac12\Delta_{\mathrm{FD}}n$\;
  Set the Jacobian columns $J_{:j}\leftarrow(e^{j,+}-e^{j,-})/\Delta_{\mathrm{FD}}$, $j=1,2$\;
  \eIf{$J$ is nonsingular}{
    Solve $J\delta=-e$; update $p_i\leftarrow p_i+\delta_i n$, $i=1,2$\;
  }{
    Hold $p_2$ fixed; move $p_1$ along $n$ to match $v_1$\;
    Hold the updated $p_1$ fixed; move $p_2$ along $n$ to match $v_2$\;
  }
}
\Return the final line if its residuals pass; otherwise failure\;
\end{algorithm}

\subsubsection{\textsc{FitCircularFacet}}
\label{appendix_fit_circular_details}

\begin{figure}[!htbp]
  \centering
  \resizebox{0.48\textwidth}{!}{\includegraphics{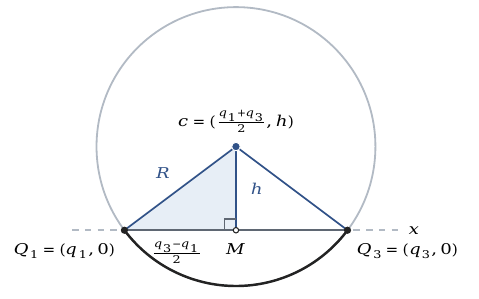}}
  \caption{\textbf{Circle-center construction used in \textsc{FitCircularFacet}.} The circle passes through $Q_1$ and $Q_3$ on the horizontal support line. Its center $c$ lies above the chord midpoint $M$, the shaded right triangle has hypotenuse $R$, the horizontal leg is $(q_3-q_1)/2$, and the vertical leg is $h$. Diagram code prepared with assistance from OpenAI Codex.}
  \label{fig:circle-center-geometry}
\end{figure}

\paragraph{Circle-center construction}
In the rotated coordinate frame, the candidate circle intersects the horizontal support line at $Q_1=(q_1,0)$ and $Q_3=(q_3,0)$, where $q_1<q_3$.
Its center lies on the perpendicular bisector of this chord, through $M=((q_1+q_3)/2,0)$.
Writing $c=((q_1+q_3)/2,h)$, the right triangle in \Cref{fig:circle-center-geometry} gives
\begin{equation}
h^2+\left(\frac{q_3-q_1}{2}\right)^2=R^2.
\label{eq:circle-center-triangle}
\end{equation}
Choosing the branch with the center above the support line therefore yields
\begin{equation}
c=\left(\frac{q_1+q_3}{2},\sqrt{R^2-\frac{(q_3-q_1)^2}{4}}\right),\qquad R\geq\frac{|q_3-q_1|}{2}.
\label{eq:circle-center-construction}
\end{equation}

\paragraph{Fitting algorithm}
For the three support cells, we define the dimensionless volume-fraction residuals $e_i=\widehat v_i-v_i$, $i=1,2,3$, where $\widehat v_i$ is the volume fraction induced by the candidate signed circle.
The circular solver alternates expanding and halving radius searches with finite-difference Newton updates of the intercepts, maintaining $|q_3-q_1|\leq2R$.
A candidate circular facet is retained only if it satisfies the residual tolerance and defines an admissible arc in the middle cell.

\begin{algorithm}[H]
\small
\DontPrintSemicolon
\caption{\textsc{FitCircularFacet}: three-cell signed circle fitting}
\label{alg:fit-circular}
\KwIn{Ordered supports $(P_i,v_i)$, $i=1,2,3$; tolerance $\varepsilon_{\mathrm{arc}}$; maximum iterations $K_{\max}$}
\KwOut{Signed circle $(c,r)$, or failure}
$\ell\leftarrow\textsc{FitLinearFacet}((P_1,v_1),(P_3,v_3),\varepsilon_{\mathrm{arc}}/10,K_{\max})$\;
\If{the line fit fails}{\Return failure\;}
Select the phase configuration using $\ell$\;
Transform the support line to the horizontal axis\;
Initialize $q_1,q_3$ at the midpoints of the outer line--cell intersections; $R\leftarrow|q_3-q_1|$\;
Restrict all intercept updates to $|q_3-q_1|\leq2R$\;
\For{$k=1,\ldots,K_{\max}$}{
  Hold $q_1,q_3$ fixed; solve $e_2=0$ in $R$ by an expanding and halving search\;
  Hold $R,q_3$ fixed; solve $e_1=0$ in $q_1$ using finite-difference Newton steps\;
  Hold $q_1,q_3$ fixed; solve $e_2=0$ in $R$ again by an expanding and halving search\;
  Hold $R,q_1$ fixed; solve $e_3=0$ in $q_3$ using finite-difference Newton steps\;
  Recompute $c$ using \Cref{eq:circle-center-construction}, then evaluate all residuals\;
  \If{$\max_i|e_i|<\varepsilon_{\mathrm{arc}}$}{
    Restore the original coordinates and phase convention; \Return $(c,r)$\;
  }
}
\Return failure\;
\end{algorithm}

\subsection{Numerical parameters}
\label{appendix_reproducibility_runtime}

Table~\ref{tab:reconstruction_parameters} defines the symbols and values of the fixed numerical parameters used throughout our static reconstruction benchmarks. Unless otherwise indicated, residuals are dimensionless volume-fraction residuals.

\begin{table}[tb]
  \centering
  \begin{minipage}{0.98\textwidth}
  \centering
  \small
  \caption{Numerical parameters used in the reported static experiments.}
  \label{tab:reconstruction_parameters}
  \begin{tabular}{p{0.30\textwidth} p{0.17\textwidth} p{0.11\textwidth} p{0.27\textwidth}}
    \hline
    \textbf{Quantity} & \textbf{Symbol} & \textbf{Value} & \textbf{Role} \\
    \hline
    General optimization tolerance & $\varepsilon_{\mathrm{opt}}$ & $10^{-10}$ & Double-precision local solves \\
    Mixed-cell classification tolerance & $\varepsilon_{\mathrm{mix}}$ & $10^{-10}$ & Classification of volume fractions as empty, mixed, or full \\
    Straight-support residual & $\varepsilon_{\mathrm{supp}}$ & $10^{-10}$ & Two-cell line solve \\
    Linear target-cell residual & $\varepsilon_{\mathrm{lin}}$ & $10^{-6}$ & Acceptance of a straight facet \\
    Circular residual & $\varepsilon_{\mathrm{arc}}$ & $10^{-5}$ & Acceptance of a three-cell arc \\
    Straight-corner residual tolerance & $\varepsilon_{\mathrm{SC}}$ & $10^{-4}$ & Straight-corner acceptance \\
    Near-parallel tangent threshold & $\tau_{\parallel}$ & $10^{-2}$ & Rejection when $|\sin\theta|\leq\tau_{\parallel}$ \\
    Curved-corner residual tolerance & $\varepsilon_{\mathrm{CC}}$ & $10^{-2}$ & Curved-corner acceptance \\
    $C^0$ volume-fraction residual tolerance & $\varepsilon_{C^0}$ & $10^{-10}$ & Conservation tolerance for local and joint refinement \\
    $C^0$ endpoint-gap tolerance & $\varepsilon_{\mathrm{gap}}$ & $10^{-8}$ & Identification of joins that remain open after refinement \\
    LVIRA angular step tolerance & $\varepsilon_{\theta}$ & $10^{-6}$ & Termination of the orientation search \\
    \hline
  \end{tabular}
  \end{minipage}
\end{table}

\clearpage
\section{Supplementary Static Benchmark Results}
\label{appendix_static_benchmarks}

\subsection{Additional Quantitative Benchmark Results}
\label{appendix_static_full_panels}

This section extends the paper's main resolution studies with perturbation sweeps and additional diagnostics.
For resolution studies, the median and interquartile range pool all $125$ combinations of perturbation magnitude and benchmark configurations.
For perturbation magnitude studies, they pool all plotted resolutions and configurations.
This gives $150$ observations per perturbation magnitude for lines, circles, and ellipses and $125$ for squares and Zalesak.

\subsubsection{Line Benchmark}

\begin{figure}[!htbp]
  \centering
  \includegraphics[width=0.95\textwidth]{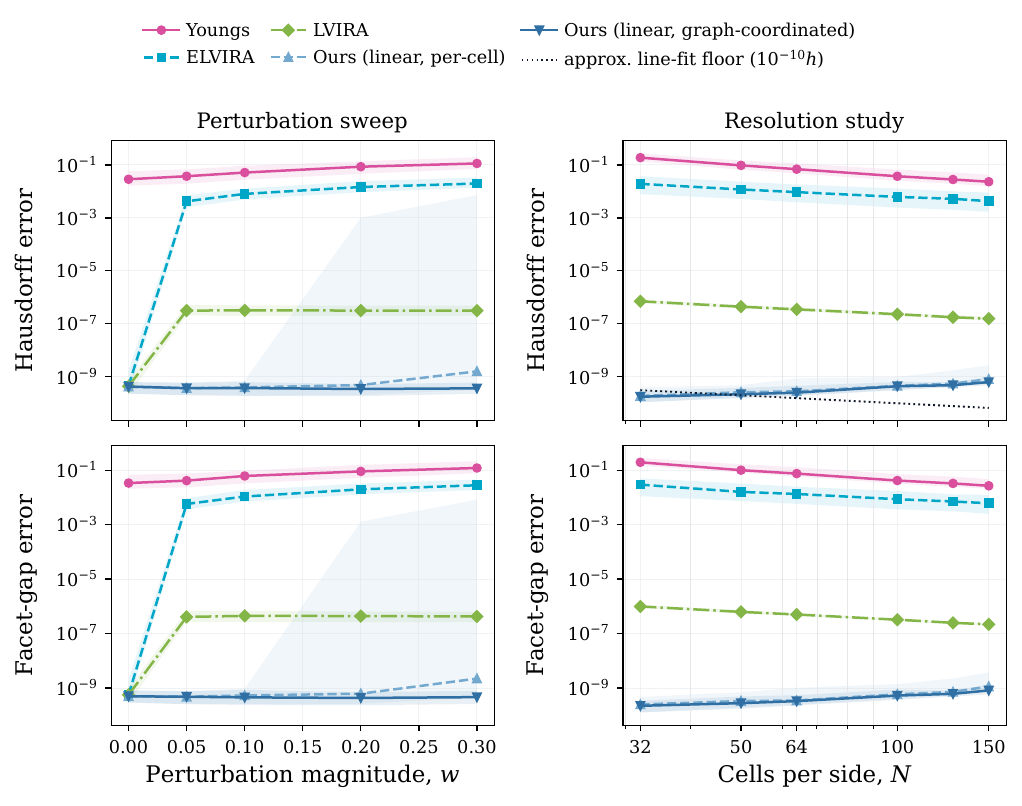}
  \caption{\textbf{Full line comparison across all methods, including the per-cell and graph-coordinated linear variants.}
  The graph-coordinated linear reconstruction maintains numerical-floor Hausdorff distance and facet gap across both sweeps.
  Both errors are at least two orders of magnitude lower than for LVIRA, the strongest classical baseline.}
  \label{fig:appendix_lines_all_methods}
\end{figure}

Graph coordination keeps line reconstruction stable as mesh perturbation increases (\Cref{fig:appendix_lines_all_methods}).
Across the resolution sweep, the graph-coordinated method has median Hausdorff distances from $1.77\times10^{-10}$ to $6.33\times10^{-10}$ and median facet gaps from $2.22\times10^{-10}$ to $8.15\times10^{-10}$.
Both errors are at least two orders of magnitude lower than those of LVIRA.
One source of the per-cell error tail is a shared-vertex conflict, in which facets incident to the same mesh vertex place that vertex on opposite sides of the reconstructed interface.
Every detected per-cell conflict has facet-gap error above $10^{-7}$.
At $w=0.3$, graph coordination reduces the fraction of cases with such a conflict from $29.3\%$ to $2.7\%$.

\clearpage
\subsubsection{Square Benchmark}

\begin{figure}[!htbp]
  \centering
  \includegraphics[width=\textwidth]{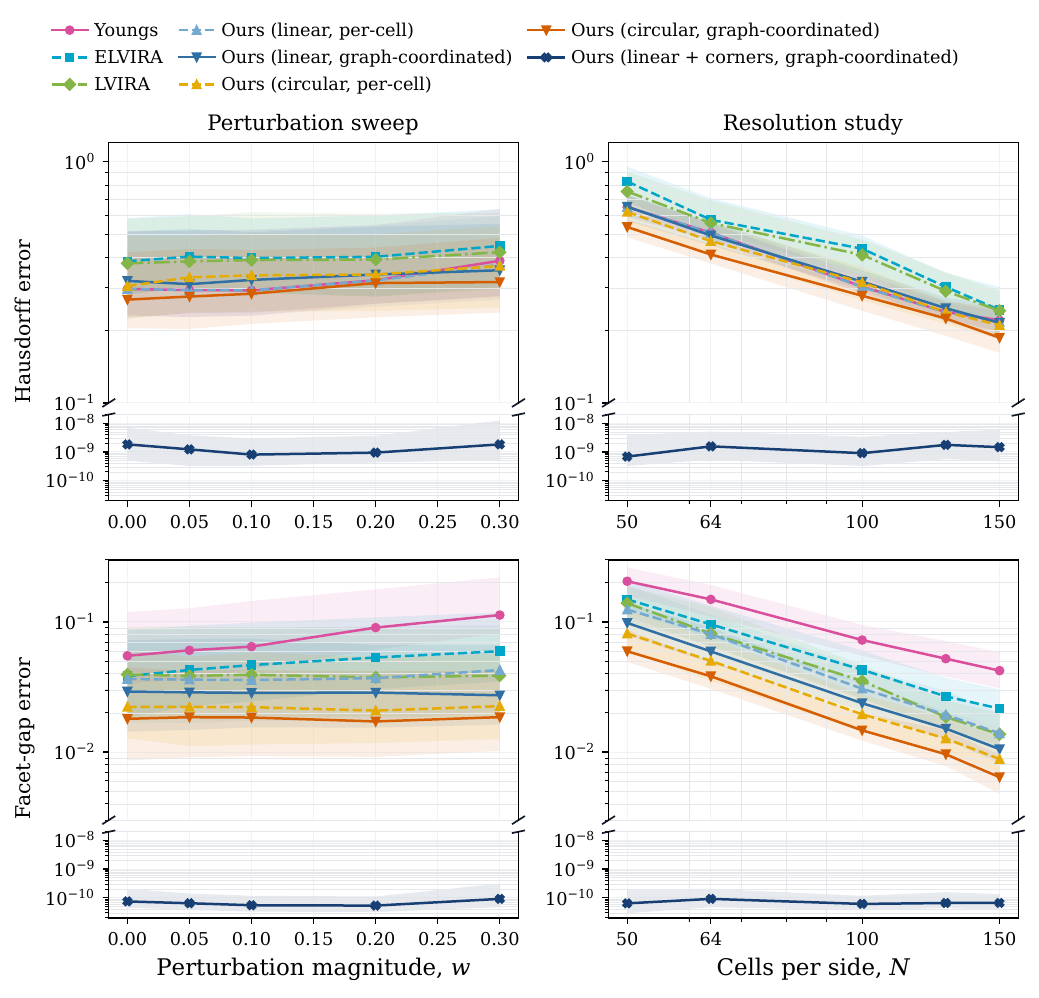}
  \caption{\textbf{Full square comparison across all methods, including per-cell and graph-coordinated linear and circular variants and the graph-coordinated corner-aware method.}
  Relative to LVIRA, the graph-coordinated linear method reduces the median Hausdorff distance by $11$--$23\%$ and the median facet gap by $18$--$33\%$ across resolutions.
  With explicit corners, our approach achieves the numerical floor.
  Broken y-axes separate the numerical floor from larger error ranges, and curves are not connected across omitted intervals.
  }
  \label{fig:appendix_squares_all_methods}
\end{figure}

Graph coordination reduces facet gaps for line-only square reconstruction, while explicit corners provide exact recovery (\Cref{fig:appendix_squares_all_methods}).
Across the resolution sweep, graph coordination reduces the median facet gap by $21$--$26\%$, and across the perturbation sweep, by $20$--$36\%$.
The per-cell and graph-coordinated linear variants have similar Hausdorff distances, but the graph-coordinated variant remains $11$--$23\%$ below LVIRA across resolutions.
Adding line--line corners lowers the median Hausdorff distance to at most $1.76\times10^{-9}$ and the median facet gap to at most $9.20\times10^{-11}$.
Both errors improve by more than eight orders of magnitude relative to the graph-coordinated linear method.

\clearpage
\subsubsection{Circle Benchmark}

\begin{figure}[!htbp]
  \centering
  \includegraphics[height=0.78\textheight,width=0.95\textwidth,keepaspectratio]{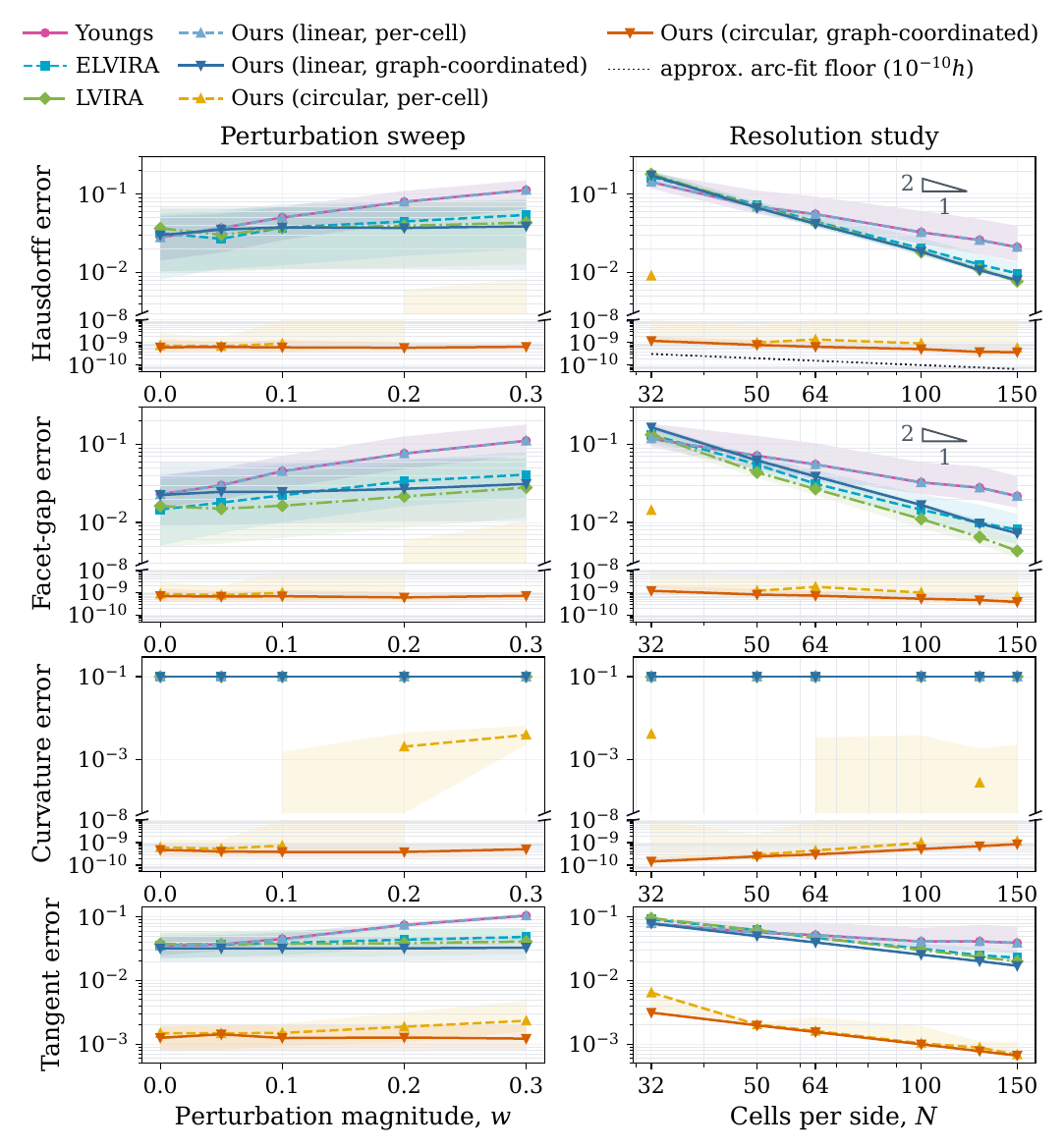}
  \caption{\textbf{Full circle comparison across all methods and diagnostics, including per-cell and graph-coordinated linear and circular variants.}
  Graph-coordinated circular facets reach the numerical floor in Hausdorff distance and facet gap.
  Among the linear methods, the graph-coordinated reconstruction has lower median Hausdorff error than ELVIRA at five of six resolutions and remains within $4\%$ of LVIRA.
  Broken y-axes separate the numerical floor from larger error ranges, and curves are not connected across omitted intervals.
  }
  \label{fig:appendix_circles_all_methods}
\end{figure}

As shown in \Cref{fig:appendix_circles_all_methods}, graph-coordinated circular facets recover the constant-curvature interface to the numerical floor, even at high mesh perturbation magnitudes.
Across the resolution sweep, their median Hausdorff distance ranges from $3.77\times10^{-10}$ to $1.20\times10^{-9}$, and their median facet gap ranges from $3.83\times10^{-10}$ to $1.20\times10^{-9}$.
At $w=0.3$, the corresponding per-cell medians rise to $2.01\times10^{-3}$ and $1.79\times10^{-3}$, respectively, while the graph-coordinated medians remain below $7.3\times10^{-10}$, an improvement of more than $2.4\times10^6$ at the largest perturbation magnitude.

\begin{figure}[!htbp]
  \centering
  \includegraphics[height=0.72\textheight,width=0.96\textwidth,keepaspectratio]{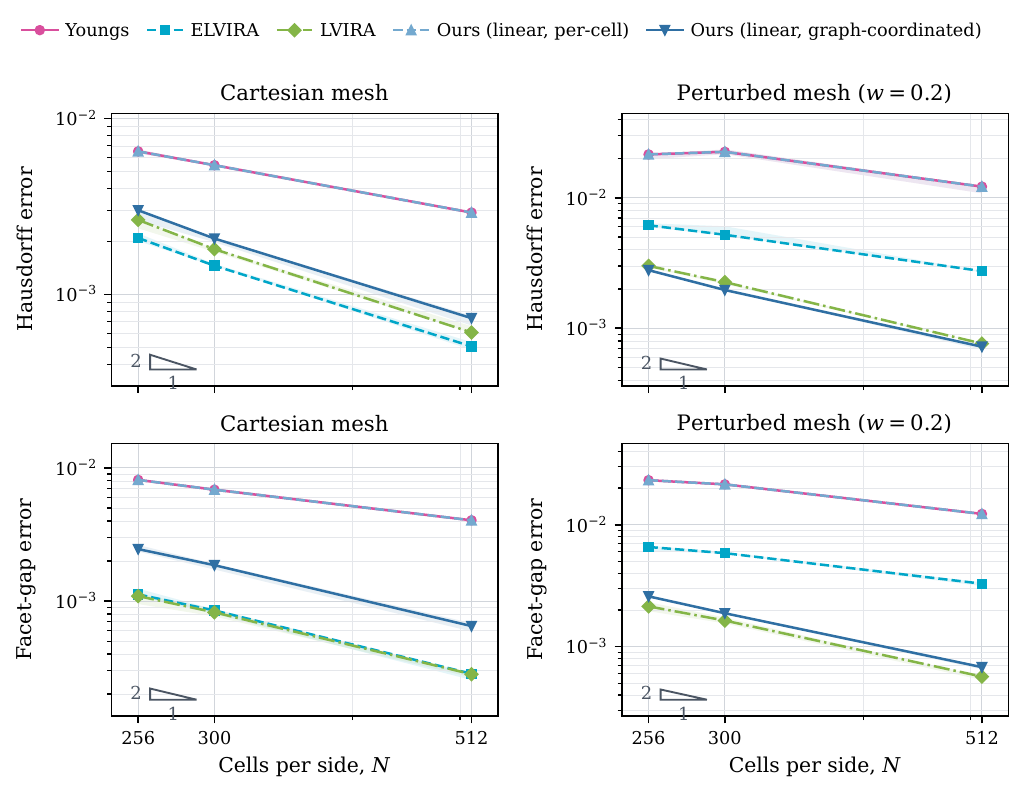}
  \caption{\textbf{Extended convergence of line-only circle reconstruction.}
  Curves show medians over five matched circle configurations, and shaded bands show the interquartile range.
  The left and right columns use Cartesian and perturbed meshes, respectively.
  Graph-coordinated linear reconstruction remains approximately second order in both metrics under mesh perturbation, while ELVIRA and the per-cell variant are approximately first order over this resolution range.}
  \label{fig:appendix_circle_linear_extended}
\end{figure}

\clearpage
The resolution study in \Cref{fig:appendix_circle_linear_extended} confirms that graph-coordinated reconstruction restricted to linear primitives remains approximately second order when applied to circular interfaces.
We evaluate five matched circle configurations at $N\in\{256,300,512\}$.
On perturbed meshes with $w=0.2$, the graph-coordinated linear method has observed orders $1.93$ in Hausdorff distance and $1.92$ in facet gap, compared with $1.99$ and $1.93$ for LVIRA.
The corresponding ELVIRA and per-cell orders are approximately first order.

\clearpage
\subsubsection{Ellipse Benchmark}

\begin{figure}[!htbp]
  \centering
  \includegraphics[height=0.78\textheight,width=0.95\textwidth,keepaspectratio]{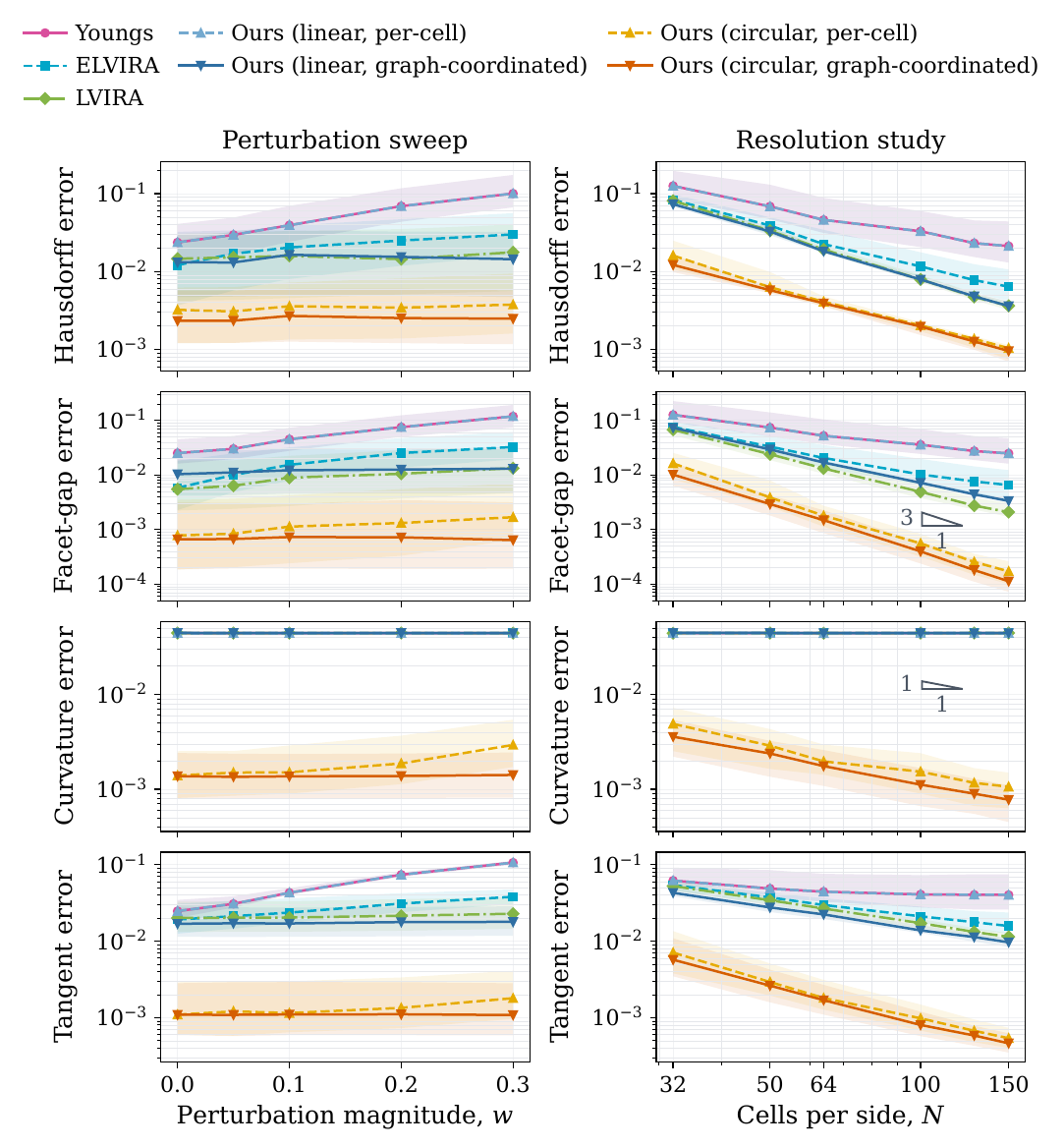}
  \caption{\textbf{Full ellipse comparison across all methods and diagnostics, including per-cell and graph-coordinated linear and circular variants.}
  Relative to LVIRA, graph-coordinated circular facets reduce the median Hausdorff distance by $3.7$--$6.7\times$ and the median facet gap by $6.7$--$19\times$ across resolutions.
  The observed convergence orders for facet gap and curvature error, estimated from log--log slopes over the resolution sweep, are $2.92$ and $1.00$, respectively.}
  \label{fig:appendix_ellipses_all_methods}
\end{figure}

As shown in \Cref{fig:appendix_ellipses_all_methods}, graph coordination improves circular reconstruction on the varying-curvature ellipse benchmark.
Across the resolution sweep, it has median Hausdorff distances from $9.53\times10^{-4}$ to $1.22\times10^{-2}$ and median facet gaps from $1.13\times10^{-4}$ to $1.02\times10^{-2}$.
These errors are $3.7$--$6.7\times$ and $6.7$--$19\times$ lower than those of LVIRA.
Relative to the per-cell circular method, graph coordination reduces the median facet gap by $17$--$38\%$ across resolutions and by $15$--$62\%$ across perturbation magnitudes. See \Cref{fig:results_ellipses_higher_order} for a comparison on Cartesian meshes.

\clearpage
\subsubsection{Zalesak Benchmark}

\begin{figure}[!htbp]
  \centering
  \includegraphics[width=\textwidth]{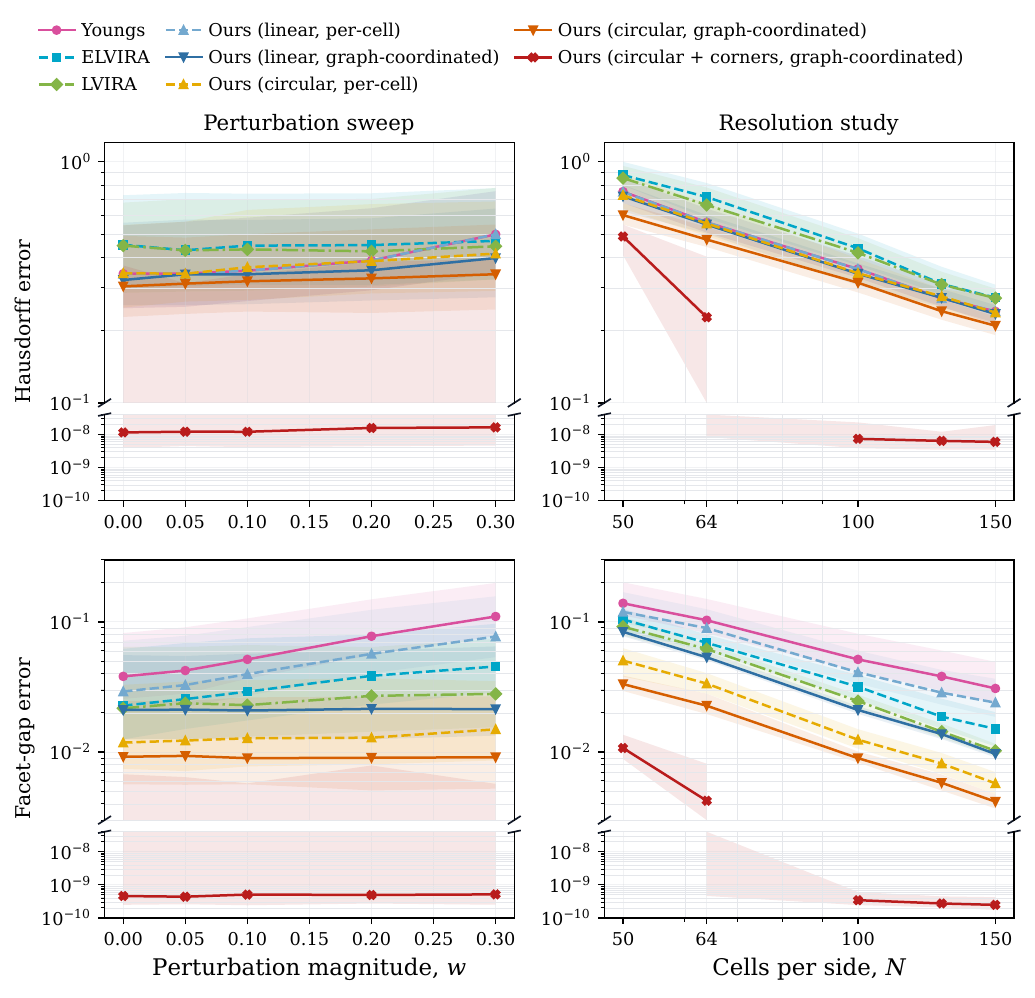}
  \caption{\textbf{Full Zalesak comparison across all methods, including per-cell and graph-coordinated circular variants and the graph-coordinated corner-aware method.}
  Relative to LVIRA, graph-coordinated circular facets reduce the median Hausdorff distance by $22$--$30\%$ and the median facet gap by $2.5$--$2.8\times$ across resolutions.
  Adding explicit corners reaches the numerical floor from $N=100$ onward.
  Broken y-axes separate the numerical floor from larger error ranges, and curves are not connected across omitted intervals.
  }
  \label{fig:appendix_zalesak_all_methods}
\end{figure}

Explicit corner facets recover the Zalesak interface to the numerical floor once the cusps are resolved (\Cref{fig:appendix_zalesak_all_methods}).
For $N=100$--$150$, the corner-aware method has median Hausdorff distances from $5.97\times10^{-9}$ to $7.37\times10^{-9}$ and median facet gaps from $2.47\times10^{-10}$ to $3.43\times10^{-10}$.
Before corners are added, graph-coordinated circular facets reduce median Hausdorff distance by $22$--$30\%$ and median facet gap by $2.5$--$2.8\times$ relative to LVIRA.
The remaining coarse-resolution error tail is associated primarily with unresolved cusp geometry rather than perturbation magnitude.
At $N=50$ and $64$, cases without accepted line--line corner cells account for around $98\%$ of the high-error cases.

\clearpage

\subsection{Multi-Neighbor Merge Example}
\label{appendix_static_usage}

\begin{figure}[t]
  \centering
  \includegraphics[width=0.84\textwidth]{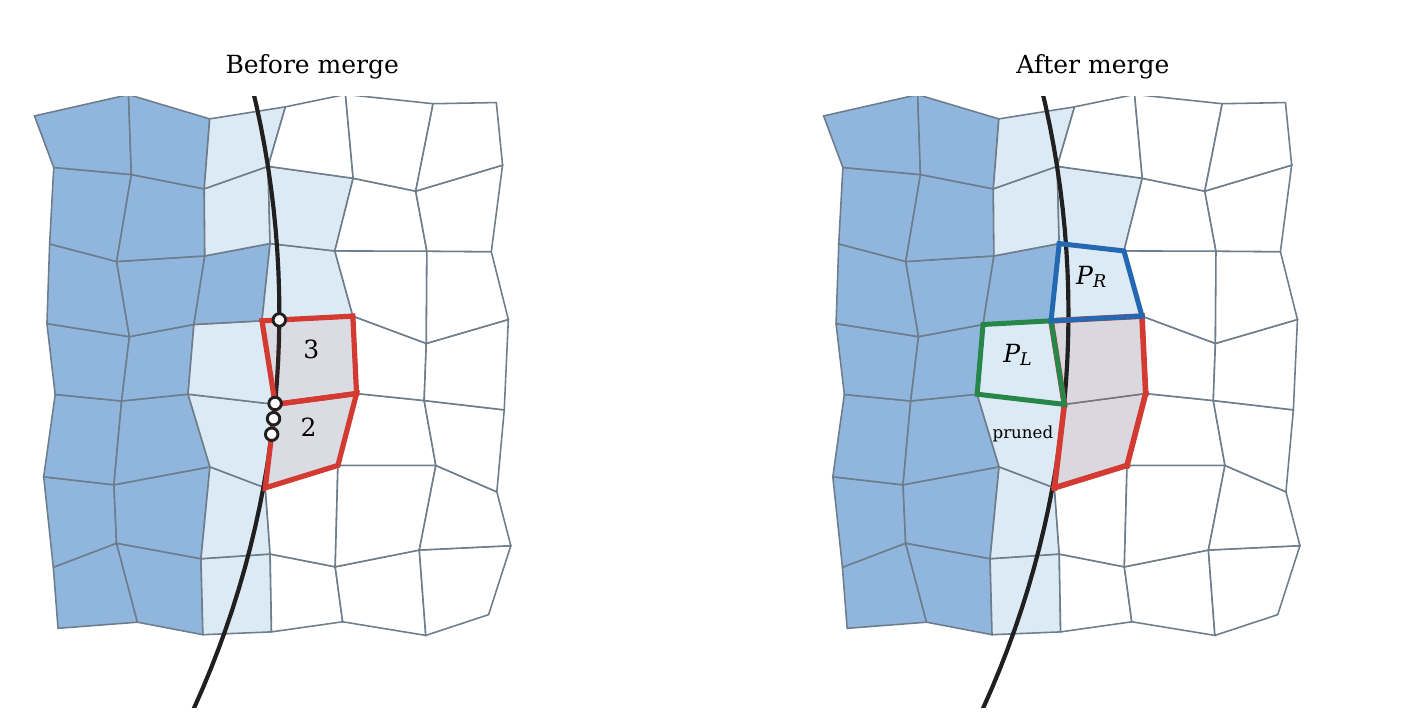}
  \caption{\textbf{Multi-neighbor merge in the circle benchmark ($N=128$, $w=0.2$).}
  Before merging, the highlighted cells have two and three candidate neighbors, respectively, and the generating interface crosses each cell boundary twice.
  After merging the pair, the graph-coordinated method selects the two exterior supports, $P_L$ and $P_R$, and prunes the unused continuation.}
  \label{fig:appendix_actual_multineighbor_merge}
\end{figure}

Most merged components contain a cell with only one initial mixed neighbor.
Figure~\ref{fig:appendix_actual_multineighbor_merge} shows the rarer case in which orientation propagation leaves three candidates.
Merging the highlighted cells converts the ambiguity into two exterior supports from which the interface can be reconstructed.

\subsection{Additional Qualitative Static Examples}

\begin{figure}[tbp]
  \centering
  \includegraphics[height=0.84\textheight,width=\textwidth,keepaspectratio]{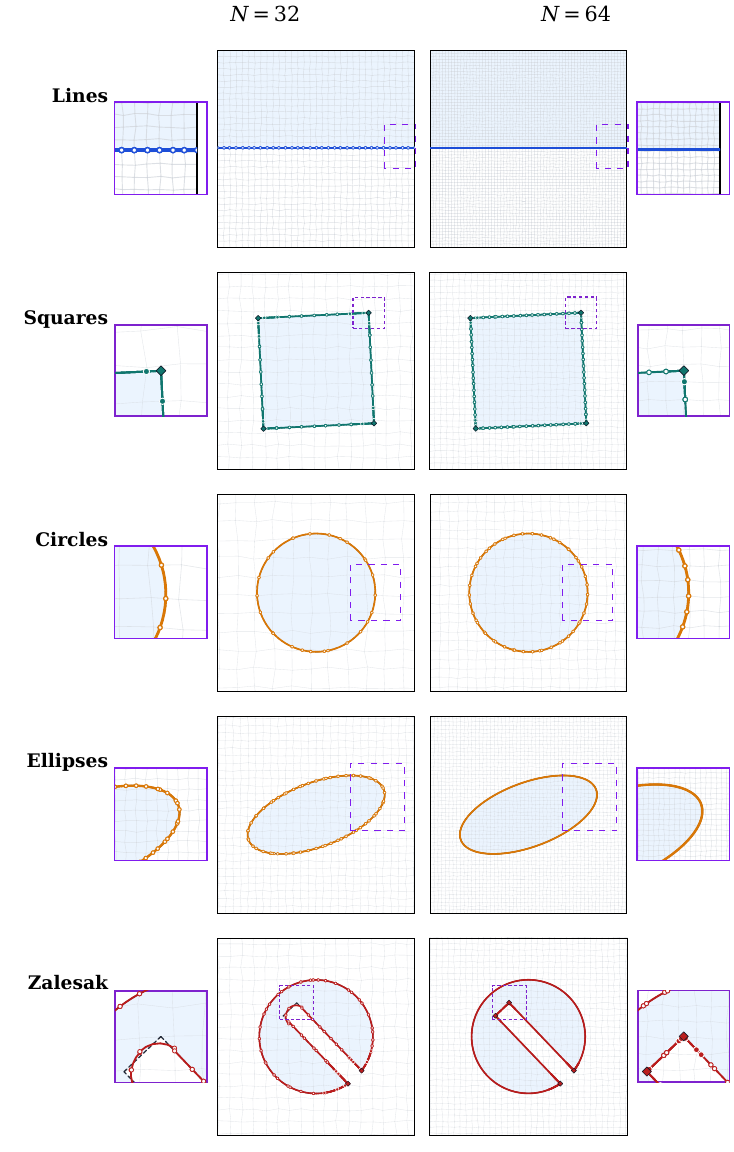}
  \caption{\textbf{Qualitative resolution study on perturbed Cartesian meshes.}
  Rows show the line, square, circle, ellipse, and Zalesak benchmarks at $N=32$ and $64$.
  The displayed reconstructions use the graph-coordinated variant of the primitive family evaluated for each benchmark in \Cref{sec:results}.
  Purple dashed boxes identify the regions enlarged in the adjacent spyglasses.}
  \label{fig:appendix_resolution_examples}
  \label{fig:appendix_line_resolution}
  \label{fig:appendix_square_resolution}
  \label{fig:appendix_circle_resolution}
  \label{fig:appendix_ellipse_resolution}
  \label{fig:appendix_zalesak_resolution}
\end{figure}

Figure~\ref{fig:appendix_resolution_examples} collects perturbed-mesh reconstructions at two representative resolutions.
The coarser column shows which geometric features are already identifiable at $N=32$, while the finer column shows the corresponding reconstruction at $N=64$.
Lines and circles are recovered to the geometric solver floor, square corners sharpen under refinement, and the varying curvature of the ellipse is followed more closely.
For the Zalesak disk, the $N=32$ slot remains slightly under-resolved because a consistent neighboring support cannot always be identified.
The slot geometry is resolved at $N=64$.

\begin{figure}[p]
  \centering
  \includegraphics[height=0.82\textheight,width=\textwidth,keepaspectratio]{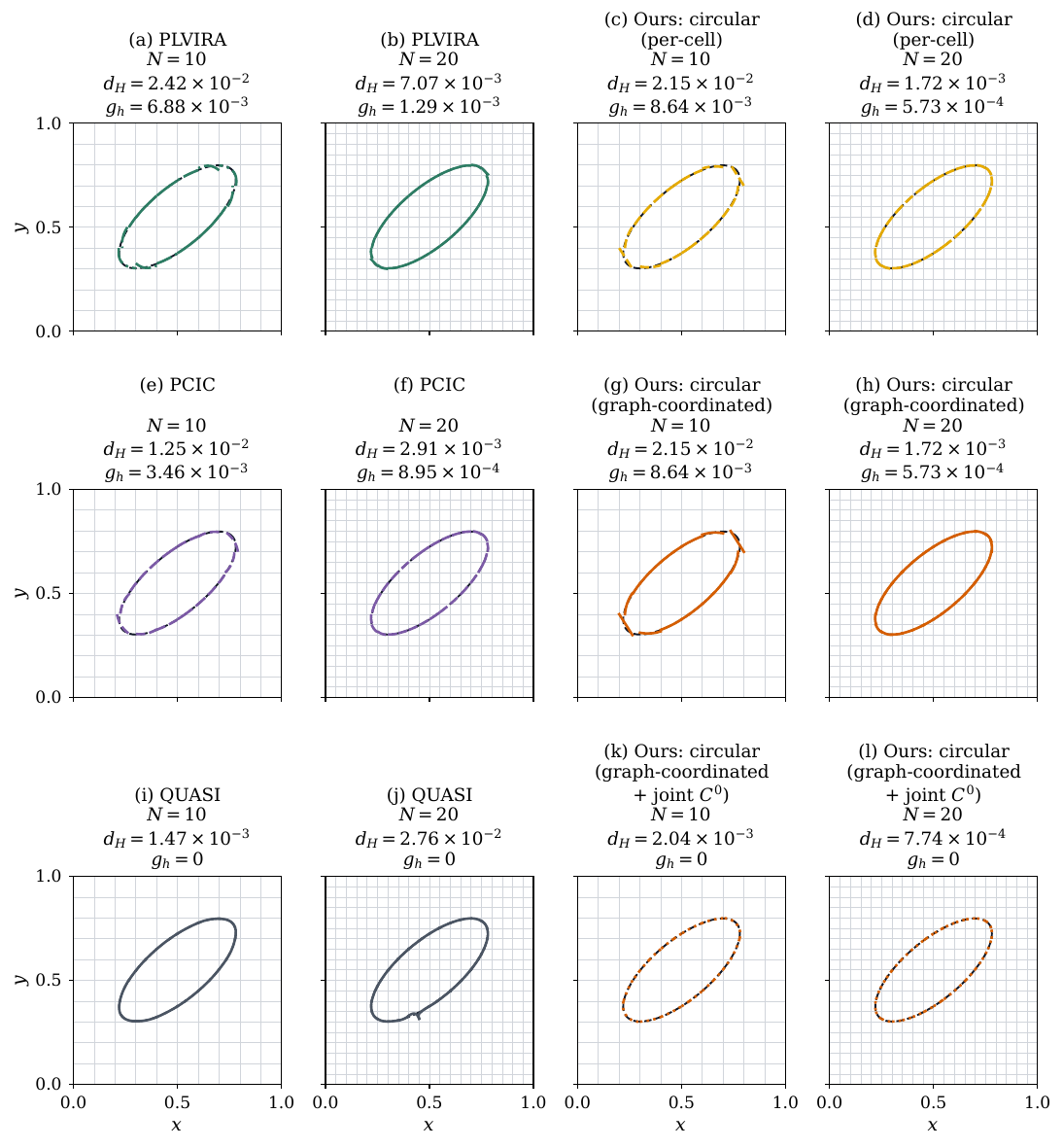}
  \caption{\textbf{Representative reconstructions for the coarse Cartesian ellipse test of Maity et al.~\citep{Maity2020}.}
  Each pair of columns shows $N=10$ and $N=20$ for one method, with the reproduced higher-order baselines on the left and our circular variants on the right.
  The dashed black curve is the generating ellipse, and the colored curves are the reconstructions.
  Each panel reports the symmetric Hausdorff distance $d_H$ and mean facet gap $g_h$ in unit-square coordinates.
  At $N=10$, QUASI and our method with joint $C^0$ refinement most closely follow the generating interface.
  At $N=20$, our circular variants give the lowest Hausdorff errors, while joint $C^0$ refinement also eliminates the facet gaps.
  On this regular Cartesian example, our per-cell and graph-coordinated circular variants choose the exact same interface orientations and thus produce identical reconstructions.}
  \label{fig:appendix_maity_ellipse}
\end{figure}

In Figure~\ref{fig:appendix_maity_ellipse}, we reconstruct an ellipse with semiaxes $a=\sqrt{0.12}$ and $b=\sqrt{0.02}$ on the unit square with resolutions $N=10$ and $N=20$, following the low-resolution test in Figure~10 of Maity et al.~\citep{Maity2020}. We visualize the reconstructions from our method using circular facets and compare to higher-order baselines.
At $N=10$, QUASI and our graph-coordinated method with joint $C^0$ refinement most closely follow the ellipse.
Their Hausdorff distances are $1.47\times10^{-3}$ and $2.04\times10^{-3}$, respectively, which are $8.5\times$ and $6.1\times$ lower than PCIC, the next-best reconstruction.
Both methods have zero facet gap.
At $N=20$, our graph-coordinated circular reconstruction has Hausdorff distance $1.72\times10^{-3}$ and facet gap $5.73\times10^{-4}$, which are $1.7\times$ and $1.6\times$ lower than PCIC, respectively.
Joint $C^0$ refinement further lowers the Hausdorff distance to $7.74\times10^{-4}$, a $3.8\times$ improvement over PCIC, and reduces the facet gap to zero.
QUASI also has zero facet gap, but a local oscillation increases its Hausdorff distance to $2.76\times10^{-2}$.

\begin{figure}[p]
  \centering
  \includegraphics[height=0.84\textheight,width=\textwidth,keepaspectratio]{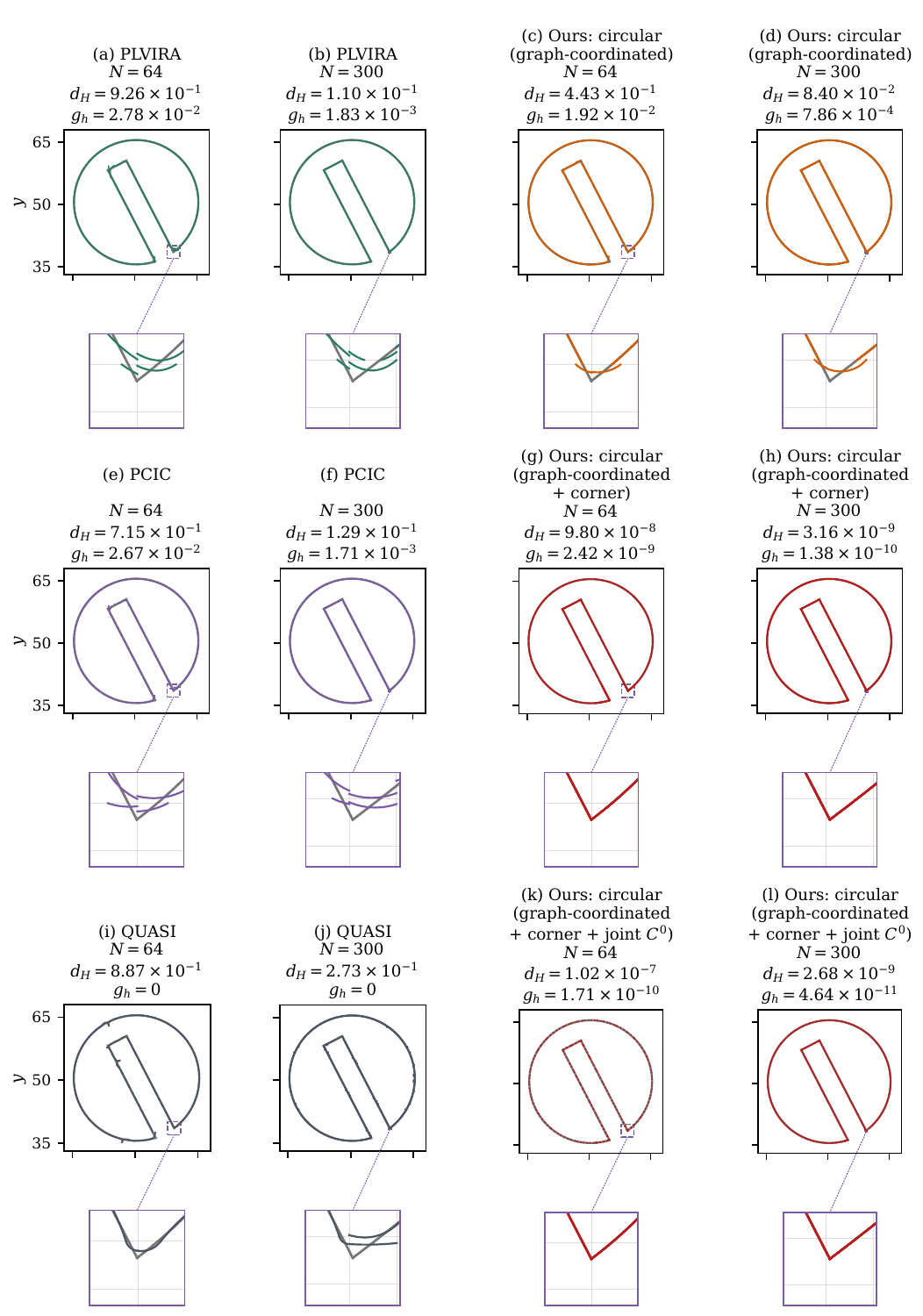}
  \caption{\textbf{Representative higher-order reconstructions of the Zalesak interface on Cartesian meshes.}
  Each pair of columns shows $N=64$ and $N=300$ for one method, with PLVIRA, PCIC, and QUASI on the left and our graph-coordinated circular, corner-aware, and joint $C^0$ variants on the right.
  Gray curves show the generating interface, and colored curves show the reconstructions.
  Purple boxes identify the regions enlarged in the separate spyglasses below, each spanning two cell widths.
  Each panel reports the symmetric Hausdorff distance $d_H$ and mean facet gap $g_h$.}
  \label{fig:appendix_zalesak_higher_order_gallery}
\end{figure}

In Figure~\ref{fig:appendix_zalesak_higher_order_gallery}, we reconstruct a representative Cartesian Zalesak configuration at $N=64$ and $300$ using PLVIRA, PCIC, QUASI, and our graph-coordinated circular variants.
The spyglasses compare the reconstruction of a sharp slot junction.
At $N=64$, the circular-only methods round the junction, whereas our corner-aware method recovers it with Hausdorff distance $9.80\times10^{-8}$.
Joint $C^0$ refinement reduces the mean facet gap from $4.24\times10^{-9}$ to $1.71\times10^{-10}$ while preserving the recovered corner geometry.
At $N=300$, the corner-aware method has Hausdorff distance $3.16\times10^{-9}$ and mean facet gap $1.38\times10^{-10}$, while the circular-only methods retain visible error at the junction.
These panels illustrate one configuration: among the $25$ configurations, the corner-aware method has Hausdorff error below $10^{-6}$ in $24$ cases at $N=100$ and all $25$ cases at $N=300$.

\clearpage

\clearpage

\clearpage

\subsection{\texorpdfstring{$C^0$}{C0} Continuity Enforcement Study}
\label{appendix_c0_study}

\begin{table}[t]
\centering
\begin{minipage}{0.78\textwidth}
  \centering
  \small
  \caption{\textbf{Residual endpoint gaps after each $C^0$ correction step.} Entries give the percentage of eligible line and arc joins whose endpoint gap remains above $\varepsilon_{\mathrm{gap}}=10^{-8}$. Explicit corner joins are excluded.}
  \label{tab:appendix_c0_stage_incidence}
  {\setlength{\tabcolsep}{10pt}
  \begin{tabular}{l|rr}
    \hline
    \textbf{Benchmark} & \shortstack{\textbf{After parallel}\\\textbf{local pass}} & \shortstack{\textbf{After joint}\\\textbf{refinement}} \\
    \hline
    \textbf{Ellipses} & $16.161\%$ & $0.0183\%$ \\
    \textbf{Zalesak}  & $3.639\%$  & $0.3715\%$ \\
    \hline
    \textbf{All}      & $10.106\%$ & $0.1891\%$ \\
    \hline
  \end{tabular}}
\end{minipage}
\end{table}

\begin{figure}[t]
  \centering
  \includegraphics[width=0.4\textwidth,trim=280 0 0 0,clip]{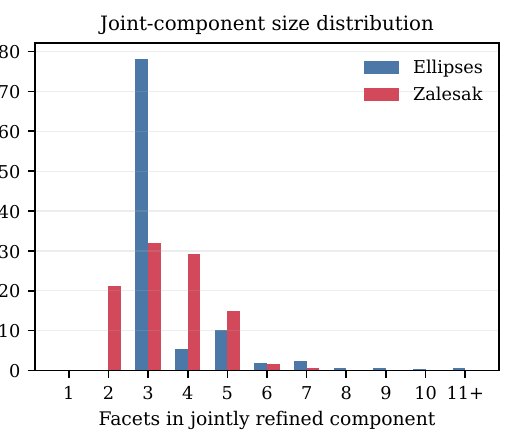}
  \caption{\textbf{Size of the connected components passed to joint $C^0$ refinement.}
  Bars show the component-size distribution separately for the ellipse and Zalesak benchmarks.
  The histogram summarizes $9{,}215$ refinement components from $1{,}375$ benchmark instances, and the bar heights report percentages rather than case counts.
  Across both benchmarks, a component contains a median of $3$ facets and a mean of $3.55$ facets, and $90\%$ contain at most $5$ facets.}
  \label{fig:appendix_c0_component_sizes}
\end{figure}

\Cref{tab:appendix_c0_stage_incidence} reports the percentage of eligible joins with residual endpoint gaps after each correction step.
\Cref{fig:appendix_c0_component_sizes} shows the sizes of the connected components passed to joint $C^0$ refinement.
These results supplement the discussion of $C^0$ refinement in \Cref{sec:results-component-effects}.

\begin{figure}[p]
  \centering
  \IfFileExists{figs/cameraready/ellipse_joint_c0_comparison.pdf}{%
    \includegraphics[height=0.72\textheight,width=\textwidth,keepaspectratio]{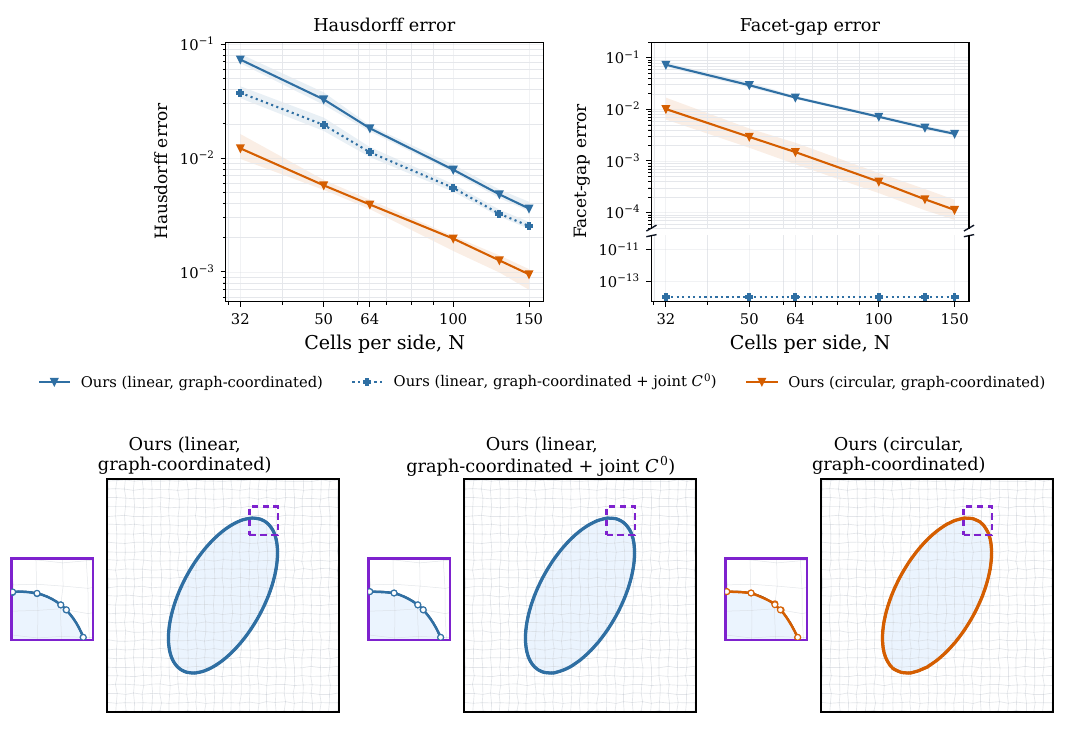}%
  }{%
    \fbox{\parbox[c][0.42\textheight][c]{0.88\textwidth}{\centering
      Ellipse joint-$C^0$ figure pending.}}
  }
  \caption{\textbf{Effect of iterative joint $C^0$ refinement on the ellipse benchmark.}
  The upper row reports median Hausdorff distance and facet gap versus cells per side.
  At each resolution, the median and interquartile range are computed directly over all $125$ combinations of perturbation magnitude and benchmark configuration.
  The lower row compares representative reconstructions, with the insets centered at a point of maximum curvature and enlarged to the left of each parent panel.
  Joint refinement closes the facet gaps while preserving the participating cells' volume constraints.
  Broken y-axes separate the numerical floor from larger error ranges, and curves are not connected across omitted intervals.
  }
  \label{fig:appendix_ellipses_c0_metrics}
  \label{fig:appendix_ellipses_c0_representative}
\end{figure}

\begin{figure}[p]
  \centering
  \IfFileExists{figs/cameraready/zalesak_joint_c0_comparison.pdf}{%
    \includegraphics[height=0.72\textheight,width=\textwidth,keepaspectratio]{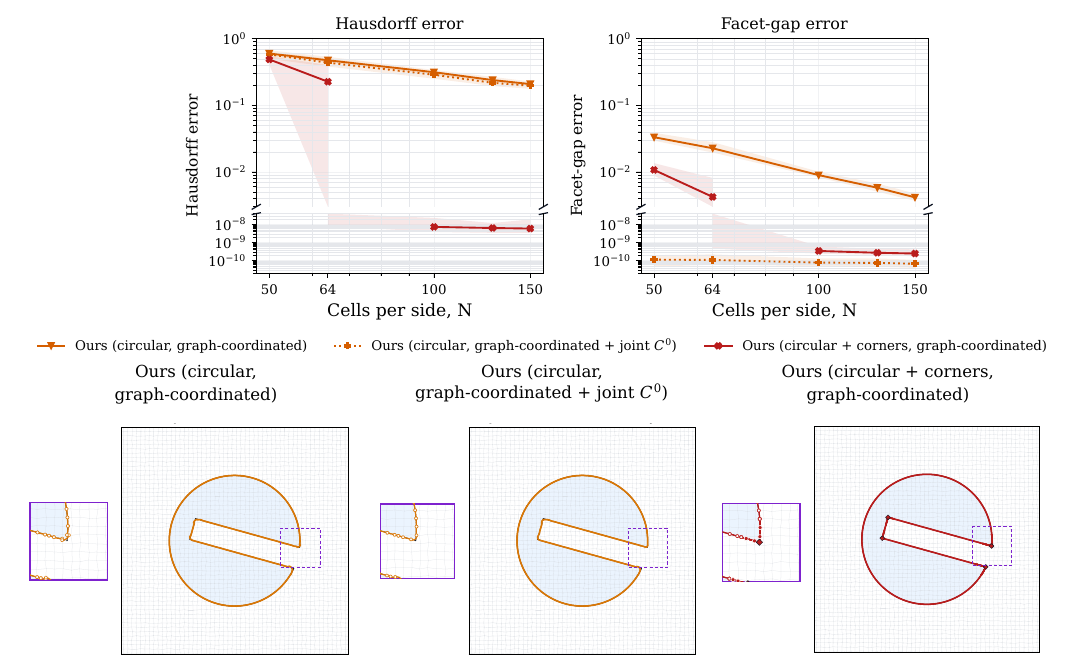}%
  }{%
    \fbox{\parbox[c][0.42\textheight][c]{0.88\textwidth}{\centering
      \textbf{Zalesak} joint-$C^0$ figure pending.}}
  }
  \caption{\textbf{Effect of iterative joint $C^0$ refinement on the Zalesak benchmark.}
  The upper row reports median Hausdorff distance and facet gap versus cells per side.
  At each resolution, the median and interquartile range are computed directly over all $125$ combinations of perturbation magnitude and benchmark configuration.
  Both error panels use the same retained ranges around the omitted logarithmic interval.
  The lower row compares representative reconstructions, with insets enlarged to the left of the parent panel.
  Joint refinement closes the facet gaps, while explicit corner primitives resolve the sharp line--arc transitions.
  Broken y-axes separate the numerical floor from larger error ranges, and curves are not connected across omitted intervals.
  }
  \label{fig:appendix_zalesak_c0_metrics}
  \label{fig:appendix_zalesak_c0_representative}
\end{figure}

Figures~\ref{fig:appendix_ellipses_c0_metrics} and~\ref{fig:appendix_zalesak_c0_metrics} supplement the aggregate results in \Cref{sec:results-component-effects} with representative reconstructions.
The iterative joint refinement adjusts neighboring endpoints and facet curvatures subject to the participating cells' volume constraints.
If no admissible state is found, the original facets are retained.

\clearpage

\clearpage

\section*{Declaration of generative AI in the manuscript preparation process}

During the preparation of this work, the authors used OpenAI Codex to assist with software development, plotting and diagram code, manuscript organization, and language editing.
Figures were generated programmatically from benchmark data or explicit geometric constructions.
The authors reviewed and edited the resulting content and take full responsibility for the content of the published article.

\bibliographystyle{unsrtnat}
\bibliography{mybibfile} 

\end{document}